\documentclass{aa}  

\usepackage{color}
\usepackage{graphicx}
\usepackage{amsmath}	
\usepackage{amssymb}	
\usepackage{mathtools}
\usepackage{txfonts}
\usepackage{booktabs}
\usepackage{natbib,twoopt}
\usepackage{float}
\usepackage[caption = false]{subfig}
\usepackage[breaklinks=true]{hyperref}
\hypersetup{
  colorlinks=true,   
  citecolor=blue,    
  urlcolor=blue,     
  linkcolor=blue,     
}
\bibpunct{(}{)}{;}{a}{}{,} 
\usepackage{lscape}

\makeatletter
\newcommandtwoopt{\citeads}[3][][]{\href{http://adsabs.harvard.edu/abs/#3}%
{\def\hyper@linkstart##1##2{}%
\let\hyper@linkend\@empty\citealp[#1][#2]{#3}}}
\newcommandtwoopt{\citepads}[3][][]{\href{http://adsabs.harvard.edu/abs/#3}%
{\def\hyper@linkstart##1##2{}%
\let\hyper@linkend\@empty\citep[#1][#2]{#3}}}
\newcommandtwoopt{\citetads}[3][][]{\href{http://adsabs.harvard.edu/abs/#3}%
{\def\hyper@linkstart##1##2{}%
\let\hyper@linkend\@empty\citet[#1][#2]{#3}}}
\newcommandtwoopt{\citeyearads}[3][][]%
{\href{http://adsabs.harvard.edu/abs/#3}
{\def\hyper@linkstart##1##2{}%
\let\hyper@linkend\@empty\citeyear[#1][#2]{#3}}}
\makeatother

\newcommand{\fo}{\log F_{\rm UV}}
\newcommand{\fx}{\log F_{\rm X}}
\newcommand{\Lo}{L_{\rm UV}}
\newcommand{\Fo}{F_{\rm UV}}
\newcommand{\Lx}{L_{\rm X}}
\newcommand{\Fx}{F_{\rm X}}

\newcommand{\aox}{\alpha_{\rm ox}}
\newcommand{\mbh}{M_{\rm BH}}
\newcommand{\lbol}{L_{\rm bol}}

\newcommand{\ledd}{\lambda_{\rm edd}}

\newcommand{\gammax}{\Gamma_{\rm X}}

\newcommand{\fmin}{F_{\rm min}}

\newcommand{\dm}{{\rm DM}}
\newcommand{\om}{\Omega_{\rm M}}
\newcommand{\ol}{\Omega_\Lambda}
\newcommand{\ebv}{E(B-V)}
    \newcommand{\xmm}{\emph{XMM--Newton}\xspace}

    \newcommand{\chandra}{\emph{Chandra}\xspace}

    \newcommand{\qsfit}{\texttt{QSFit}\xspace}
   \newcommand{\xspec}{\textsc{xspec}\xspace}
   \newcommand{\emcee}{\textsc{emcee}\xspace}
   
\DeclareRobustCommand{\ion}[2]{%
\relax\ifmmode
\ifx\testbx\f@series
{\mathbf{#1\,\mathsc{#2}}}\else
{\mathrm{#1\,\mathsc{#2}}}\fi
\else\textup{#1\,{\mdseries\textsc{#2}}}%
\fi}

\DeclareCaptionFormat{cont}{#1 (cont.)#2#3\par}

\newcommand{\rev}[1]{{ #1}}
\newcommand{\revs}[1]{{ #1}}

\begin{document}

   \title{Quasars as standard candles}
   \subtitle{III. Validation of a new sample for cosmological studies}

   \author{E. Lusso\inst{1,2}\thanks{\email{elisabeta.lusso@unifi.it}}, G. Risaliti\inst{1,2}, E. Nardini\inst{1,2}, G. Bargiacchi\inst{1}, M. Benetti\inst{3,4,5}, S. Bisogni\inst{6}, S. Capozziello\inst{3,4,5,7,8}, F. Civano\inst{9}, L. Eggleston\inst{10}, M. Elvis\inst{9}, G. Fabbiano\inst{9}, R. Gilli\inst{11}, A. Marconi\inst{1,2}, M. Paolillo\inst{3,4,13}, E. Piedipalumbo\inst{3,4}, F. Salvestrini\inst{12,11}, M. Signorini\inst{1} \and C. Vignali\inst{12,11}}
\institute{
$^{1}$Dipartimento di Fisica e Astronomia, Universit\`a di Firenze, via G. Sansone 1, 50019 Sesto Fiorentino, Firenze, Italy\\
$^{2}$INAF -- Osservatorio Astrofisico di Arcetri, 50125 Florence, Italy\\
$^{3}$Dipartimento di Fisica, Universit\`a degli studi di Napoli Federico II, via Cinthia, 80126 Napoli, Italy\\
$^{4}$INFN -- Sezione di Napoli, via Cinthia 9, 80126 Napoli, Italy\\
$^{5}$Scuola Superiore Meridionale, Universit\`a di Napoli Federico II, Largo San Marcellino 10, 80138 Napoli, Italy\\
$^{6}$INAF -- Istituto di Astrofisica Spaziale e Fisica Cosmica Milano, via Corti 12, 20133 Milano, Italy \\
$^{7}$Gran Sasso Science Institute, Via F. Crispi 7, I-67100, L'Aquila, Italy\\
$^{8}$Laboratory for Theoretical Cosmology, Tomsk State University of Control Systems and Radioelectronics (TUSUR), 634050 Tomsk, Russia\\
$^{9}$Center for Astrophysics | Harvard \& Smithsonian, 60 Garden Street, Cambridge, MA 02138, USA \\
$^{10}$Yale University, Department of Computer Science, 51 Prospect St, New Haven, CT 06511\\
$^{11}$INAF -- Osservatorio di Astrofisica e Scienza dello Spazio di Bologna, via Gobetti 93/3 - 40129 Bologna - Italy\\
$^{12}$Dipartimento di Astronomia, Universit\`a degli Studi di Bologna, via Gobetti 93/2, 40129 Bologna, Italy\\
$^{13}$INAF -- Osservatorio Astronomico di Capodimonte, via Moiariello 16, 80131 Napoli, Italy\\
}
  \titlerunning{Quasars as standard candles III}
  \authorrunning{Lusso, E. et al.}

   \date{Received 13 July 2020 / Accepted 19 August 2020}

 
  \abstract{We present a new catalogue of $\sim$2,400 optically selected quasars with spectroscopic redshifts and X-ray observations from either \chandra or \xmm. The sample can be used to investigate the non-linear relation between the UV and X-ray luminosity of quasars, and to build a Hubble diagram up to redshift $z\sim7.5$. We selected sources that are neither reddened by dust in the optical/UV nor obscured by gas in the X-rays, and whose X-ray fluxes are free from flux-limit related biases. After checking for any possible systematics, we confirm, in agreement with our previous works, that (i) the X-ray to UV relation provides distance estimates matching those from supernovae up to $z\sim1.5$, and (ii) its slope shows no redshift evolution up to $z\sim5$. We provide a full description of the methodology for testing cosmological models, further supporting a trend whereby the Hubble diagram of quasars is well reproduced by the standard flat $\Lambda$CDM model up to $z\sim1.5$--2, but strong deviations emerge at higher redshifts. Since we have minimized all non-negligible systematic effects, and proven the stability of the $\Lx-\Lo$ relation at high redshifts, we conclude that an evolution of the expansion rate of the Universe should be considered as a possible explanation for the observed deviation, rather than some systematic (redshift-dependent) effect associated with high-redshift quasars.} 


   \maketitle
%

\section{Introduction}

Quasars are the most luminous and persistent energy sources in our Universe. As they can be observed up to redshift $\simeq7.5$ \citepads{banados2018}, when the age of the Universe was less than $\simeq$700 million years, quasars bear an extraordinary potential as cosmological probes. 
Several techniques making use of empirical correlations between quasar properties have been proposed to determine cosmological parameters such as the dark matter ($\om$) and dark energy ($\ol$) content of the Universe.
Examples include the relation between the continuum luminosity and the  emission-line equivalent width \citepads{baldwin1977}, or with the radius of the quasar broad-line region established via reverberation mapping \citepads{Watson2011}. Another luminosity distance estimator combines the correlation between the quasar X-ray variability amplitude and its black hole (BH) mass \citepads{lafranca2014}. None the less, these correlations are affected by too large a dispersion (up to $\simeq$0.6 dex), and are typically applicable over a limited redshift range. 
Other promising methods employ geometric distances through, again, the broad-line region radius via reverberation mapping \citepads{elviskarovska2002}, the luminosity of super-Eddington accreting quasars \citepads{wang2013}, the eigenvector formalism in the quasar main sequence \citepads{marzianisulentic2014}, \rev{or the combination of spectroastrometry and reverberation mapping \citep{wang2020}}. All these techniques still need some level of refinement, \rev{and/or much higher sample statistics}, to be competitive as cosmological tools.

Since 2015, our group has been developing a new technique that hinges upon the observed non-linear relation between the ultraviolet (at 2500 \AA, $\Lo$) and the X-ray (at 2 keV, $\Lx$) emission in quasars (e.g.~\citeads{avnitananbaum79}, $\Lx\propto\Lo^\gamma$, with $\gamma\simeq0.6$) to provide an independent measurement of their distances, thus turning quasars into {\it standardizable} candles and extending the distance modulus--redshift relation (or the so-called {\it Hubble-Lema\^itre diagram}) of supernovae Ia to a redshift range still poorly explored ($z>2$; \citeads{rl15}). 

The applicability of our technique is based upon the understanding that most of the observed dispersion in the $\Lx-\Lo$ relation is not intrinsic to the relation itself but due to observational issues, such as X-ray absorption by gas, UV extinction by dust, calibration uncertainties in the X-rays \citepads{2019AN....340..267L}, variability, and selection biases associated with the flux limits of the different samples. In fact, with an optimal selection of {\it clean} sources (i.e. where we can measure the {\it intrinsic} UV and X-ray quasar emission), the dispersion drops from 0.4 to $\simeq$0.2 dex \citepads{lr16,lr17}.

A key consequence of this technique is that the $\Lx-\Lo$ relation must be the manifestation of a universal mechanism at work in the quasar engines, although the details on the physical process originating this relation are still unknown \citepads[e.g.][]{1991ApJ...380L..51H,1993ApJ...413..507H,1994ApJ...432L..95H,nicastro2000,2003MNRAS.341.1051M,lr17,arcodia2019}. 

The main results of our work are that {\it (1)} the distance modulus-redshift relation of quasars at $z < 1.4$ is in agreement with that of supernovae Ia and with the concordance $\Lambda$CDM model \citep{rl15,rl19,lusso19}, yet {\it (2)} a deviation from $\Lambda$CDM emerges at higher redshift, with a statistical significance of about 4$\sigma$. If we interpret the latter result by considering an evolution of the dark energy equation of state in the form $w(z)=w_0+w_a\times z/(1+z)$, the data suggest that the dark energy density is increasing with time \citep{rl19,lusso19}. 

As our approach may still have shortcomings, we need to demonstrate that the observed deviation from $\Lambda$CDM at redshift $>2$ is neither driven by systematics in the quasar sample selection nor by the procedure adopted to fit the quasar Hubble-Lema\^itre diagram.
To build a quasar sample that can be utilised for cosmological purposes, both X-ray and UV data are required to cover the rest-frame 2 keV and 2500 \AA. 
The choice of these monochromatic luminosities is rather arbitrary, and mostly based on historical reasons. It is possible that the $\Lx-\Lo$ relation is tighter with a different choice of the indicators of UV and X-ray emission (see e.g. \citeads{young2010}). A careful analysis of this issue may also provide new insights on the physical process responsible for this relation. In the present analysis we will not investigate the possible alternatives, but we will focus on demonstrating that the commonly used relation is calibrated in a robust way and can be safely used as a tool to determine quasar distances.
In this third paper of our series, we thus mainly concentrate on the quasar sample selection, whilst we defer a detailed analysis of the cosmographic fitting technique we adopted in a forthcoming publication.

At the time of writing, the most extended spectroscopic coverage in the UV is given by the \textit{Sloan Digital Sky Survey} \citepads{paris2018}, which supplies more than $\sim$500,000 quasars with spectroscopic redshift up to $z\sim7$. This sample needs to be cross-matched with the current X-ray catalogues, namely the {\it Chandra X-ray Catalogue} \citep[CXC2.0,][]{evans2010} and the 4XMM Data Release 9 \citep{webb2020}, which contain all the X-ray sources detected by the \chandra and \xmm observatories that are publicly available in the respective archives. These data need to be complemented by dedicated pointed observations of active galactic nuclei\footnote{In the following we will refer to AGN and quasars indistinctly.} (AGN) at both low ($z<0.1$) and high ($z>3$) redshifts to extend the coverage and increase the sample statistics in the distance modulus--redshift relation. 

The main aims of this manuscript are to discuss in detail all the criteria required to select a homogeneous sample of quasars for cosmological purposes from the above archives, and to present the key steps in fitting the Hubble-Lema\^itre diagram that can be adopted to reproduce our results. 
Specifically, we will examine the effects on the sample selection and on the UV and X-ray flux measurements of {\it (1)} dust extinction and host-galaxy contamination, {\it (2)} gas absorption in the X-rays, and {\it (3)} Eddington bias.
We will identify the quasars that can be used for testing cosmological models, and investigate in depth all the possible systematics in the quasar Hubble-Lema\^itre diagram as a function of the contaminants mentioned above. 

Since a detailed spectroscopic UV and X-ray analysis can be carried out only for a relatively small number of sources, our latest quasar sample presented here still relies also on broadband photometry in both UV and X-rays, as most of the parameters currently employed in our works, e.g. monochromatic UV and X-ray fluxes, UV colours and X-ray slopes, are derived from the photometric spectral energy distribution (SED) of our sources. In the future, we plan to gradually move towards a full spectroscopic analysis, as spectroscopy can deliver cleaner measurements of the relevant parameters.

The paper is constructed as follows. In Section~\ref{The data set} we present the different data sets employed to build the main quasar sample and the procedures adopted to measure the UV fluxes and slopes from the photometry. Section~\ref{SED compilation} discusses how the photometric quasar SEDs are constructed. Section~\ref{2keVflux} is devoted to the presentation of our technique to compute the monochromatic X-ray emission and the photon index from the catalogued broadband flux values. In Section~\ref{Selection of a clean quasar sample} we discuss the several quality filters employed to select a homogeneous sample of quasars for cosmological purposes, whilst in Section~\ref{Analysis of the relation with redshift} we verify that the $\Lx-\Lo$ relation for the final ``best'' quasar sample does not evolve with redshift. 
Section~\ref{The Hubble diagram} presents the main steps adopted to fit the Hubble diagram using a model independent technique (i.e. cosmography), whilst in Section~\ref{Cosmological fits of the Hubble diagram} we fit the Hubble diagram with the most commonly used $\Lambda$CDM extension to test our fitting technique and to verify how different choices regarding the fitting method and the quasar sample selection affect the final results. In Section~\ref{Study of systematics in the Hubble diagram} we carry out an in-depth investigation on possible remaining systematics in the residuals of the quasar Hubble diagram, as a function of the parameters involved in the selection of the sample.
Finally, we summarise our work and main results in Section~\ref{Conclusions}.

Although we mainly use fluxes, whenever luminosity values are reported we have assumed a standard flat $\Lambda$CDM cosmology with $\om=0.3$ and $H_0=70$ km s$^{-1}$ Mpc$^{-1}$.

\section{The data set}
\label{The data set}
\begin{figure}
   \centering
   \includegraphics[width=\linewidth]{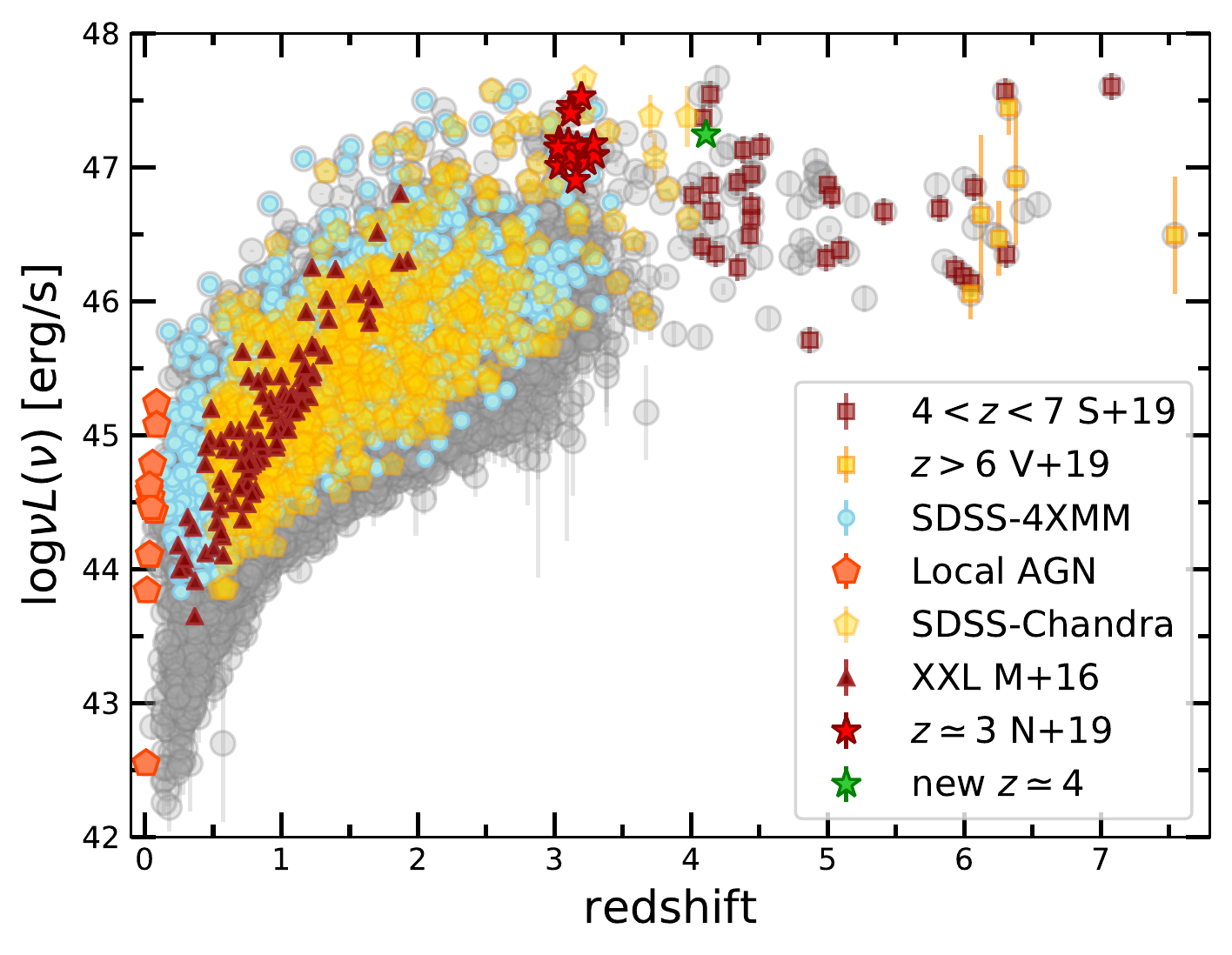}
   \caption{Distribution of luminosities at rest-frame 2500 \AA\ as a function of redshift for the main (grey points) and the selected (cleaned) samples (see \S\,\ref{samplefin}). Brown and yellow squares: high-$z$ sample (\citeads{salvestrini2019,vito2019}, see \S\,\ref{high redshift sample}), cyan points: SDSS--4XMM (\S\,\ref{SDSS-4XMM}), brown triangles: XMM--XXL (\citeads{menzel2016}, \S\,\ref{XXL}), orange pentagons: the local AGN sample (\S\,\ref{local}), red stars: $z\simeq3$ quasar sample (\citeads{nardini2019}, green star: new $z\simeq4$ quasar (\S\,\ref{z4}), gold pentagons: SDSS--\chandra (\S\,\ref{SDSS-Chandra}).}
              \label{loz}
    \end{figure}

The broad-line quasar sample we considered for our analysis has been assembled by combining seven different samples from both the literature and the public archives. The former group includes the samples at $z\simeq3.0$--3.3 by \citetads{nardini2019}, $4<z<7$ by \citetads{salvestrini2019}, $z>6$ by \citetads{vito2019}, and the XMM--XXL North quasar sample published by \citetads{menzel2016}. We then complemented this collection by including quasars from a cross-match of optical (i.e. the {\it Sloan Digital Sky Survey}) and X-ray public catalogues (i.e. \xmm and \chandra), which we refer to as SDSS--4XMM and SDSS--\chandra samples hereafter. Finally, we also added a local subset of AGN with UV (i.e. {\it International Ultraviolet Explorer}) data and X-ray archival information. The same order in which these samples have been introduced above is adopted as an order of priority to take into account all the possible overlaps. 
X-ray fluxes coming from pointed observations and medium/deep surveys (i.e. XMM--XXL) have been considered first, as they are generally more reliable. 
The main parent sample is composed by $\sim$19,000 objects spanning the redshift range $0.009<z<7.52$. In Figure~\ref{loz} we present the distribution of rest-frame 2500 \AA\ luminosities as a function of redshift for all the different quasar subsamples. A summary of the sample statistics is provided in Table~\ref{tbl1}, whilst below we describe in detail of how each sub-sample has been constructed. 
\begin{table}
\caption{Summary of sample statistics.}              
\label{tbl1}      
\centering                                      
\begin{tabular}{lcccc}          
\hline\hline                        
Sample & Initial & Main & Selected & Ref \\    
             & (1) & (2) & (3) \\
\hline                                   
    \xmm $z\simeq3$ &     29 &      29 &    14 & 1 \\      
    \xmm $z\simeq4$ &      1 &       1 &     1 & 2 \\      
    High-$z$        &     64 &      64 &    35 & 3 \\
    XXL             &    840 &     840 &   106 & 4 \\
    SDSS--4XMM      & 13,800 &   9,252 & 1,644 & 5 \\
    SDSS--\chandra  &  7,036 &   2,392 &   608 & 6 \\
    Local AGN       &     15 &      15 &    13 & 7\\
\hline                                             
    Total           & 21,785 &  12,593 & 2,421 & \\
\hline                                             
\end{tabular}
\tablefoot{
(1)~These number counts refer to the sample statistics before correcting for overlaps amongst the subsamples.
(2)~Sample statistics after accounting for overlaps and the quality pre-selection described in \S\,\ref{Selection of a clean quasar sample}. The order of priority decreases from the top to the bottom row. 
(3)~Sample statistics in the final cleaned quasar sample after the various filtering steps: see \S~\ref{Selection of a clean quasar sample} for details.
References for the various samples: 1: \citetads{nardini2019}; 2: see \S\,\ref{z4}; 3: \citetads{salvestrini2019} and \citetads{vito2019}; 4: \citetads{menzel2016}; 5: see \S\,\ref{SDSS-4XMM}; 6: Bisogni et al., to be submitted; 7: \S\,\ref{local}.
}
\end{table}

\subsection{The SDSS--4XMM sample}
\label{SDSS-4XMM}
The bulk of the data belongs to the {\it Sloan Digital Sky Survey} quasar catalogue, Data Release 14 (\citeads{paris2018}; SDSS--DR14 hereafter). The catalogue contains 526,356 optically selected quasars detected over 9376 deg$^2$ with robust identification and spectroscopic redshift. 
Firstly, we removed from the catalogue all quasars flagged as broad absorption lines (BALs, where sources with the BI\_CIV=0 flag are non-BALs) and kept all the objects with a measurement in all the SDSS magnitudes. 
This preliminary selection leads to 503,744 quasars. 

\rev{We note that the BAL classification in the SDSS--DR14 quasar catalogue is based on a fully automated detection procedure on \ion{C}{iv} absorption troughs for sources at $z > 1.57$. Hence, a number of BAL quasars might still be included in this preliminary sample. BAL quasars are often found in galaxies with red optical/UV colours and hard X-ray spectra \citepads[e.g.][]{gallagher2006}, the latter suggesting that their relative X-ray weakness could be primarily due to gas absorption. The selection criteria discussed in Section~\ref{Selection of a clean quasar sample} efficiently remove red/X-ray absorbed quasars, possibly excluding most unclassified BALs from the final sample.}

The photometric rest-frame spectral energy distribution (SED), whose derivation is discussed in Section~\ref{SED compilation}, is then used to define the parameters required to exclude radio-loud, dust-absorbed or host-galaxy contaminated sources. Also the rest-frame monochromatic luminosities at $2500$ \AA\ are obtained from the photometric SEDs. 
For comparison purposes, as in our previous works on this topic, we discard bright radio quasars through the same radio loudness parameter, $R$, as that used in \citetads{shen2011}, which is defined as the ratio of the rest-frame fluxes at 6 cm and 2500 \AA\ (i.e. $R=L_{\nu,6\rm{cm}}/L_{\nu,2500\AA}$). A quasar is then classified as radio-loud if $R>10$.
We computed $R$ for the 17,561 objects with a FIRST detection, 16,315 of which are indeed radio-loud and have been therefore excluded from our sample. 

To further remove powerful radio-loud quasars we considered the catalogue published by \citetads{2016MNRAS.462.2631M}, which is currently the largest available \rev{Mid-Infrared (WISE), X-ray (3XMM) and Radio (FIRST+NVSS) collection (MIXR)} of AGN and star-forming galaxies: 2,753 sources, 918 of which are considered radio-loud based on multiwavelength diagnostics (we refer to their paper for details). 
We excluded 349 quasars in our sample flagged as radio-loud in the MIXR catalogue within a matching radius of 3 arcsec. 
%
This yields 487,080 SDSS radio-quiet quasars with a $\Lo$ measurement. 

This SDSS quasar sample is then cross-matched with the latest \xmm source catalogue 4XMM--DR9 \citep{webb2020}. 
4XMM--DR9 is the fourth generation catalogue of serendipitous X-ray sources, 
which contains 810,795 detections (550,124 unique X-ray sources) made publicly available by 2018 December 18\footnote{http://xmmssc.irap.omp.eu/Catalogue/4XMM-DR9/4XMM-DR9\_Catalogue\_User\_Guide.html}. 
The net sky area covered (taking into account overlaps between observations) is $\sim$1152 deg$^2$, for a net exposure time of $\geq$1 ksec.

To select reliable X-ray detections, we have applied the following quality cuts in the 4XMM--DR9 catalogue: SUM\_FLAG$<$3 (low level of spurious detections), OBJ\_CLASS$\leq$3 (quality classification of the whole observation\footnote{For more details the reader should refer to the 4XMM catalogue user guide.}) and EP\_TIME$>$0 (EPIC exposure time available). These filters lead to 692,815 X-ray detections.
We have adopted a maximum separation of 3 arcsec to provide optical classification and spectroscopic redshift for all the cross-matched objects. 
This yields 22,196 \xmm observations: 13,858 unique sources (3,976 of which have $\geq2$ observations) covering the redshift range $0.056<z<4.306$. 

Following the results presented by \citet[][LR16 hereafter; see their Section~4]{lr16}, we decided to average all X-ray observations for sources with multiple detections that meet our selection cuts, including that associated with the Eddington bias (see \S~\ref{eddbias} for details). In this way, we reduce the effect of X-ray variability on the dispersion ($\sim0.12$ dex, see \S~4 in LR16) by using only unbiased detections.

For each \xmm observation, we have computed the EPIC sensitivity ($5\sigma$ minimum detectable flux) at 2 keV following a similar approach as in LR16. We first estimated the minimum detectable flux in the soft band for both pn and MOS as a function of the on-time\footnote{The total good (after flares removal) exposure time (in seconds) of the CCD over which the source is detected.} exposure following the relations plotted in Figure~3 by \citet{2001A&A...365L..51W}. The total MOS on-time exposure is the one with the largest exposure value between the two individual cameras, MOS1 and MOS2. We then corrected this sensitivity for the pn and MOS, using the same vignetting correction for both cameras at the energy of 1.5 keV, as a function of their respective off-axis angles, where the smaller value between the two individual cameras is again assumed for the MOS. 
The sensitivity at 2 keV ($\fmin$) is then estimated for both pn and MOS, assuming a photon index of 1.7, following the same approach as in LR16. Finally, we have prioritized the pn sensitivity flux values over the MOS when available. 

\subsection{The SDSS--\chandra sample}
\label{SDSS-Chandra}
To further increase the statistics, we also cross-matched the SDSS--DR14 quasar catalogue with the second release of the {\it Chandra Source Catalog} (CSC2.0). The CSC2.0\footnote{http://cxc.harvard.edu/csc2/} \citepads{evans2010} contains $\sim$315,000 X-ray sources observed in 10,382 \chandra ACIS and HRC-I imaging observations publicly released prior to 2015. A cross-match of these two catalogues, with a matching radius of 3 arcsec, leads to 7,036 unique objects.
The detailed analysis of this sample will be presented in a forthcoming publication (Bisogni et al., to be submitted). Briefly, from this sample we excluded radio-loud and BAL quasars following the same approach adopted with the SDSS--4XMM sample. SEDs were also compiled for all the quasars (see \S~\ref{SED compilation}), which were then used to estimate both $\Lo$ and optical colours and thus select objects with low levels of dust reddening and host-galaxy contamination.

CSC2.0 provides photometric information and data products for each source, already reduced and ready to use for spectroscopic analysis\footnote{All the X-ray info can be downloaded from the CSCview application http://cda.harvard.edu/cscview/}. We selected all the AGN with at least one measure of the flux in the soft band and with an off-axis angle $<$10 arcmin (3,569 quasar observations, 2,392 single quasars). We performed a full spectral analysis with the \xspec v.12.10.1b X-ray spectral fitting package \citep{arnaud1996}. For each observation, we assumed a model consisting of a power law with Galactic absorption, as provided by the CSC 2.0 catalogue at the source location. The spectral analysis provides us with the rest-frame flux at 2 keV and its uncertainty. 
\rev{Overall, the \chandra data have a reasonable signal-to-noise ratio ($S/N>5$ in the soft band) that ensures uncertainties on $\Fx$ on the order of 0.15 dex or better.}

Flux limits are estimated for any given \chandra observation by computing the percentage of net counts to deduce the significance of the source detection, and a factor that takes into account the level of background, $P_{\rm bkg}$. 
The 0.5--2 keV and 2--7 keV fluxes are then multiplied by $P_{\rm bkg}$ to obtain an approximate value of the background flux in these energy bands. The flux limit in each energy band is then estimated from the background flux by assuming a minimum signal-to-noise ratio of 3. The flux limit at 2 keV is finally inferred by interpolation (or extrapolation) of the band flux limits, depending on the redshift of the source.

\subsection{The XMM--XXL sample}
\label{XXL}
We also considered the AGN sample published by \citetads{menzel2016} from the equatorial subregion of the \xmm XXL survey (XMM--XXL, PI: Pierre), i.e XMM--XXL North (in the following we will refer to this sample as XXL for simplicity), which overlaps with the SDSS--DR8 imaging survey.
XMM--XXL North is a medium-depth (10 ks per pointing) X-ray survey distributed around the area of the original 11 deg$^2$ XMM-LSS survey. The total catalogue contains 2,570 X-ray AGN with optical counterparts, spectroscopic redshifts and emission lines information. From the main sample, we considered only the AGN classified as (point-like) optically unobscured (flagged as BLAGN1; 1,353 sources). To have consistent measurements of optical/UV luminosities and redshifts amongst the different samples, we cross-matched the XXL BLAGN1 with the SDSS--DR14 quasar catalogue (with 3 arcsec matching radius) finding 1,067 objects. We have then included only the AGN with available SDSS photometry and classified as non-BAL, leading to 915 AGN. For this sample, we compiled the photometric SEDs following the same approach as in Section~\ref{SED compilation}, and computed luminosities at various rest-frame wavelengths (e.g. 2500 \AA, 1450 \AA, 6 cm), optical/UV colours ($\Gamma_1$, $\Gamma_2$, see \S~\ref{dusthost} for details) and radio loudness. 
The latter parameter further excludes 75 AGN, leading to a final sample of 840 sources.

   \begin{figure*}[!]
   \centering
   \includegraphics[width=0.49\linewidth]{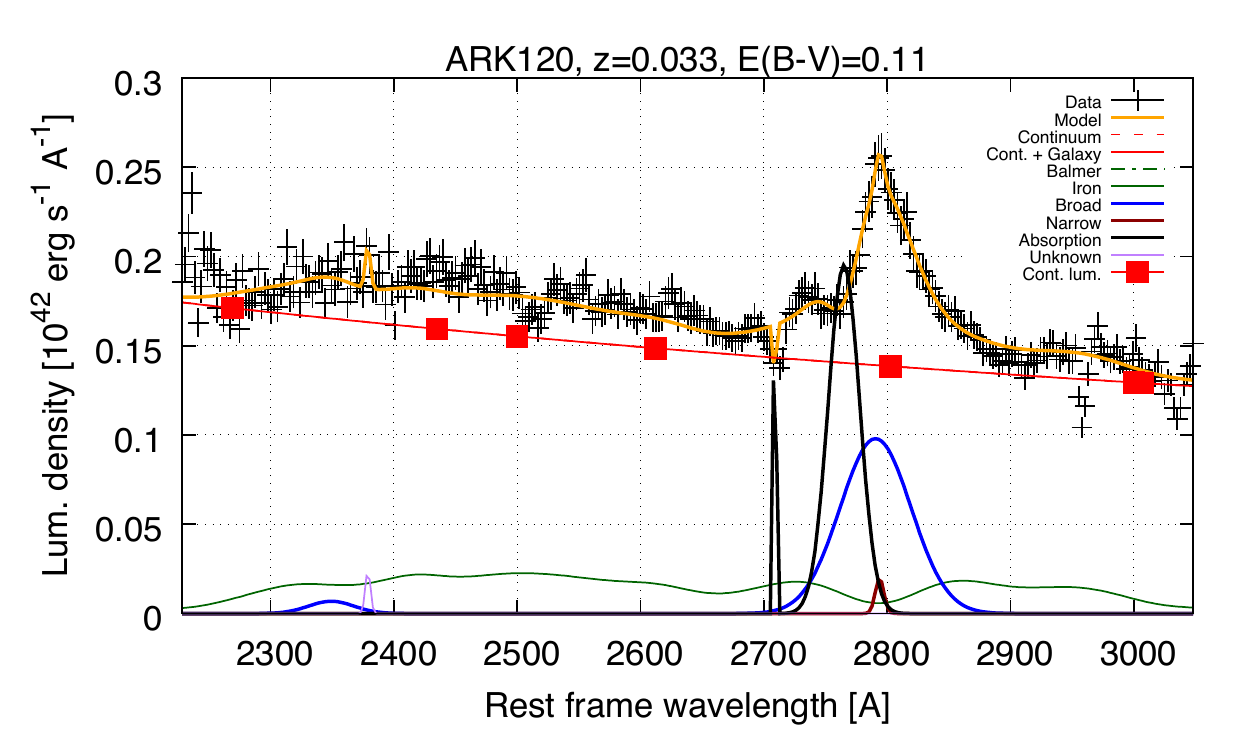}
   \includegraphics[width=0.46\linewidth]{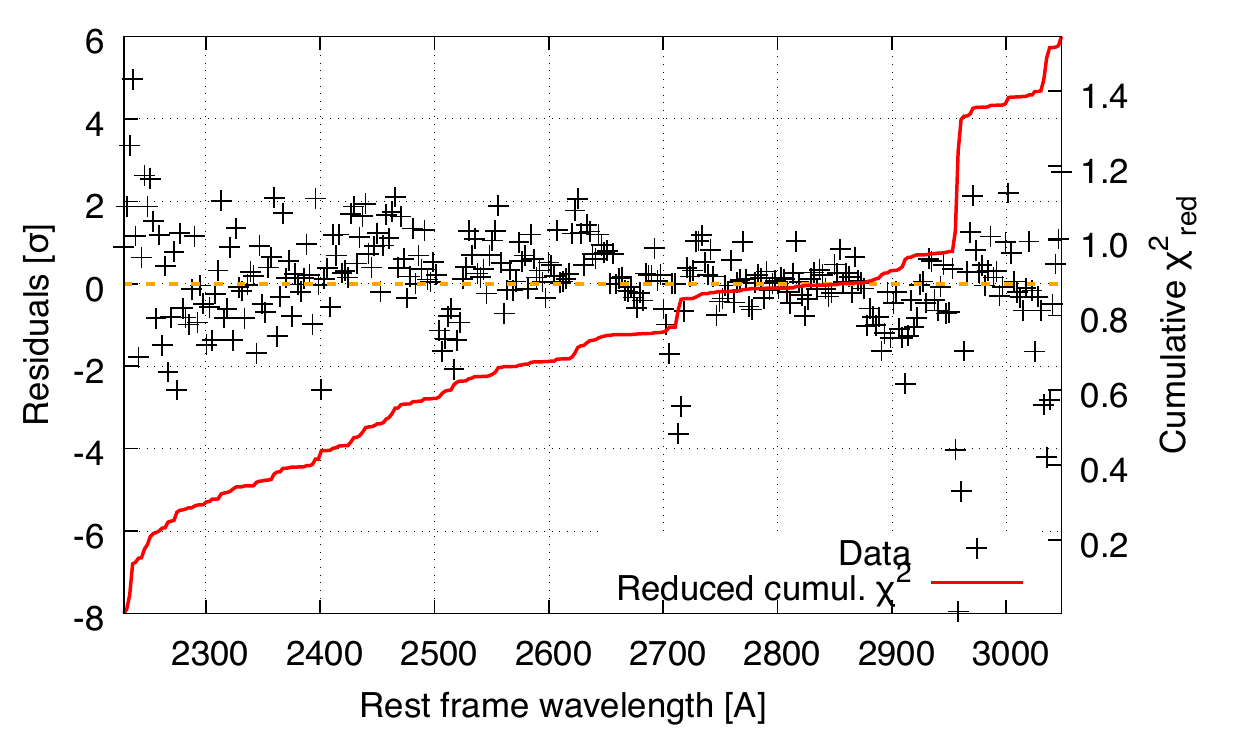}
   \caption{Example of UV spectral fitting. Best-fit model (left panel, orange curve) and residuals (right panel, in units of $1\sigma$ uncertainties on the data and cumulative reduced $\chi^2$) are presented for Ark 120. Redshift and Galactic extinction at the source location are shown on top of the plot. The main model components are plotted with different colours: the solid red line is the sum of continuum and host galaxy components; the black solid line is the absorption component; the dot-dashed green line is the Balmer component; the solid green line is the UV iron template; the sum of all broad and narrow emission-line components is shown with blue and brown lines, respectively. 
The red square symbols are the continuum luminosities estimated by \qsfit.}
              \label{ark120}
    \end{figure*}

\subsection{The $z\sim3$ quasar sample}
\label{z3}
We included a sample of 29 bright ($\lbol>10^{47}$ erg s$^{-1}$) quasars at $z\simeq3$  with X-ray observations obtained from an extensive campaign performed with \xmm (cycle 16, proposal ID: 080395, PI: Risaliti). This campaign targeted 30 quasars\footnote{One quasar in this sample turned out to be radio-loud, although not flagged as such in the SDSS--DR7 catalogue, so we exclude it from the present analysis.} in the $z=3.0$--3.3 redshift range for a total exposure of 1.13 Ms. This sample, selected in the optical from the SDSS Data Release 7 to be representative of the most luminous, intrinsically blue quasar population, boasts by construction a remarkable degree of homogeneity in terms of optical/UV properties. The X-ray data have been extensively analysed: the interested reader should refer to \citetads{nardini2019} for details.

\subsection{New $z\sim4$ quasar}
\label{z4}
We also included one new optically-selected SDSS quasar at $z=4.109$, J074711.14$+$273903.3, whose X-ray observation was obtained as part of a proposed large programme with \xmm (cycle 18, proposal ID: 084497, PI: Lusso). This is the only target actually observed from its parent sample, which consisted of 19 quasars in the $z\simeq4$ redshift range for a total exposure of 1.34 Ms. The sample had been selected in the optical from the SDSS Data Release 14 with the same criteria described in Section~\ref{z3}. The X-ray spectrum of this quasar has been analysed following the same procedure as presented by \citetads{nardini2019}, and we decided to include this source in the current sample as a proof of feasibility for future campaigns.

\subsection{The high redshift sample}
\label{high redshift sample} 
To improve the coverage at high redshifts, we considered two additional samples of $z>4$ quasars with pointed X-ray observations published by \citetads{salvestrini2019} and \citetads{vito2019}. 

The \citetads{salvestrini2019} quasar sample consists of 53 objects in the redshift range $4.01 < z < 7.08$, which benefit from a moderate-quality coverage in the UV and X-ray energy bands.
Of the 53 quasars, 47 objects were observed with \chandra and 9 with \xmm. The galaxies ULAS J1120$+$0641, SDSS J114816.7$+$525150.4 and SDSS 1030$+$0524 have been observed by both satellites. The authors performed a full X-ray spectral analysis of the archival data, we thus refer to their paper for details. 
The majority of the quasars in this sample (33 out of 53) have $\Lo$ measurements from the SDSS--DR7 quasar catalogue. For the remaining quasars, the $\Lo$ values are computed by extrapolating the UV spectra to longer wavelengths with a fixed continuum slope (see their Section 4 and Appendix B for further details).

\citetads{vito2019} published a sample of 25 quasars at $z>6$ with either archival data (15 objects) or new \chandra observations (10 sources), which were selected to have virial black-hole mass estimates from \ion{Mg}{ii} line spectroscopy. All the X-ray data were reprocessed by the authors (see their Section 3.1), whilst the $\Lo$ values were computed from the 1450 \AA\ magnitude assuming a power-law spectrum ($F_\nu \propto \nu^{-\alpha}$) with $\alpha=-0.3$ (see their Section 4.1). 
We excluded from their sample 3 BAL candidates, 1 weak line quasar, all sources with an upper limit in $\aox$ (i.e. X-ray undetected) and all radio-loud sources, for a total of 9 quasars. For the remaining 16 sources, we found five overlaps with the \citetads{salvestrini2019} sample, so the final number of quasars included from \citetads{vito2019} is 11 sources. This sample also contains the highest-redshift quasar observed so far, i.e. ULAS J134208.10$+$092838.61 at $z=7.54$ \citepads{banados2018}.

\begin{table*}
\caption{Properties of the local AGN sample.}              
\label{tbl:local}      
\centering                                      
\begin{tabular}{lcccc}          
\hline\hline                        
 Name & $z$ & $\fo$ & $\fx$ & Ref. \\    
\hline                                   
		Ark\,120              & 0.0327   &$-$24.88$\pm0.01$  &$-$28.37$\pm$0.002 & 1  \\
		Mrk\,841              & 0.0364   &$-$25.46$\pm0.01$  &$-$29.15$\pm$0.006& 1  \\
		NGC\,4593             & 0.0090   &$-$25.79$\pm0.01$  &$-$28.38$\pm$0.002& 1  \\
		HE\,1029$-$1401       & 0.0858   &$-$25.10$\pm0.01$  &$-$28.77$\pm$0.005& 1  \\
		ESO\,141$-$G055       & 0.0371   &$-$25.10$\pm0.01$  &$-$28.43$\pm$0.005& 2  \\
		IRAS\,09149$-$6206    & 0.0573   &$-$25.19$\pm0.01$  &$-$28.86$\pm$0.011& 3  \\
		HE\,1143$-$181        & 0.0329   &$-$25.37$\pm0.01$  &$-$28.57$\pm$0.005& 1  \\
		NGC\,7469             & 0.0163   &$-$25.02$\pm0.01$  &$-$28.55$\pm$0.003& 1  \\
		Mrk\,205              & 0.0708   &$-$25.73$\pm0.01$  &$-$29.33$\pm$0.016& 1  \\
		Mrk\,926              & 0.0469   &$-$25.25$\pm0.02$  &$-$28.54$\pm$0.007& 1  \\
		Fairall\,9            & 0.0470   &$-$25.34$\pm0.01$  &$-$29.01$\pm$0.016& 1  \\
		Mrk\,1383             & 0.0866   &$-$25.28$\pm0.01$  &$-$29.02$\pm$0.017& 1  \\
		Mrk\,509              & 0.0344   &$-$24.89$\pm0.01$  &$-$28.43$\pm$0.003& 1  \\
		Mrk\,478              & 0.0791   &$-$25.52$\pm0.01$  &$-$29.51$\pm$0.023& 1  \\
		Mrk\,352              & 0.0149   &$-$26.84$\pm0.01$  &$-$29.03$\pm$0.011& 1  \\
\hline                                             
\end{tabular}
\tablefoot{UV and X-ray fluxes are in units of log(erg~s$^{-1}$cm$^{-2}$Hz$^{-1}$). References for the X-ray data: 1: \citetads{bianchi09}; 2: \citetads{demarco09}; 3: \citetads{ricci2017}.
}
\end{table*}

\subsection{The local quasar sample}
\label{local}
To anchor the normalization of the quasar Hubble diagram with Type Ia supernovae, we need to extend the coverage at very low redshifts ($0.009<z<0.1$). We searched for all the local AGN with ultraviolet data from the {\it International Ultraviolet Explorer} (IUE) in the Mikulski Archive for Space Telescopes (MAST). We chose to use the reduced spectra from the long-wavelength prime (LWP) camera of IUE, which spans the wavelength interval 1845--2980 \AA, thus always covering the rest-frame 2500 \AA\ at the redshifts of interest. We then considered all AGN with X-ray data available in the \xmm archive or the with X-ray flux values in the literature, finding 17 objects, 11 of which with $\geq$2 UV spectra (although the majority consists of consecutive observations). In this sample, NGC 1566 and NGC 7603 are well known highly variable/changing look sources, so we excluded them from the starting sample.
Multiple UV spectra for the remaining AGN have been stacked, verifying that the inclusion of non consecutive observations does not change the final composite for each AGN. 

We then carried out a detailed spectral fitting of all the UV spectra using the publicly available \qsfit package \citepads{calderone2017}.  
We modelled each spectrum as follows: the \ion{Mg}{ii} emission line is reproduced by a combination of a broad (with a full-width at half-maximum, FWHM, larger than $2000$ km/s) and a narrow (FWHM$<2000$ km/s) profile, whilst the continuum includes the contributions from the host galaxy, the iron complex, the Balmer continuum and the AGN continuum. 
Spectra are also corrected for Galactic extinction using the parametrization by \citetads{CCM1989} and \citetads{odonnel1994}, with a total selective extinction $A(V)/E(B-V) = 3.1$ \citepads{calderone2017}. The rest-frame 2500 \AA\ luminosity is finally measured from the AGN continuum component only. An example of a UV spectral fit on one of the objects in the local AGN sample is shown in Figure~\ref{ark120}. 

The X-ray information (soft and hard fluxes, photon index) has been taken from the literature. Most of the sources in the local sample have been drawn from the CAIXA catalogue, which consists of radio-quiet, X-ray unobscured ($N_{\rm H}<2\times10^{22}$ cm$^{-2}$) AGN observed by \xmm in targeted observations \citepads{bianchi09}. For two AGN, ESO 141-G055 and IRAS 09149$-$6206, the X-ray fluxes and $\gammax$ values are given by \citetads{demarco09} and \citetads{ricci2017}, respectively.
The rest-frame 2 keV monochromatic flux is then estimated following the procedure described in Section~\ref{2keVflux}.

A summary of the properties of the local AGN sample is provided in Table~\ref{tbl:local}.

\section{The quasar SED compilation}
\label{SED compilation}
   \begin{figure*}
   \centering
   \includegraphics[width=0.49\linewidth]{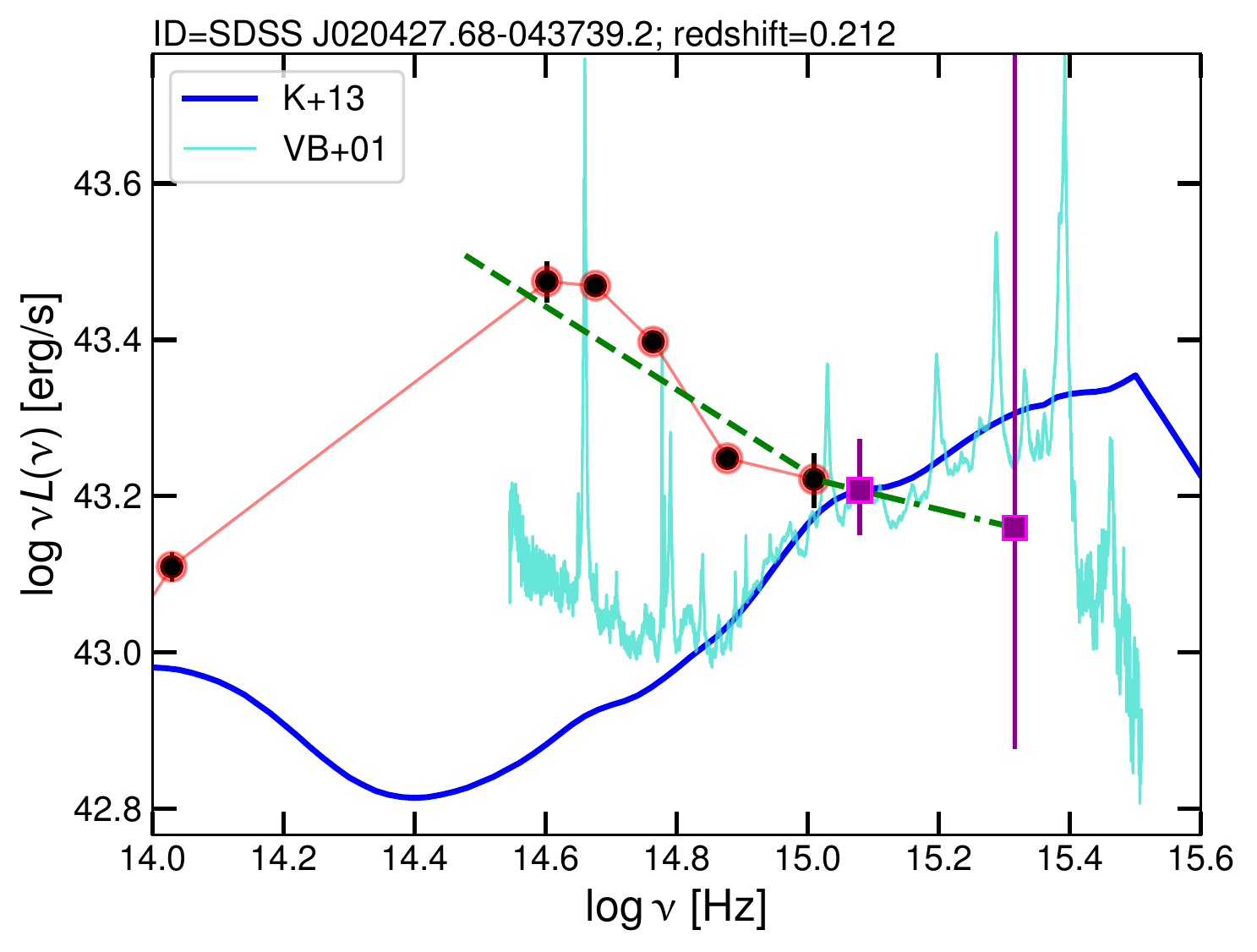}
   \includegraphics[width=0.49\linewidth]{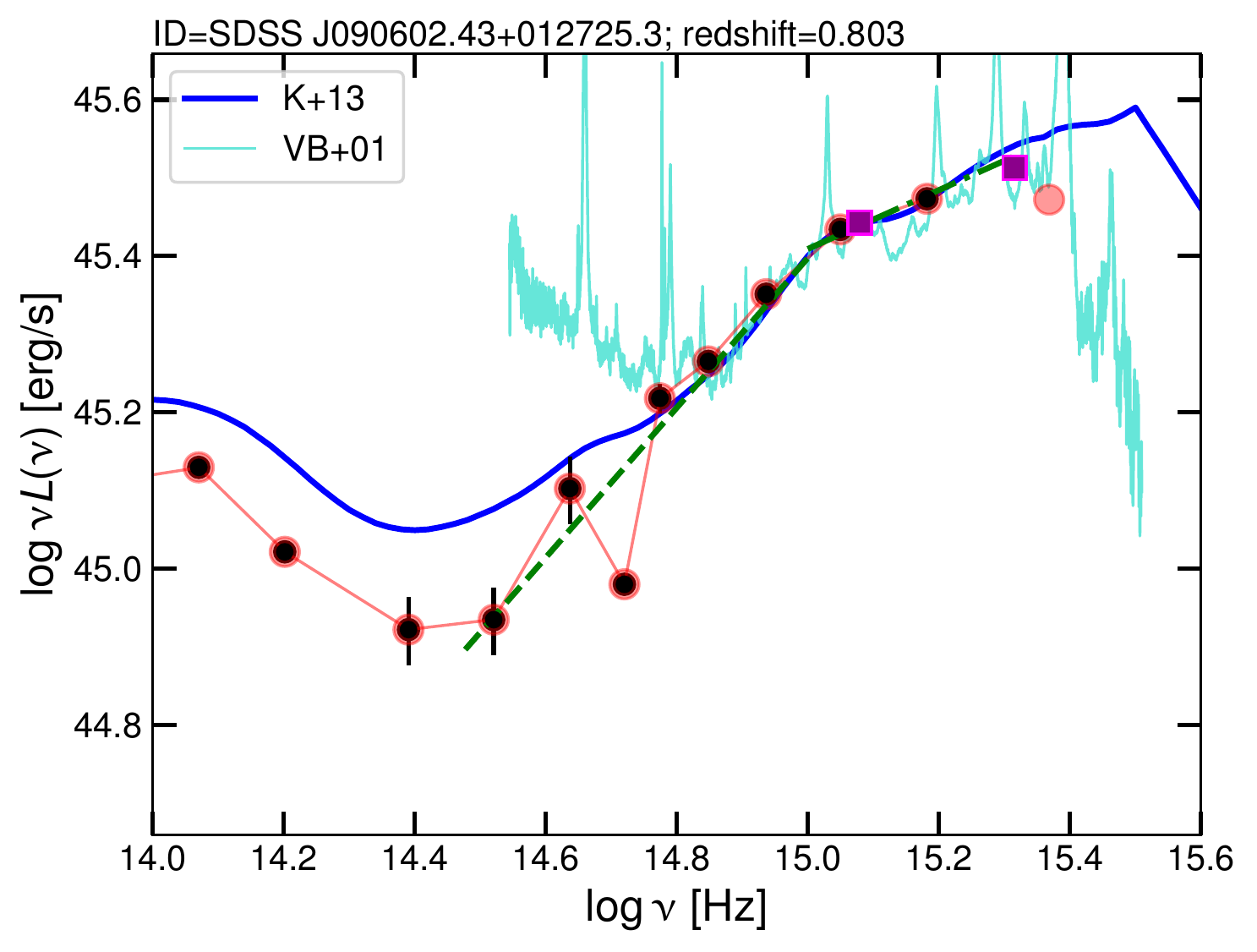}
   \caption{Examples of AGN SEDs. The red circles mark all the available photometry from the SDSS catalogue, whilst the ones used to construct the SEDs are highlighted with black circles. The magenta squares represent the luminosities at 2500 \AA\ and 1450 \AA. The cyan line is the composite SDSS quasar spectrum from \citetads{vandenberk2001}, whilst the blue solid line is the average SDSS quasar SED from \citetads{2013ApJS..206....4K}. Both composites are normalised to 2500 \AA. The green dashed and dot-dashed lines represent the two near-infrared/optical slopes $\Gamma_1$ and $\Gamma_2$ in the 0.3--1 $\mu$m and 1450--3000~\AA\ range (rest frame), respectively. The SED in the left panel is representative of an AGN that does not fulfil our selection criteria (see \S~\ref{Selection of a clean quasar sample}), as it is affected by both host-galaxy contamination and dust absorption in the UV. The SED in the right panel, instead, is characteristic of an object with low levels of contaminants, which therefore belongs to the clean AGN sample.}
              \label{seds}
    \end{figure*}
To compile the quasar SEDs, we used all the multicolour information as reported in the SDSS--DR14 catalogue. The catalogue includes multiwavelength data from radio to UV: 
the FIRST survey in the radio \citepads{becker1995}, the Wide-Field Infrared Survey (WISE, \citeads{write2010}) in the mid-infrared, the Two Micron All Sky Survey (2MASS, \citeads{cutri2003,2006AJ....131.1163S}) and the UKIRT Infrared Deep Sky Survey (UKIDSS; \citeads{lawrence2007}) in the near-infrared, and the Galaxy Evolution Explorer (GALEX, \citeads{martin2005}) survey in the UV.
Galactic reddening has been properly taken into account by utilising the selective attenuation of the stellar continuum $k(\lambda)$ from \citet{F99}, whilst Galactic extinction is estimated from \citet{schlegel98} for each object in the SDSS catalogue.
For each source, we computed the observed flux and the corresponding frequency in all the available bands. The data used in the SED computation were blueshifted to the rest-frame, and no K-correction was applied. All the rest-frame luminosities were then determined from a first-order polynomial between two adjacent points. 
At wavelengths bluer than about 1400 \AA, we expect significant absorption by the intergalactic medium (IGM) in the continuum ($\sim$10\% between the \ion{Ly}{$\alpha$} and \ion{C}{iv} emission lines, see \citeads{lusso2015} for details). Hence, when computing the relevant parameters, we excluded from the SED all the rest-frame data at $\lambda<1500$\AA.
The rest-frame monochromatic luminosities are finally obtained by interpolation whenever the reference frequency is covered by the photometric SED. Otherwise, the value is extrapolated by considering the slope between the luminosity values at the closest frequencies. 
Thanks to this broad photometric coverage, we can compute the rest-frame luminosity at 2500 \AA\ ($\Lo$) via interpolation for the majority of the SDSS quasars. Indeed, we were not able to estimate $\Lo$ due to a sparse photometric SED coverage (i.e. when the SED is composed by a single rest-frame data point) for only 130 quasars.

Uncertainties on monochromatic luminosities ($L_\nu\propto \nu^{-\gamma}$) from the interpolation (extrapolation) between two values $L_1$ and $L_2$ are computed as:
\begin{equation}
\label{uncertainties}
\delta L = \sqrt{\left( \frac{\partial L}{\partial L_1}\right)^2 (\delta L_1)^2 + \left( \frac{\partial L}{\partial L_2}\right)^2 (\delta L_2)^2}.
\end{equation}  
Examples of photometric SEDs for two quasars at different redshifts in the SDSS--4XMM sample are shown in Figure~\ref{seds}. The red circles in the figure mark all the available photometry from the SDSS--DR14 catalogue, whilst the ones used to construct the SEDs are highlighted with black circles. The magenta squares represent the luminosities at 2500 \AA\ and 1450 \AA. The cyan and blue solid lines are the composite SDSS quasar spectrum from \citetads{vandenberk2001} and the average SDSS quasar SED  from \citetads{2013ApJS..206....4K}, respectively. Both composites are shown for reference, for the AGN continuum plus line emission and continuum only, and are normalised to 2500 \AA. 
The AGN SED in the left panel shows a case of both host-galaxy contamination and dust absorption. This source, indeed, does not fulfil our selection criteria (described in detail in \S~\ref{Selection of a clean quasar sample}), as opposed to the object in the right panel, which represents an object with low levels of contaminants.

   \begin{figure}
   \centering
   \includegraphics[width=\linewidth]{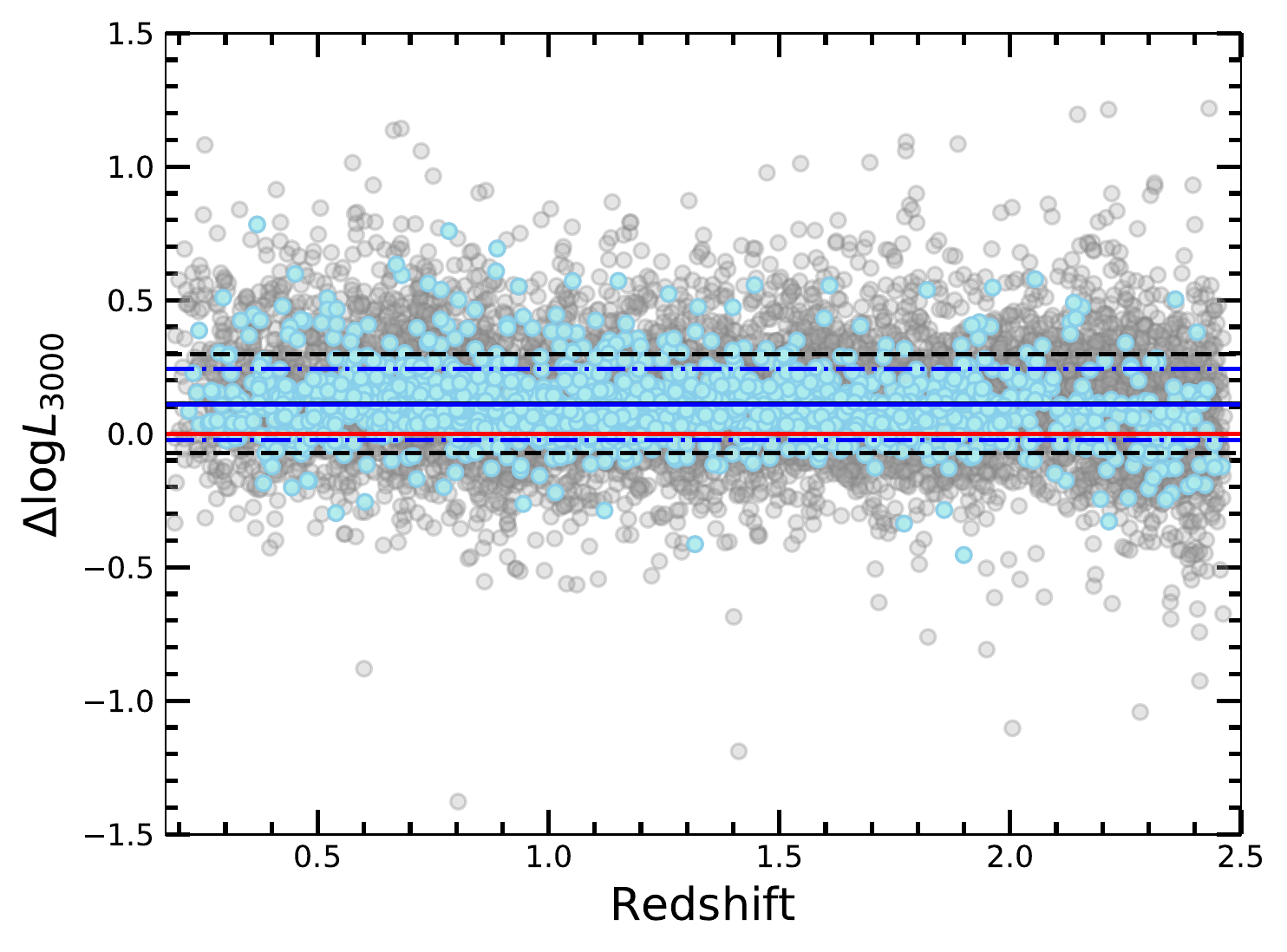}
   \caption{\rev{Comparison between the photometric SDSS luminosity values at 3000 \AA\ and the spectroscopic ones (i.e. $\Delta \log L_{3000}=\log L_{3000,\,\rm photo}-\log L_{3000,\,\rm spectro}$) as a function of redshift for all the quasars in our sample with available 3000 \AA\ monochromatic luminosities from \citetads{rakshit2020}. \revs{Grey and cyan points represent the initial ($\sim12,000$) and selected ($\sim1,500$) quasar samples from SDSS--4XMM, respectively. The black and blue lines mark the mean (solid) and $1\sigma$ dispersion (dashed and dot-dashed) of $\Delta \log L_{3000}$ for the initial and selected samples, respectively (black and blue solid lines overlap at $\langle\Delta \log L_{3000}\rangle=0.11$). The red solid line is the $\Delta \log L_{3000}=0$.}}}
              \label{fluxuvcheck}
    \end{figure}
\rev{
\subsection{On the use of photometric rest-frame 2500 \AA\ fluxes}
\label{uvfluxcheck}
In this work, we focus on rest-frame 2500 \AA\ monochromatic fluxes as derived from photometry for two main reasons. The first one is based on the physics of the $\Lx-\Lo$ relation, whilst the other on the fact that broadband photometry allows us to build much bigger samples over a larger redshift and luminosity range than spectroscopy alone. 

Concerning the former, the 2500 \AA\ monochromatic flux has been adopted since the first studies on the topic, yet its choice was mainly based upon observational considerations. Indeed, the rest-frame 2500 \AA\ at $z\simeq1$ corresponds to the observed $V$ band, which was available for a significant number of sources. Additionally, this rest-frame UV wavelength is less affected by host-galaxy contamination (dominant at low luminosity for $\lambda>4000$ \AA) and intergalactic absorption ($\lambda<1450$ \AA) than other regions of the AGN SED, thus representing the ideal proxy of the intrinsic disc emission. This notwithstanding, the photometric 2500 \AA\ flux might be contaminated by strong \ion{Fe}{ii} line emission (see Figure~\ref{ark120}), which can introduce systematic uncertainties on the photometrically derived $\Fo$ values (e.g. \citeads{netzer2019}). 

\citetads{rakshit2020} recently published spectroscopic measurements for more than 500,000 quasars selected from the SDSS--DR14 quasar catalog. They performed a homogeneous analysis of the SDSS spectra to estimate the continuum and line properties (e.g. \ion{H}{$\alpha$}, \ion{H}{$\beta$}, \ion{Mg}{ii}, \ion{C}{iv}, and \ion{Ly}{$\alpha$}) of these sources. This catalogue also provides a measurement of the 3000 \AA\ luminosity, which is the closet wavelength to the one of our interest. We have therefore estimated the 3000 \AA\ monochromatic luminosities from the photometric SEDs for all the SDSS--4XMM quasars in the initial sample of 13,800 quasars, similarly to what we have done at 2500 \AA. Figure~\ref{fluxuvcheck} shows the comparison between the photometric and the spectroscopic SDSS luminosity values (i.e. $\Delta \log L_{3000}=\log L_{3000,\,\rm photo}-\log L_{3000,\,\rm spectro}$) as a function of redshift for the objects within the SDSS--4XMM sample with a good quality 3000 \AA\ monochromatic luminosity value (QUALITY\_L3000=0) available from spectroscopy, \revs{i.e. $\simeq12,000$ quasars (where $z\simeq2.5$ represents the higher redshift for which the rest-frame 3000 \AA\ is covered by the BOSS spectrograph). The $\Delta \log L_{3000}$ distribution shows a mean at 0.1 dex, a dispersion around the mean of 0.18 dex, and no trend with redshift. Although a systematic offset in the $\Delta \log L_{3000}$ measurements is expected, as our $L_{3000}$ could be contaminated by the \ion{Fe}{ii} emission, this is reassuringly small (only a flux factor of $\sim$1.3). We also note that, any redshift independent offset in the UV fluxes would not be an issue, since it is balanced out in the cross-calibration between the Hubble diagram of supernovae and quasars.

Figure~\ref{fluxuvcheck} also shows the $\Delta \log L_{3000}$ distribution as a function of redshift for the 1,473 quasars in the clean SDSS--4XMM sample (1,644 total sources, see Section~\ref{Selection of a clean quasar sample} for details) with a measurement of $L_{3000}$. The average $\Delta \log L_{3000}$ is perfectly consistent with the initial sample, the dispersion around the mean is 0.13 dex, and again there is no trend with redshift, implying that our selection criteria have the only effect of singling out the most reliable luminosity measurements as proxy of the nuclear emission.}

We finally note that, regardless of the details of the physical mechanism driving the $\Lx-\Lo$ relation, the characteristic UV flux wavelength should be the one most closely related to the global emission of accretion disc, and might not be precisely the rest-frame 2500 \AA\ (see Section 6 in \citeads{rl19}). It is even possible that nuclear emission should be combined with other AGN spectral properties (e.g. emission-line FWHM, continuum slope; see \citeads{lr17}). We are currently exploring these possibilities, and results will be published in a forthcoming work.

Since we are still far from grasping the nature of the $\Lx-\Lo$ relation, our photometric fluxes are seen as a more conservative representation of the broadband disc emission, capturing the ``true'' dependence between disc and X-ray corona in AGN in a statistical sense. 
}

   \begin{figure}
   \centering
   \includegraphics[width=\linewidth]{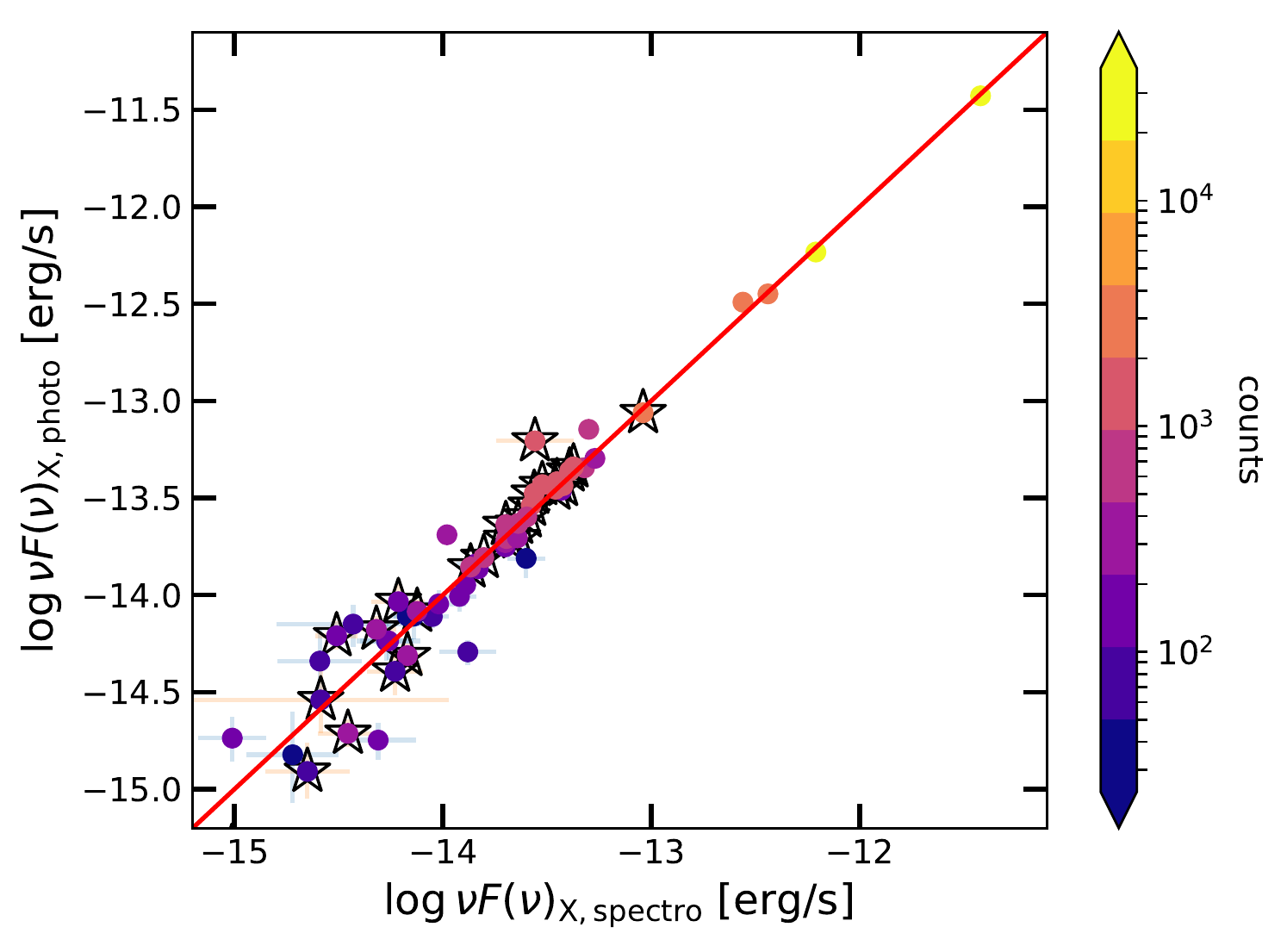}
   \includegraphics[width=\linewidth]{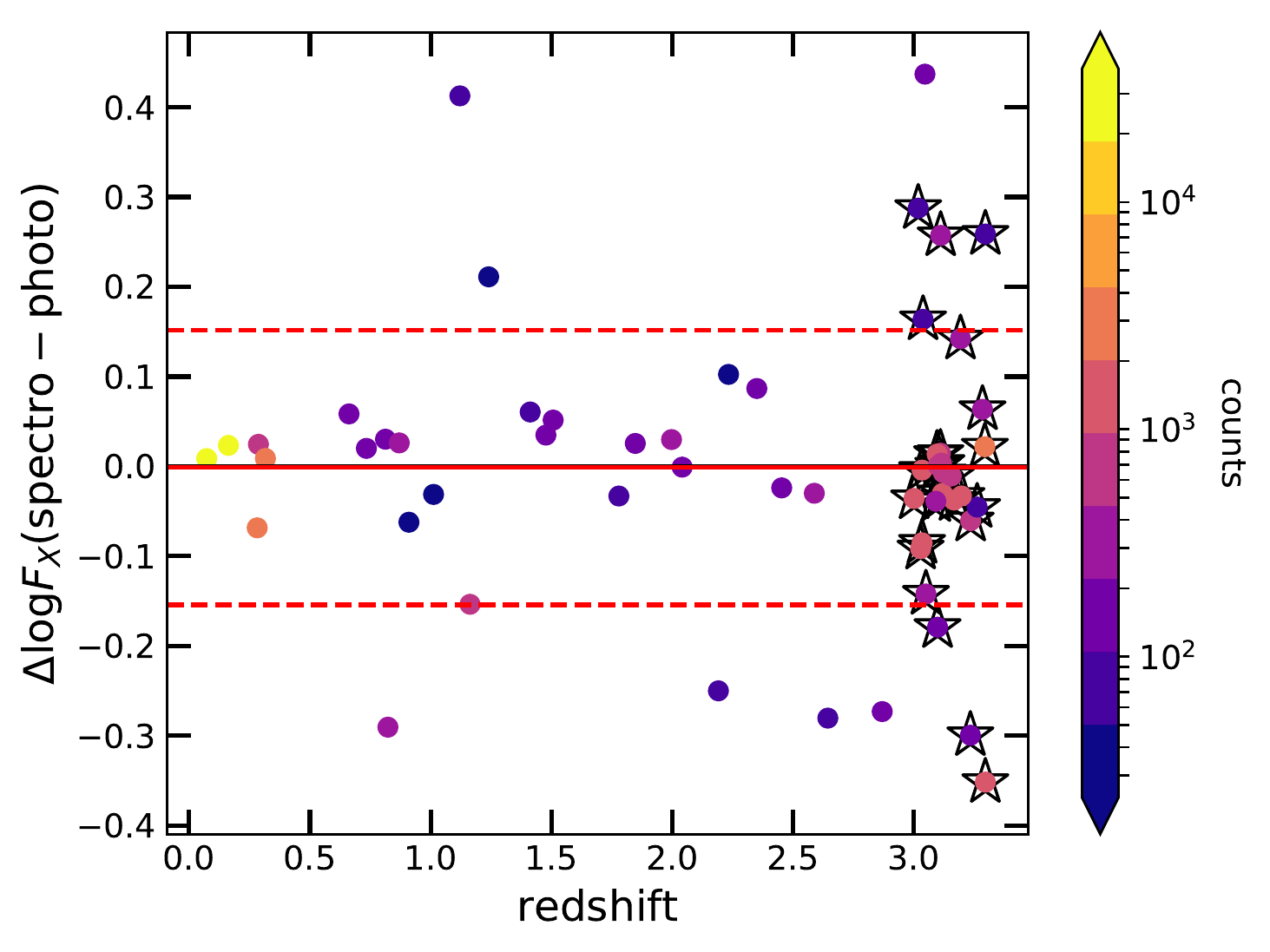}
   \caption{Top panel: comparison between the spectroscopic and photometric 2 keV monochromatic fluxes for 30 quasars randomly extracted from the SDSS--4XMM sample \rev{and the quasars of the $z\simeq3$ sample \citep[][marked with open star symbols]{nardini2019}}. Bottom panel: difference between the spectroscopic and photometric 2 keV monochromatic fluxes as a function of redshift. The mean $\Delta\log\Fx= \log F_{\rm X,spectro}-\log F_{\rm X,photo}$ value and its 1\,$\sigma$ dispersion are shown with the red and dashed lines, respectively. Points are colour-coded by the number of net counts. The $\Delta\log\Fx$ distribution is scattered around $\Delta\log\Fx\sim0$ with no clear trend with redshift. 
   }
              \label{fluxcheck}
    \end{figure}
   \begin{figure}
   \centering
   \includegraphics[width=\linewidth]{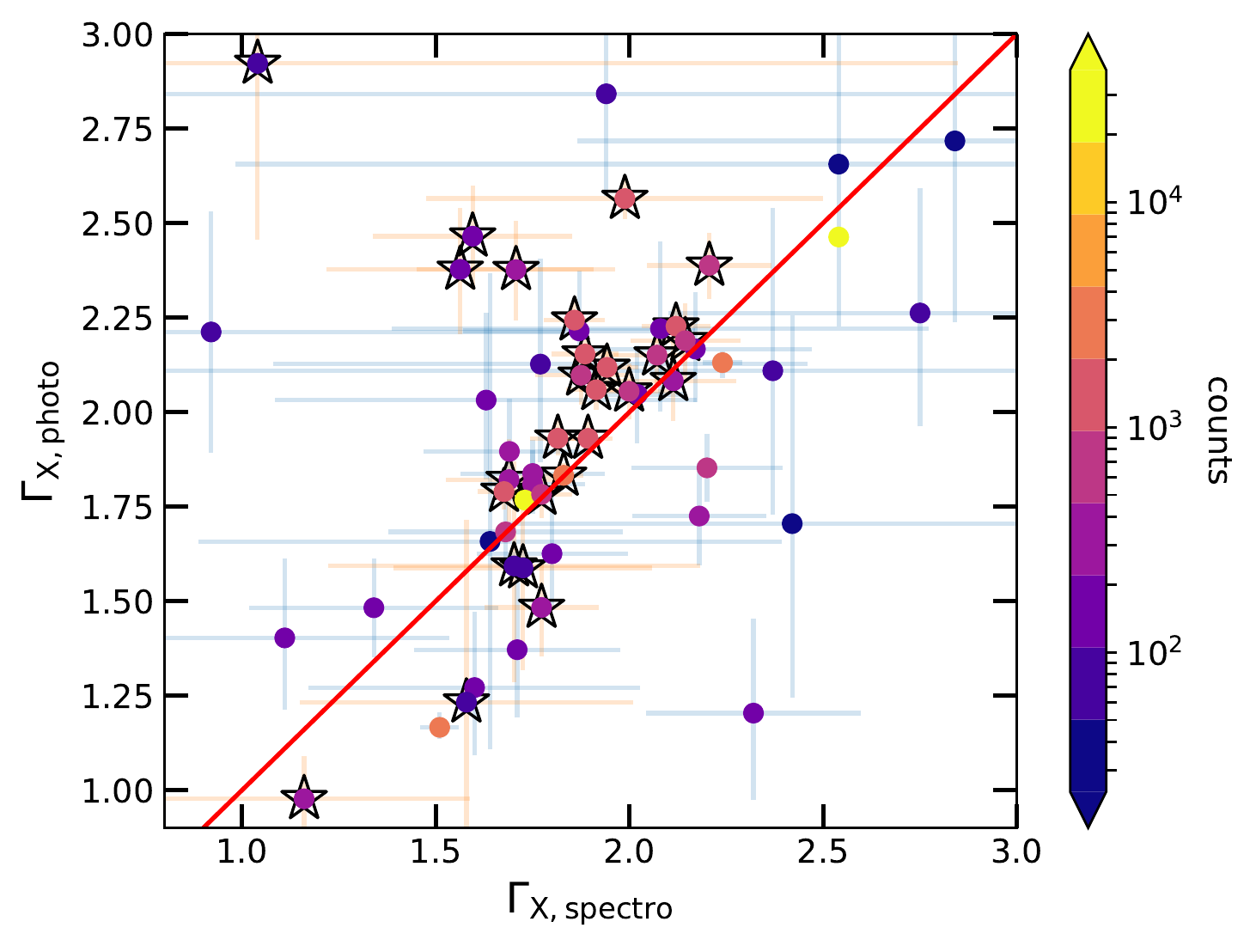}
   \includegraphics[width=\linewidth]{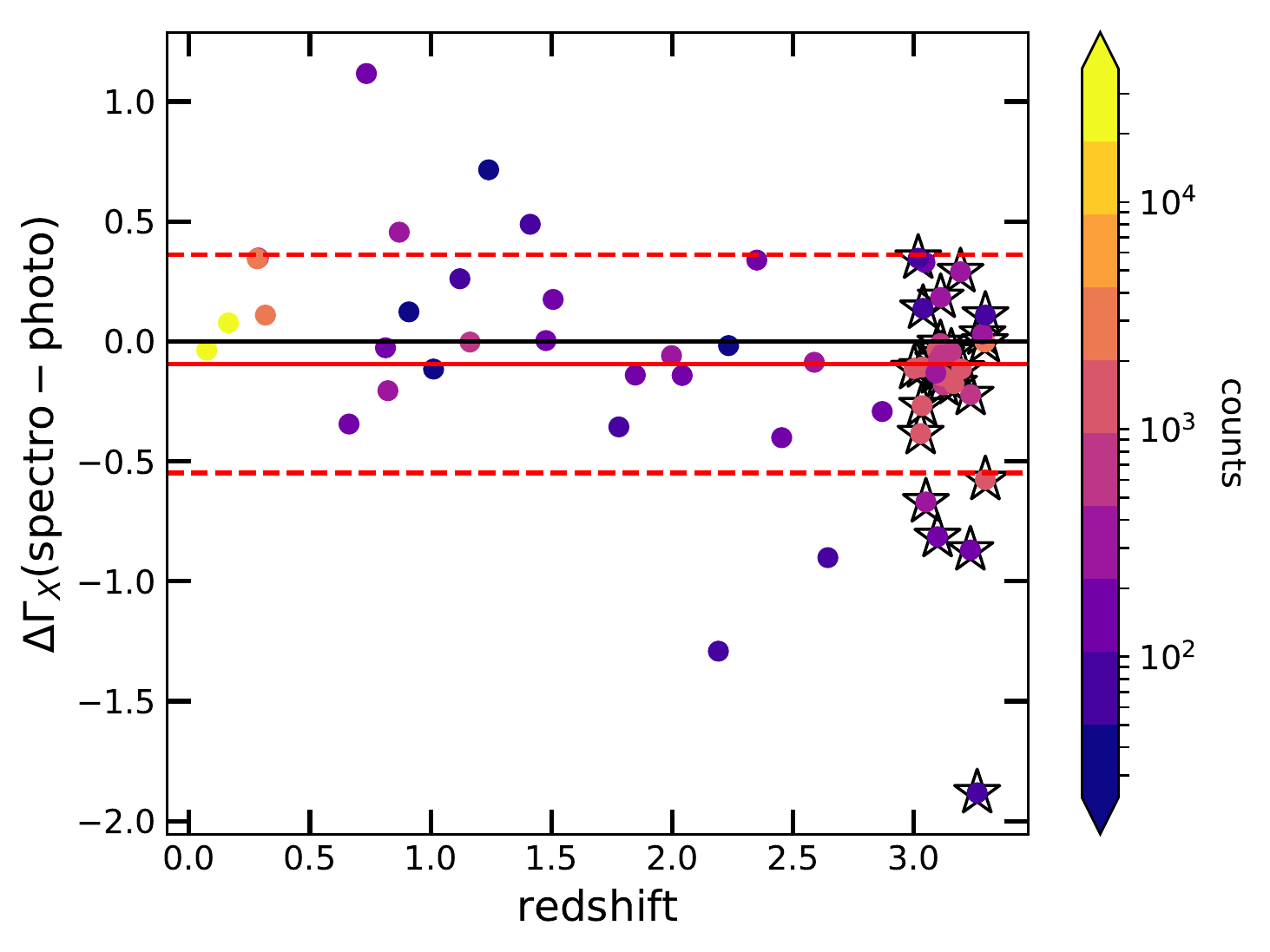}
   \caption{Top panel: comparison between the spectroscopic and photometric photon indices for 30 quasars randomly extracted from the SDSS--4XMM sample \rev{and the quasars of the $z\simeq3$ sample \citep[][marked with open star symbols]{nardini2019}}. Bottom panel: difference between the spectroscopic and photometric photon index values as a function of redshift. The mean $\Delta\gammax=\Gamma_{\rm X,spectro}-\Gamma_{\rm X,photo}$ value and its 1\,$\sigma$ dispersion are shown with the red and dashed lines, respectively. Points are colour-coded by the number of net counts. Although the $\gammax$ distribution along the one-to-one relation is rather scattered, the $\Delta \gammax$ does not seem to show a clear trend with redshift. 
   }
              \label{gammacheck}
    \end{figure}
\section{The rest-frame 2 keV monochromatic flux}
\label{2keVflux} 
Given the large source statistics in the SDSS--4XMM and XXL samples, a detailed X-ray spectral analysis of all the quasars is impractical. Therefore, to compute the rest-frame 2 keV monochromatic flux ($\Fx$), we follow the same approach as described in \citetads{rl19}. For the SDSS--4XMM sample, we derived the rest-frame 2 keV fluxes and the relative (photometric) photon indices, $\gammax$ (along with their 1$\sigma$ uncertainties), from the tabulated 0.5--2 keV (soft, $F_{\rm S}$) and 2--12 keV (hard, $F_{\rm H}$) fluxes reported in the 4XMM--DR9 serendipitous source catalogue. These band-integrated fluxes are blueshifted to the rest-frame by considering a pivot energy value of 1 keV ($E_{\rm S}$) and 3.45 keV ($E_{\rm H}$), respectively, and by assuming the same photon index used to derive the fluxes in the 4XMM catalogue (i.e. $\gammax=1.42$). For the soft band, the monochromatic flux at $E_{\rm S}$ is then:
\begin{equation}
\label{soft}
F_E(E_S)=F_{\rm S}\frac{(2-\gammax) E_{\rm S}^{1-\gammax}}{(2\,\rm keV)^{2-\gammax}-(0.5\,\rm keV)^{2-\gammax}},
\end{equation}
in units of erg s$^{-1}$ cm$^{-2}$ keV$^{-1}$. An equivalent expression holds for the hard band, with the obvious modifications. Flux values are corrected for Galactic absorption.

The photometric photon index is then estimated from the slope of the power law connecting the two soft and hard monochromatic fluxes at the rest-frame energies corresponding to the observed pivot points. 
The rest-frame photometric 2 keV flux (and its uncertainty) is interpolated (or extrapolated) based on such a power law. 

To justify the employed soft and hard pivot energy values and to ensure that our photometric $\Fx$ values are accurate, we performed on the one hand several simulations, and on the other hand full X-ray spectral fitting of a number of random objects at different redshifts. 

Regarding the former approach, we simulated a high-quality power-law spectrum, assuming both a typical average background and calibrations for \xmm, with the same photon index assumed by the 4XMM--DR9 catalogue. 
We fitted the data in the soft band with a power law parametrized
as $F(E)=F(E_0)(E/E_0)^{-\gammax}$, with $F(E_0)$ and $\gammax$ as free parameters, with $E_0$ ranging from 0.5 to 1.5 keV in steps of 0.05 keV. 
In each case, we derived the $F(E_0)-\gammax$ confidence contours. In general, the covariance between these two parameters is non-zero (i.e. the contours are elongated and tilted), so we explored which value of energy $E_{\rm S}$ returns covariance zero between $F(E_{\rm S})$ and $\gammax$ (i.e. the contours become 2D Gaussians). This \textit{pivot energy} represents the energy value dividing the soft band in two regions having the same statistical weight. As such, this value is not located at exactly the centre of the energy band because of the dependence of the effective area on energy. 

As a result of the $F(E_{\rm S})-\gammax$ zero-covariance, our photometric $\Fx$ values are independent of the specific $\gammax$ assumed in the 4XMM--DR9 catalogue. This also implies that, even if our photometric $\gammax$ deviates from the {\it true intrinsic} value, the resulting $\Fx$ will be accurate in any case.  
Finally, the relative error on the monochromatic flux, $\Delta F(E)/F(E)$, at the pivot energy is the same as the one of the band flux, whereby the absolute value of $\Delta F(E)$ at the pivot energy is the smallest possible.

The same procedure is also applied to the XXL sample using their catalogued soft and hard band fluxes. 

In parallel, we performed a full spectral analysis on a number of random objects. The top panels in Figures~\ref{fluxcheck} and~\ref{gammacheck} present the comparison between the inferred spectroscopic and photometric $\Fx$ and $\gammax$ values, respectively, for 30 random quasars in the SDSS--4XMM sample. \rev{We also considered for this comparison the 27 sources in the $z\simeq3$ quasar sample \citepads{nardini2019} that have an entry in the 4XMM--DR9 catalogue}\footnote{\rev{Out of the 30 objects in the $z\simeq3$ sample, one had no public data on 2018 December 18, and two (J0945+23 and J1159+31; see Section 4.3\ in \citealt{nardini2019}) are not detected in 4XMM--DR9.}.} The points are colour-coded by their number of net counts. \rev{Whilst the values of $\gammax$ display a large scatter (up to $\sim$0.46 dex), our photometric $\Fx$ values are in remarkable agreement with the spectroscopic ones (with a scatter of just $\sim$0.15 dex). The most obvious outlier in the bottom panel of Figure~\ref{gammacheck} is J1425+54, a marginally detected $z\simeq3$ quasar with $22\pm13$ net counts in the pn (see Table~1\ in \citeads{nardini2019}; for the same camera, 4XMM--DR9 gives a consistent number of counts). The observed soft flux reported in the 4XMM--DR9 catalogue for this object is $(2.0\pm0.5)\times10^{-15}$ erg s$^{-1}$ cm$^{-2}$, whilst it is virtually undetected in the observed hard band, with a S/N of $\sim$0.5 and $F_{\rm H}=(9\pm20)\times10^{-16}$ erg s$^{-1}$ cm$^{-2}$. None the less, even with a nominally large discrepancy between the spectroscopic and photometric $\gammax$ values\footnote{\rev{Interestingly, J1425+54 would not have met the selection criterion on $\gammax$ in either case (see \S\,\ref{xrayabs})}.}, the $\Fx$ estimates are well within a factor of 1.2.}

\rev{Overall, we have a consistency within a flux factor of 1.6 for about 80\% of the sample (only 12/57 quasars lie outside $|\Delta\log\Fx|=0.2$) and, as expected, the higher the number of counts, the better the agreement, with the most deviant points having less than 100 counts\footnote{Note that this is the typical threshold for spectral analysis to return reasonably accurate results (see e.g. \citealt{nardini2019}).}.}

We have also investigated whether the difference between spectroscopic and photometric fluxes ($\Delta\Fx=F_{\rm X,spectro}-F_{\rm X,photo}$) and photon indices ($\Delta\gammax=\Gamma_{\rm X,spectro}-\Gamma_{\rm X,photo}$) displays any trend with redshift. The bottom panels of Figures~\ref{fluxcheck} and \ref{gammacheck} show such distributions and, despite the limited statistics, both $\Delta\Fx$ and $\Delta \gammax$  are scattered around zero with no clear trend.

The rather poor comparison between $\Gamma_{\rm X,spectro}$ and $\Gamma_{\rm X,photo}$ shown in Figure~\ref{gammacheck} might cast some doubts on the reliability of the photon indices derived from the broadband (soft and hard) fluxes. However, we believe that our technique of computing photometric $\gammax$ values can be safely employed for large sample of quasars and that it provides robust results, for the following reasons: {\it (1)} the spectroscopic and photometric X-ray fluxes are in very good agreement, meaning that our distance measures are not strongly affected by the use of photometric $\gammax$ values; and {\it (2)} we performed a series of checks by varying the photometric $\gammax$ range employed to select the final sample, finding that our main results are not significantly modified (see \S\,\ref{xrayabs} for further details).
Summarizing, our $\Gamma_{\rm X,photo}$ may not be correct on an object-by-object basis, but they are reliable in a {\it statistical sense} for large enough samples.

\rev{
\subsection{X-ray non-detected quasars}
\label{X-ray non-detected quasars}
Quasar samples that include X-ray non-detections are likely to be unbiased, but the analysis of both the $\Lo-\Lx$ and the distance modulus--redshift relations is far from straightforward, since it strongly depends on the weights assumed in the fitting algorithm. In the case of flux-limited surveys, objects with an expected emission (based on the observed $\Lo-\Lx$ relation) close to the flux limit will be observed only in case of positive fluctuations, and this effect is likely redshift-dependent (see \S~\ref{eddbias}). Considering only detections might thus introduce some bias in the $\Lo-\Lx$ relation, and this should be more relevant to the X--rays, since the relative observed flux interval is narrower than in the UV.  

\citetads{lr16} investigated the effect of the inclusion of X-ray non-detections in the study of the $\Lx-\Lo$ relation for an optically selected sample of quasars, whose selection was very similar to the one employed in the present analysis. Their main conclusion was that there were no statistically significant variations on slope, intercept and dispersion (within their uncertainties) between X-ray detected and censored quasar samples across the different selection steps, with the slope being rather constant around 0.6. Further analysis was performed in \citetads{rl19} (see Section~3 of their Supplementary Material), where the fraction of X-ray non-detected quasars was on the order of 2\% in their final cleaned sample. Such a fraction of censored data has negligible statistical weight in the fitting procedure, so their inclusion does not change the results of the statistical analysis.

Here, we have adopted a similar strategy as the one presented by  \citetads{rl19} to obtain a sample where biases are minimised even without the inclusion of non-detections, which is discussed at length in Section~\ref{eddbias}. 
Additionally, we explored whether any possible remaining bias in our X-ray detected quasar sample is present in the residuals of the quasar Hubble diagram (see \S~\ref{residuals-eddbias}). 
All these checks motivated us to analyse the Hubble diagram where non-detections are neglected.
}


\section{Selection of a clean quasar sample}
\label{Selection of a clean quasar sample}
Our aim is to select a subsample with accurate estimates of $\Lo$ and $\Lx$, covering a redshift range as wide as possible, by removing systematic effects and low-quality measurements. For the latter, we applied a couple of preliminary filters that ensure good measurement quality. These filters mainly involve the X-ray data, since these affected by larger uncertainties. Specifically, we considered only soft and hard flux measurements with a relative error smaller than 1 (i.e. a minimum S/N of 1 on both band fluxes): $\Delta F_{\rm S}/F_{\rm S}<1$ and $\Delta F_{\rm H}/F_{\rm H}<1$. A similar filter is currently not applied to UV fluxes since the S/N at these wavelengths is typically much higher than 1. Overall, these two filters exclude about 30\% of the X-ray detections in the initial sample.

The main possible sources of contamination/systematic error are: dust reddening and host-galaxy contamination in the optical/UV, gas absorption in the X-rays, and {\it Eddington bias} associated with the flux limit of the X-ray observations. Here we briefly discuss each of these effects, and describe the filters we applied to obtain the final `best' sample for a cosmological analysis.

\subsection{Dust reddening and host-galaxy contamination}
\label{dusthost}
   \begin{figure}
   \centering
   \includegraphics[width=\linewidth]{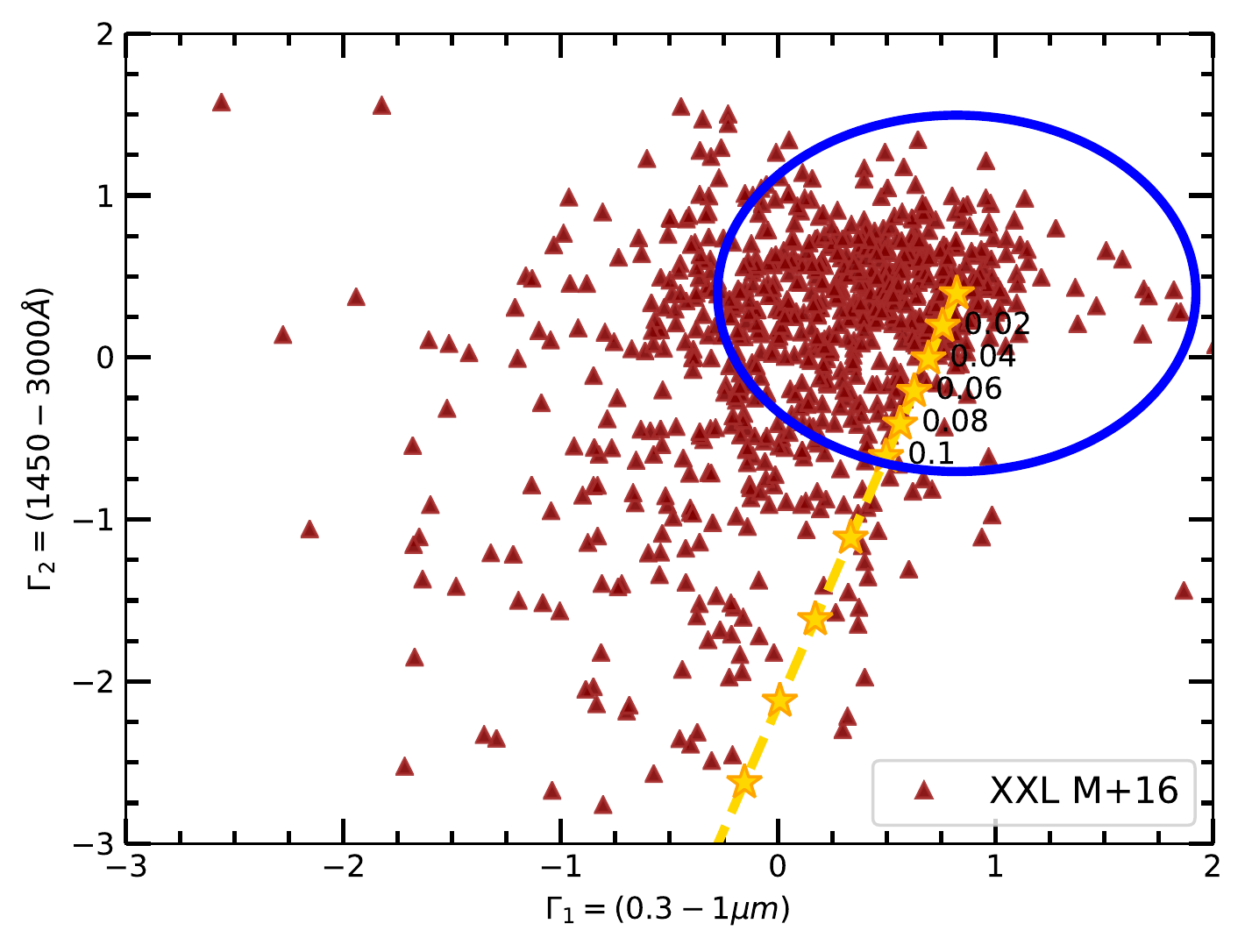}
   \caption{Example of the $\Gamma_1-\Gamma_2$ distribution for the XXL quasar sample, where $\Gamma_1$ and $\Gamma_2$ are the slopes of a power law in the $\log(\nu)-\log(\nu L_\nu)$ plane in the 0.3--1~$\mu$m and 1450--3000~\AA\ intervals, respectively (see \S\,\ref{dusthost}). The stars represent the $\Gamma_{1}-\Gamma_{2}$ values of the quasar SED by \citet{2006AJ....131.2766R} with increasing dust reddening (following the extinction law of \citealt{prevot84}), with $\ebv$ in the range 0--0.3. We selected all the quasars inside the blue circle (i.e., with minimum host-galaxy and dust reddening contamination).}
              \label{redqso}
    \end{figure}
To retain the quasars with minimum levels of dust reddening and host galaxy contamination, we follow a similar approach to the one presented in our previous works \citepads{rl15,lr16,rl19}. We used the rest-frame photometric SEDs discussed in Section~\ref{SED compilation} to compute, for each object, the slope $\Gamma_1$ of a $\log(\nu)-\log(\nu L_\nu)$ power law in the rest frame 0.3--1 $\mu$m range, and the analogous slope $\Gamma_2$ in the 1450--3000~\AA\ range (see also \citeads{hao2013}). Figure~\ref{seds} shows two examples, where the green dashed and dot-dashed lines represent the 0.3--1 $\mu$m and 1450--3000~\AA (rest frame) slopes $\Gamma_1$ and $\Gamma_2$, respectively. 
The wavelength intervals for these slopes are chosen based on the fact that the SED of an {\it intrinsically} blue quasar is very different from the one of an inactive galaxy or a dust-reddened source. The intrinsic SED of a quasar presents a dip around 1\,$\mu$m, where the galaxy has the peak of the emission from the passive stellar population (e.g. \citeads{1994ApJS...95....1E,2006AJ....131.2766R,2012ApJ...759....6E,2013ApJS..206....4K}). Dust reddening is wavelength dependent and the UV portion of the quasar SED will be attenuated differentially.
These two concurrent factors impact on the quasar SED shape, allowing us to define a set of slopes that single out the majority of quasars with minimum levels of both host-galaxy emission and dust reddening (see Figure~1 in \citeads{hao2013}).

The $\Gamma_1-\Gamma_2$ distribution for the XXL subset of quasars is shown, as an example, in Figure~\ref{redqso}. We assumed a standard SMC extinction law $k(\lambda)$ after \citet{prevot84}, with $R_V=3.1$ (as appropriate for unobscured AGN; \citealt{2004AJ....128.1112H,salvato09}), to estimate the $\Gamma_1-\Gamma_2$ correlation as a function of extinction, parametrised by the colour excess $\ebv$. We obtained the red dashed line shown in Figure~\ref{redqso}, where the starting point corresponds to the SED of \citet[i.e. $\Gamma_1=0.82$, $\Gamma_2=0.40$]{2006AJ....131.2766R} with zero extinction. The distribution of $\Gamma_1-\Gamma_2$ towards low values along the red dashed line is indicative of possible dust reddening, whilst sources towards more negative $\Gamma_1$ values are objects with possible host-galaxy contamination. 
The $\Gamma_1-\Gamma_2$ plane is also very useful to identify unusual SEDs or SEDs characterised by bad photometry, which are then excluded from the sample.

We selected all the sources in the ($\Gamma_{1}$, $\Gamma_{2}$) plane within a circle centred at the reference values for a standard quasar SED (see \citeads{rl15,lr16,rl19} for further details), with a radius corresponding to a reddening $\ebv\simeq0.1$. 

We note that our quasar selection based on photometry could still be affected by some contamination from the light of the host, especially in low redshift ($z\la 0.7$) AGN, whose flux values at 2500 \AA\ are located at the edge of the SDSS photometric coverage. Hence, any uncertainties in the estimate of the quasar UV continuum from the optical can make the 2500~\AA\ monochromatic fluxes less reliable and possibly overestimated. Low-redshift AGN are on average less luminous in the optical/UV, with $\lbol\la10^{44}$ erg s$^{-1}$, thus the contrast between nuclear continuum and host-galaxy emission is smaller with respect to higher luminosity objects. Moreover, the data quality of low-redshift/low-luminosity AGN is, on average, lower. 
Host-galaxy contamination can be minimised through a source-by-source spectral fitting, as we did for the local AGN sample, but this procedure is rather time consuming for samples of several hundred thousands of objects. 
We will further discuss possible issues for cosmology related to our selection in Section~\ref{Cosmological fits of the Hubble diagram}.

\subsection{X-ray absorption} 
\label{xrayabs}
Since X-ray fluxes may contain some level of absorption, which is naturally heavier in the soft band, we included only X-ray detections with a photon index $\gammax$ that falls within a range representative of unobscured quasars. For the majority of the sample, we adopted the following selection criterion, which also takes into account the uncertainties on $\gammax$, i.e. $\gammax - \delta \gammax \geq \Gamma_{\rm X, min} $ and $\gammax \leq \Gamma_{\rm X, max}$.
The values of $\Gamma_{\rm X, min}$ and $\Gamma_{\rm X, max}$ are chosen based on two considerations: the average $\gammax$ within that interval should roughly correspond to $\gammax\sim2$ with a dispersion of 0.2--0.3 (consistent with e.g. \citealt{2009ApJS..183...17Y}), and the $\Lx-\Lo$ relation should not present any systematic deviation from the assumed {\it true} slope of 0.6 (within uncertainties).

We thus proceeded as follows. We evaluated the $\Fx-\Fo$ relation in narrow redshift bins (so the effect of cosmology is negligible) for different choices of $\Gamma_{\rm X, min}$ and $\Gamma_{\rm X, max}$. For the SDSS--4XMM sample, we started by assuming a reasonable $\Gamma_{\rm X, max}$ of 2.8 and a varying $\Gamma_{\rm X, min}$ in the interval 1.4--1.9 with steps of 0.1, and converged to a $\Gamma_{\rm X, min}=1.7$.  
\rev{We checked that a smaller value of $\Gamma_{\rm X, min}$ (i.e. 1.6) would not change the results of our analysis, but we prefer to be conservative, even at the expenses of sample statistics.}
For the SDSS-4XMM sample, we selected only X-ray observations with a photon index satisfying the condition $\gammax-\delta\gammax>1.7$, and excluded the (few) objects with $\gammax>2.8$. The latter filter on $\gammax$ is needed to avoid strong outliers ($\sim$5\%) which may be due to observational issues such as incorrect background subtraction in one of the two bands. This $\gammax$ interval roughly corresponds to an average $\gammax\sim2.1-2.2$ and a dispersion of 0.3. The same selection is applied to all the other subsamples at $z<4$. 

For the higher redshift ($z>4$) sample, such a stringent criterion on $\gammax$ would exclude the majority of the objects, given their higher uncertainties. We thus decided to simply select all the objects with $\gammax\geq1.7$.

Given the observed $\gammax$ range (up to 2.8), some soft-excess \citepads[e.g.]{sobolewskadone2007,gliozzi2020} contribution for low-$z$ quasars might be still present. 
We have thus repeated the analysis further imposing an upper limit to the $\gammax$ range of 2.5, but, besides losing statistics, our results are not affected.

\subsection{Eddington bias}
\label{eddbias}
Owing to X-ray variability, AGN with an average X-ray intensity close to the flux limit of the observation will be observed only in case of a positive fluctuation. This introduces a systematic, redshift-dependent bias towards high fluxes, known as {\it Eddington bias}, which has the effect to flatten the $\Lx-\Lo$ relation.

To reduce this bias, we excluded all X-ray detections below a threshold defined as $\kappa$ times the intrinsic dispersion of the $\Lx-\Lo$ relation (LR16; \citeads{rl19}), specifically:
\begin{equation}
\label{fthr}
\log F_{2\,\rm keV,\,exp} - \log \fmin < \kappa \delta,
\end{equation}
where $F_{2\,\rm keV,\,exp}$ is the monochromatic flux at 2 keV expected from the observed rest-frame quasar flux at 2500 \AA\ with the assumption of a {\it true} $\gamma$ of 0.6, and it is calculated as follows:
\begin{equation}
\label{fexp}
\log F_{2\,\rm keV,\,exp} =(\gamma-1)\log(4\pi) + (2\gamma-2)\log d_{\rm L} + \gamma\log\Fo + \beta,
\end{equation}
where $d_{\rm L}$ is the luminosity distance calculated for each redshift with a fixed cosmology, and the parameter $\beta$ represents the pivot point of the non-linear relation in luminosities, $\beta=26.5-30.5\gamma\simeq8.2$\footnote{The value of the luminosity normalizations are chosen based on the average values for the entire sample.}.
$\fmin$ in equation~(\ref{fthr}) is obtained as detailed in Sections~\ref{SDSS-4XMM} and ~\ref{SDSS-Chandra}, whilst the product $\kappa \delta$ is a value estimated for all the subsamples that we constructed from archives (SDSS--4XMM, SDSS--\chandra) or surveys (XXL).

Specifically, we first computed the flux limit of each X-ray observation, for both the SDSS--4XMM and SDSS--\chandra samples (see Sections \ref{SDSS-4XMM} and \ref{SDSS-Chandra}). 
We minimised the Eddington bias by including only X-ray detections for which the minimum detectable flux $\fmin$ in that given observation is lower than the expected X-ray flux $F_{2\,\rm keV,\,exp}$ by a factor that is proportional to the intrinsic dispersion in the $\Lx-\Lo$ relation (we refer to Appendix~A in LR16 and \citeads{rl19}). On average, the minimum detectable monochromatic fluxes at 2 keV are approximately $4.6\times10^{-32}$ erg s$^{-1}$ cm$^{-2}$ Hz$^{-1}$ and $3\times10^{-32}$ erg s$^{-1}$ cm$^{-2}$ Hz$^{-1}$ for the SDSS--4XMM and SDSS--\chandra samples, respectively. However, we caution that these values should not be considered as the ``survey limiting fluxes'', since both the 4XMM and CSC2.0 catalogues are not proper flux-limited samples, but rather a collection of all X-ray observations performed over a certain period. It is thus not trivial to estimate the expected minimum flux for these catalogues. 
The XXL sample is, instead, a ``standard'' flux-limited sample, so we applied a soft-band flux threshold to the data ($F_{\rm S}>10^{15}$ erg s$^{-1}$ cm$^{-2}$), which corresponds to a flux limit at 2 keV of $5\times10^{-32}$ erg s$^{-1}$ cm$^{-2}$ Hz$^{-1}$. We considered $\kappa\delta=0.9$ for SDSS--4XMM and XXL, whilst we used $\kappa\delta=0.5$ for the SDSS--\chandra sample.
All the other subsamples rely on pointed observations, so we did not apply any flux threshold to the data.

In principle, the effects of this bias could be further reduced if also non-detections were considered. Yet, this would not only complicate the statistical analysis, but also make the estimate of the intrinsic dispersion of the observed relations (e.g. $\Lo-\Lx$, Hubble diagram) much more uncertain. Moreover, we have shown that there is no significant variation in both the slope and the intercept of the $\Lo-\Lx$ correlation (within their uncertainties) among censored and detected samples once the Eddington bias is taken into account (see Appendix~A in LR16).
We therefore decided to include only detections in this work. This choice implies that we have to be very conservative in the correction for the Eddington bias, at the expense of sample statistics.

It is worth noting that our procedure to minimise the Eddington bias is slightly circular: we need the $\Lx-\Lo$ relation (i.e. we assumed $\gamma=0.6$) in order to estimate the `expected' X-ray flux. Yet, our simulations show that we are able to retrieve the assumed cosmology (using different input values for $\om$ and $\ol$), when the selection criteria are applied to mock quasar samples.

\subsection{The final cleaned sample}
\label{samplefin}
Summarizing, we applied a series of selection criteria to filter all the data that are likely contaminated by dust reddening, host-galaxy contamination, and X-ray absorption, or affected by the Eddington bias. We first selected all quasars within a circle centered at $(\Gamma_1,\Gamma_2)=(0.82,0.4)$, i.e. $\ebv=0$, and with a radius such as:
\begin{equation}
\sqrt{(\Gamma_1-0.82)^2 + (\Gamma_2-0.40)^2}\leq1.1,
\end{equation}
which corresponds to an $\ebv\la0.1$.
The equation above filters out all quasar SEDs that show reddening in the UV, significant host-galaxy contamination in the near-infrared, as well as bad photometry (see \S\,\ref{dusthost}). We then applied an additional cut to keep only the X-ray observations where photon indices are indicative of low levels of X-ray absorption, and to exclude the X-ray data characterized by peculiar photon indices, especially at low/moderate redshifts (see \S\,\ref{xrayabs}). Specifically, we required that:
\begin{equation}
\begin{cases} \gammax+\delta\gammax\geq 1.7 \mbox{ and } \gammax\leq2.8, & \mbox{if } z<4 \\ \gammax\geq 1.7, &  \mbox{if } z\geq4.  \end{cases}
\end{equation}
To correct for the Eddington bias, we further selected all observations that satisfy equation~(\ref{fthr}) where the product $\kappa\delta$ is 0.9 for the SDSS--4XMM and XXL subsamples, and 0.5 for SDSS--\chandra. Pointed observations are available for the local, $z\simeq3$, and high-redshift samples (see \S\,\ref{eddbias}).
For any quasar, all the multiple X-ray observations that survive the filters above are finally averaged to minimize the effects of X-ray variability (e.g. LR16, see also \citeads{2019AN....340..267L}).

The final cleaned sample is composed by 2,421 quasars spanning a redshift interval $0.009\leq z\leq7.52$, with a mean (median) redshift of 1.442 (1.295). Table~\ref{tbl1} summarizes the statistics of each subsample, whilst a more detailed summary of the various subsamples after a given selection is provided in Table~\ref{tbl:sel}. 
The main UV and X-ray properties of the final sample are presented in Table~\ref{tbl:sample}.

\begin{table*}
\caption{Properties of the final quasar sample}              
\label{tbl:sample}      
\centering                                      
\begin{tabular}{lcccccccc}          
\hline\hline                        
 Name & ra & dec & $z$ & $\fo$ & $\fx$ & Group & $\gammax$ & $\rm DM$ \\    
\hline                                   
  030341.04$-$002321.9 & 45.92103 & $-$0.38942 & 3.235 & $-$27.00 $\pm$ 0.04 & $-$31.38 $\pm$ 0.04 & 1 & 1.87 $\pm$ 0.12 & 46.42 $\pm$ 0.31\\
  030449.85$-$000813.4 & 46.20775 & $-$0.13708 & 3.296 & $-$26.98 $\pm$ 0.01 & $-$31.24 $\pm$ 0.03 & 1 & 1.99 $\pm$ 0.29 & 45.62 $\pm$ 0.20\\
  090508.88$+$305757.3 & 136.28702 & 30.96593 & 3.034 & $-$26.97 $\pm$ 0.01 & $-$31.14 $\pm$ 0.02 & 1 & 2.12 $\pm$ 0.09 & 44.90 $\pm$ 0.17\\
\hline                                             
\end{tabular}
\tablefoot{This table is presented in its entirety in the electronic edition; a portion is shown here for guidance. \\Fluxes are in units of log(erg~s$^{-1}$cm$^{-2}$). The Group column flags the different subsamples: 1 = \xmm $z\simeq3$ sample, 2 = new \xmm $z\simeq4$ quasar, 3 = High$-z$ sample, 4 = XXL, 5 = SDSS -- 4XMM, 6 = SDSS -- \chandra, 7 = local AGN. \revs{The column $\rm DM$ reports the distance moduli (with uncertainties) to reproduce the top panel of Figure~\ref{hubbleclean}.}}
\end{table*}

   \begin{figure}
   \centering
   \includegraphics[width=\linewidth]{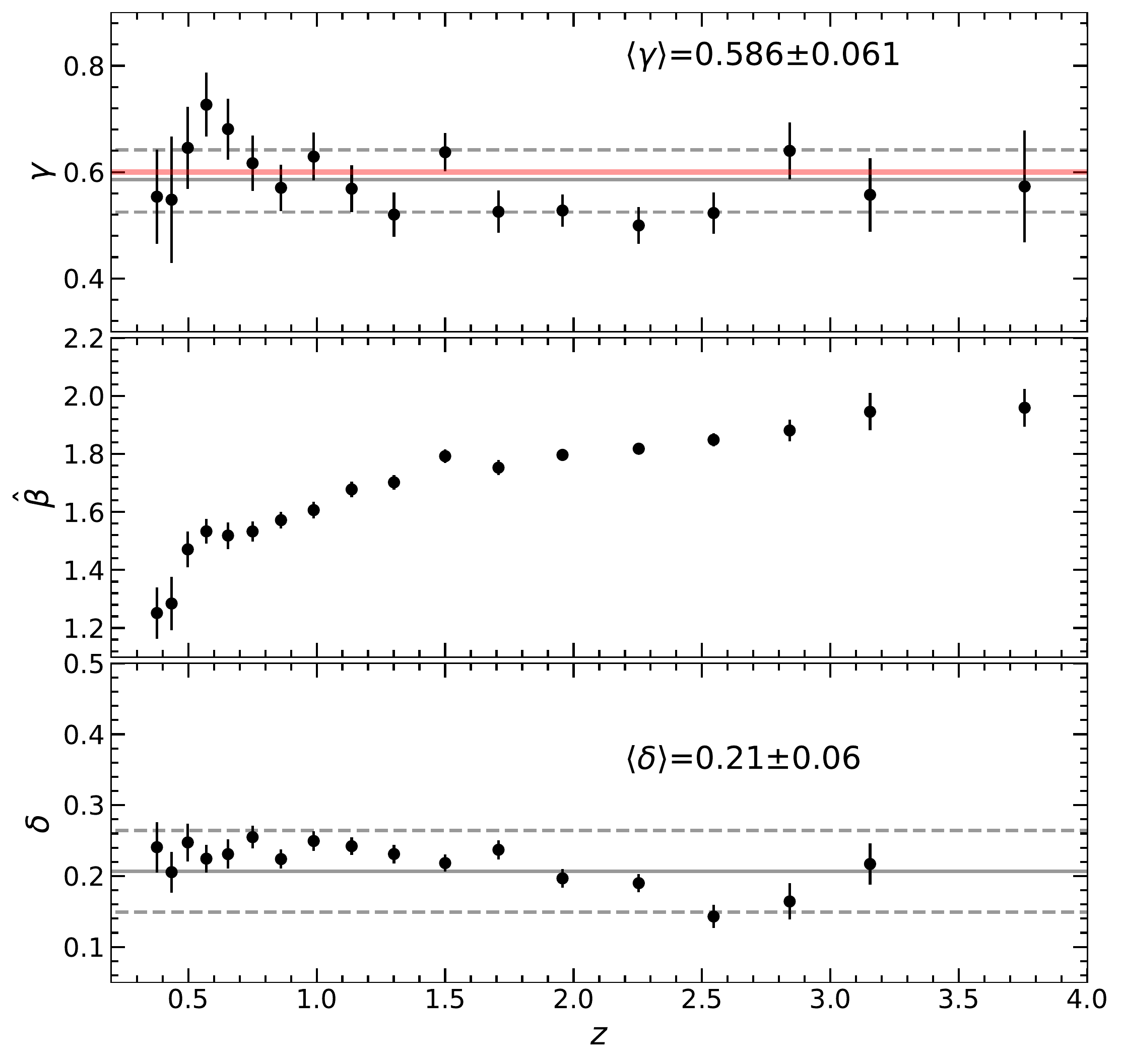}
   \caption{\rev{Redshift evolution of the slope $\gamma$, the \revs{intercept $\hat\beta$ and the dispersion $\delta$ of the $\Fx-\Fo$ relation}. To perform the regression fit, X-ray and UV fluxes have been normalised to $10^{28}$ and $10^{30}$ erg s$^{-1}$ cm$^{-2}$ Hz$^{-1}$, respectively. The data points in each panel are averages in narrow redshift bins ($\Delta z\simeq0.06-0.45$). Error bars represent the 1\,$\sigma$ uncertainty on the mean in each bin. The grey solid and dashed lines are the means and 1\,$\sigma$ uncertainties, respectively, on the slope ($\gamma$) and on the dispersion ($\delta$). The red line marks $\gamma=0.6$.} 
   }
              \label{fig:gdz}
    \end{figure}

\section{Analysis of the $\Fx-\Fo$ relation with redshift}
\label{Analysis of the relation with redshift}
Before building the Hubble diagram, we need to check whether the $\Fx-\Fo$ relation for the clean quasar sample shows any trend with redshift. We thus divided the sample in narrow redshift bins, with a variable step $\Delta z=0.06-0.45$ within the redshift range 0.45--4 to have enough statistics. The redshift step is chosen to ensure that the dispersion in distances over each interval is smaller than the one of the relation in luminosities. In this way, we can consider fluxes as proxies of luminosities.

The best-fit parameters (slope, intercept and dispersion) of the $\Fx-\Fo$ relation and their uncertainties are shown in Figure~\ref{fig:gdz}, whilst all the fits of the $\Fx-\Fo$ relation in the chosen redshift bins are presented in Figure~\ref{fig:fxfuv}. They are obtained through the Python package \emcee \citep{2013PASP..125..306F}, which is a pure-Python implementation of Goodman \& Weare's affine invariant Markov chain Monte Carlo (MCMC) ensemble sampler. 
\rev{To perform the regression fit, X-ray and UV fluxes were normalised to $10^{28}$ and $10^{30}$ erg s$^{-1}$ cm$^{-2}$ Hz$^{-1}$, respectively.}
On average, the $\Fx-\Fo$ slope does not show any clear trend with redshift within the analysed interval. \revs{Conversely, the trend of the intercept $\hat\beta$ of the normalized $\Fx-\Fo$ relation observed in the middle panel of Figure~\ref{fig:gdz} just reflects the overall shape of the quasar Hubble diagram (see Section~\ref{The Hubble diagram}). We note that, the trend of $\hat\beta$ with redshift, is not exactly the same as the one in Figure~\ref{hubbleclean} since such a parameter does not have a simple direct proportional dependence on the distance modulus (equation~\ref{dl}) because of the different dependence between slope and normalization in each redshift bin.}

The sample statistics is so sparse at redshift higher than 4 that we cannot provide a meaningful fit of the relation. None the less, we have checked that the data points at $z>4$ do not show any trend with redshift in the residuals of the Hubble diagram (see Section~\ref{Study of systematics in the Hubble diagram}).
In fact, these data points are extremely useful to set the shape of the Hubble diagram, thus providing better constraints on the measurements of the expansion rate of the Universe.
We thus confirm that the slope of the X-ray to UV relation shows no redshift evolution up to $z\sim4$, in agreement with our previous works (e.g. \citeads{rl15,lr16,lr17,rl19,lusso19}; see also \citeads{salvestrini2019} for a high redshift analysis).

\section{The quasar Hubble diagram}
\label{The Hubble diagram}
To fit the Hubble diagram we first need to derive the distance modulus for each object. We start by computing the luminosity distance (e.g. see \citeads{rl15,rl19}) as:
\begin{multline}
\label{dl}
\log d_{\rm L} = \frac{\left[\log\Fx -\beta -\gamma(\log\Fo+27.5) \right]}{2(\gamma-1)} +\\-\frac{1}{2}\log(4\pi) + 28.5,
\end{multline}
where $\Fx$ and $\Fo$ are the flux densities (in erg s$^{-1}$ cm$^{-2}$ Hz$^{-1}$). $\Fo$ is normalised to the (logarithmic) value of 27.5 in the equation above, whilst $d_{\rm L}$ is in units of cm and is normalised to 28.5 (in logarithm). The slope of the $\Fx-\Fo$ relation, $\gamma$, is a free parameter, and so is the intercept $\beta$\footnote{\revs{The intercept $\beta$ of the $\Lx-\Lo$ relation is related to the one of the $\Fx-\Fo$, $\hat\beta$ (see \S\,\ref{Analysis of the relation with redshift}), as $\hat\beta(z)=2(\gamma-1)\log d_{\rm L}(z) + (\gamma-1)\log 4\pi + \beta$.}}. The distance modulus, DM, is thus:
\begin{equation}
\dm = 5 \log d_{\rm L} - 5 \log (10\,{\rm pc}),
\end{equation}
and the uncertainty on $\dm$, $d\dm$, is:
\begin{multline}
\label{dDM}
d\dm = \frac{5}{2(\gamma-1)} \left[ \left(d\log\Fx\right)^2 + \left(\gamma d\log\Fo\right)^2 + \left(d\beta\right)^2 + \right. \\
\left. + \left( \frac{d\gamma \left[\beta+\log\Fo+27.5-\log\Fx\right]}{\gamma-1} \right)^2\right]^{1/2},
\end{multline}
where $d\log\Fx$ and $d\log\Fo$ are the logarithmic uncertainties on $\Fx$ and $\Fo$, respectively. Equation~\ref{dDM} assumes that all the parameters are independent, and takes into account also the uncertainties on $\beta$ and $\gamma$.
The fitted likelihood function, $LF$, is then defined as:
\begin{equation}
\label{lf}
\ln LF = - \frac{1}{2} \sum_i^N\left(\frac{(y_i-\psi_i)^2}{s_i^2} - \ln s^2_i\right)
\end{equation}
where $N$ is the number of sources, $s_i^2 = d y_i^2 +\gamma^2 d x_i^2 + \exp(2\ln\delta)$ takes into account the uncertainties on both the $x_i$ ($\log \Fo$) and $y_i$ ($\log \Fx$) parameters of the fitted relation, whilst $\delta$ represents its intrinsic dispersion. The variable $\psi$ is the modelled X-ray monochromatic flux, defined as:
\begin{equation}
\label{model}
\psi = \log F_{\rm X,\,mod} = \beta + \gamma(\log\Fo+27.5) +2(\gamma-1)(\log d_{\rm L,\,mod} -28.5),
\end{equation}
and is dependent upon the data, the redshift and the cosmological model assumed for the distances (e.g. $\Lambda$CDM, $w$CDM or a polynomial function). 
We fitted the data with a luminosity distance described by a fifth-grade polynomial of $\log(1+z)$, where the cosmographic function is:
\begin{equation}
  \begin{gathered}
  \label{dl_mod}
d(z)_{\rm L,\,mod}=k \ln(10)\frac{c}{H_0}[\log(1+z)+a_2\log^2(1+z)+\\
        +a_3\log^3(1+z)+a_4\log^4(1+z)+a_5\log^5(1+z)] +\\
        +\emph{O}[\log^6(1+z)],
  \end{gathered}
\end{equation}
where $k, a_2, a_3$, $a_4$ and $a_5$ are free parameters. 

For any analysis that involves a detailed test of cosmological models, we should cross-calibrate quasar distances making use of the {\it distance ladder} through Type Ia supernove. In fact, the $\dm$ values of quasars are not {\it absolute}, thus a cross-calibration parameter ($k$) is needed. The parameter $k$ should be fit separately for SNe Ia and quasars (i.e. $k$ is a rigid shift of the quasar Hubble diagram to match the one of supernovae). 

   \begin{figure}
   \centering
   \includegraphics[width=\linewidth]{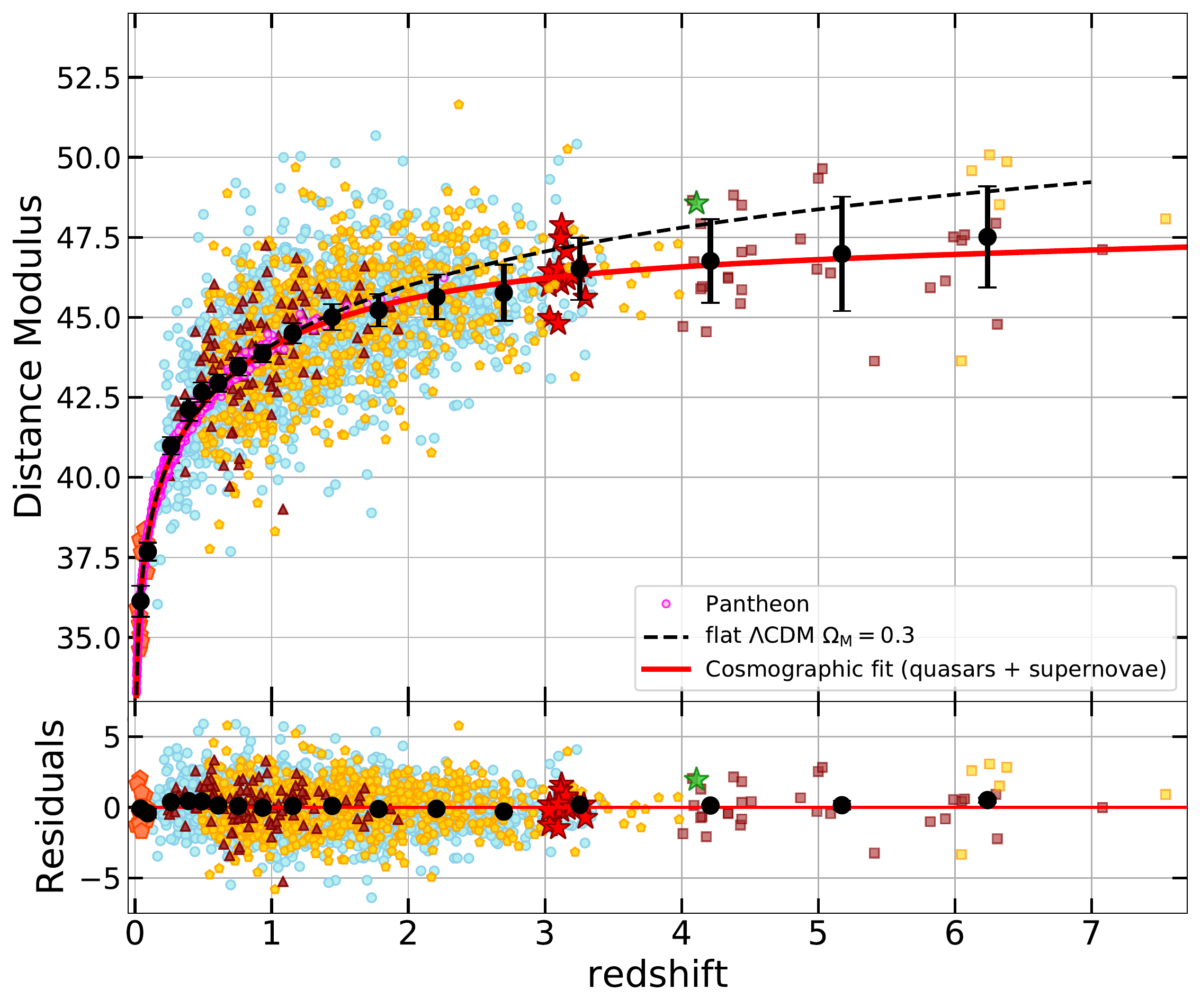}   
   \includegraphics[width=\linewidth]{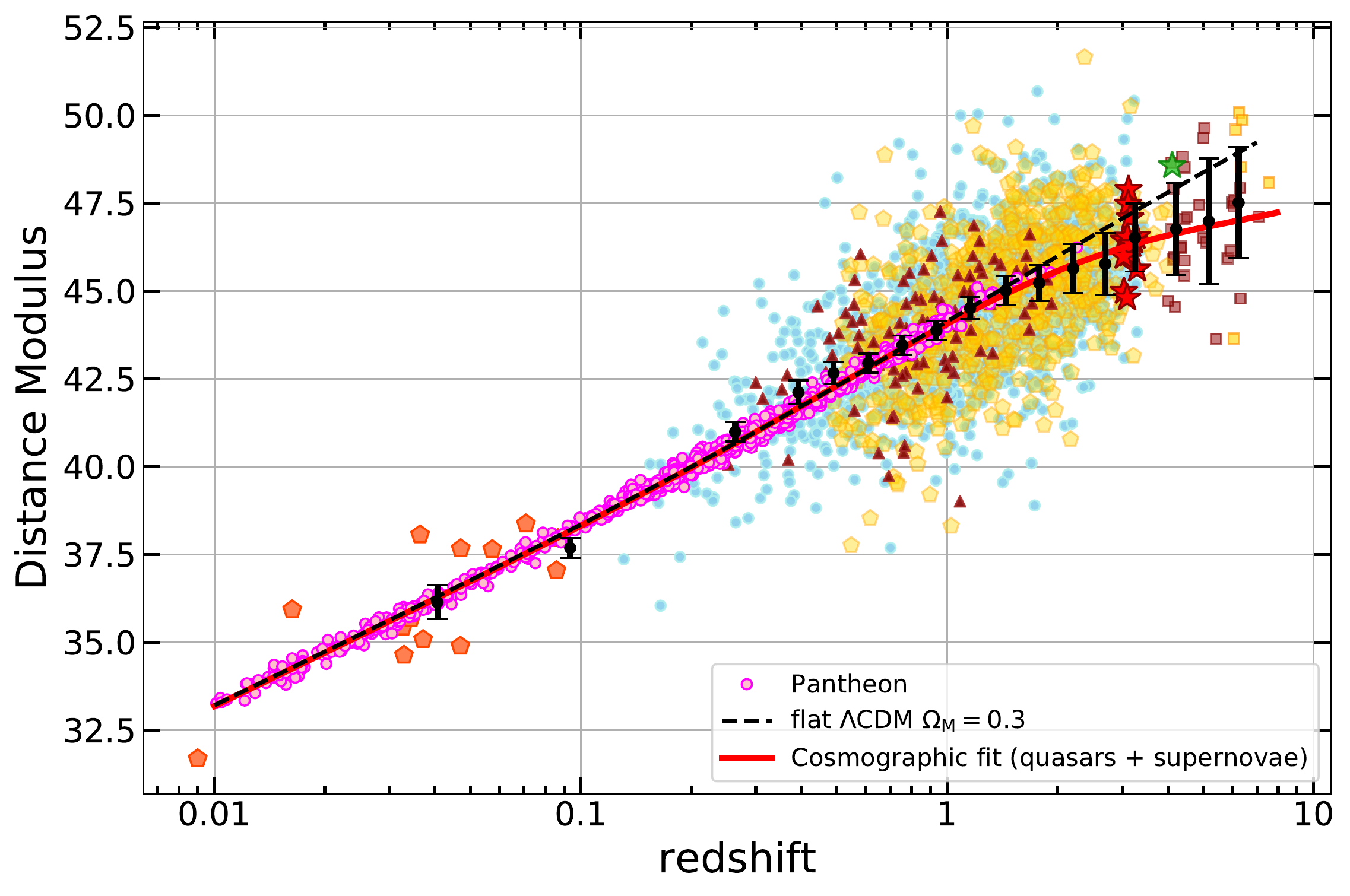}
   \caption{Top panel: distance modulus--redshift relation (Hubble diagram) for the clean quasar sample and Type Ia supernovae ({\it Pantheon}, magenta points). Symbol keys are the same as in Figure~\ref{loz}. The red line represents a fifth order cosmographic fit of the data, whilst the black points are averages (along with their uncertainties) of the distance moduli in narrow (logarithmic) redshift intervals. The dashed black line shows a flat $\Lambda$CDM model fit with $\om= 0.3$. The middle panel shows the residuals with respect to the cosmographic fit and the black points are the averages of the residuals over the same redshift intervals. Bottom panel: distance modulus--redshift relation plotted with a logarithmic horizontal axis scale to better visualise the agreement between Type Ia SNe and quasars in the low-redshift range. 
}
              \label{hubbleclean}
    \end{figure}
Whilst in our previous works we kept $\gamma$ fixed, in this analysis we have marginalized over the slope $\gamma$ of the $\Lx-\Lo$ relation. The latter approach is preferred to check whether any degeneracy of the slope with the other parameters is present, and whether the statistical significance of the deviation from the $\Lambda$CDM model can be affected by the assumption of a $\gamma$ value that slightly deviates from the {\it true} one. The marginalization on $\gamma$ is a more conservative procedure, hence it might reduce the significance of the deviation with respect to the same MCMC analysis with $\gamma$ fixed. Therefore, if a statistical deviation persists even allowing for a variable $\gamma$, its significance should be considered as an indicative lower limit with respect to the case where $\gamma$ is fixed. 

We finally note that the Hubble constant $H_0$ in equation~(\ref{dl_mod}) is degenerate with the $k$ parameter, so it can assume any arbitrary value and was fixed to $H_0=70$~km s$^{-1}$ Mpc$^{-1}$ (see also \citeads{lusso19}).

Figure~\ref{hubbleclean} shows the Hubble diagram for the clean quasar sample, combined with the most updated compilation of Type Ia supernovae from the {\it Pantheon} survey \citepads{scolnic2018}. 
The best MCMC cosmographic fit is shown with the red line, whilst black points are the averages (along with their uncertainties) of the distance modulus in narrow (logarithmic) redshift intervals, plotted for clarity purposes only. The residuals are displayed in the \rev{middle} panel with the same symbols, and do not reveal any apparent trend with redshift. 
The MCMC fit assumes uniform priors on the parameters. More details on our cosmographic technique will be provided in a companion publication (Bargiacchi et al., in preparation). 

We confirm that, while the Hubble diagram of quasars is well reproduced by a standard flat $\Lambda$CDM model (with $\om=0.3$) up to $z\sim1.5$, as shown in the \rev{top} panel of Figure~\ref{hubbleclean}, a statistically significant deviation emerges at higher redshifts, in agreement with our previous works (e.g. \citeads{rl15,rl19,lusso19}) and other works on the same topic (e.g. \citeads{divalentino2020}). 

The detailed discussion of the cosmological implications of this deviation and its statistical significance is not the main aim of this analysis. Here we want to focus on the study of possible systematic effects that could drive this deviation instead.


\section{Cosmological fits of the Hubble diagram}
\label{Cosmological fits of the Hubble diagram}
In this Section we want to test our quasar sample and our method by fitting a ``physical'' cosmological model. Our aim is not to fully explore the consequences of our new Hubble diagram for the determination of cosmological parameters, or for the tests of different cosmological models, which will be presented in subsequent papers. Here we only intend to verify how different choices regarding the fitting method and the quasar subsample affect the final results. We choose to perform these tests with a flat $w_0w_a$CDM model, which is the simplest and most commonly used extension of the standard $\Lambda$CDM model, where the parameter $w$ of the equation of state of the dark energy is assumed to vary with redshift according to the parametrization $w(z)=w_0+w_a\times(1-a)$, where $a=1/(1+z)$. 
\begin{figure}
 \centering
  \includegraphics[width=\linewidth]{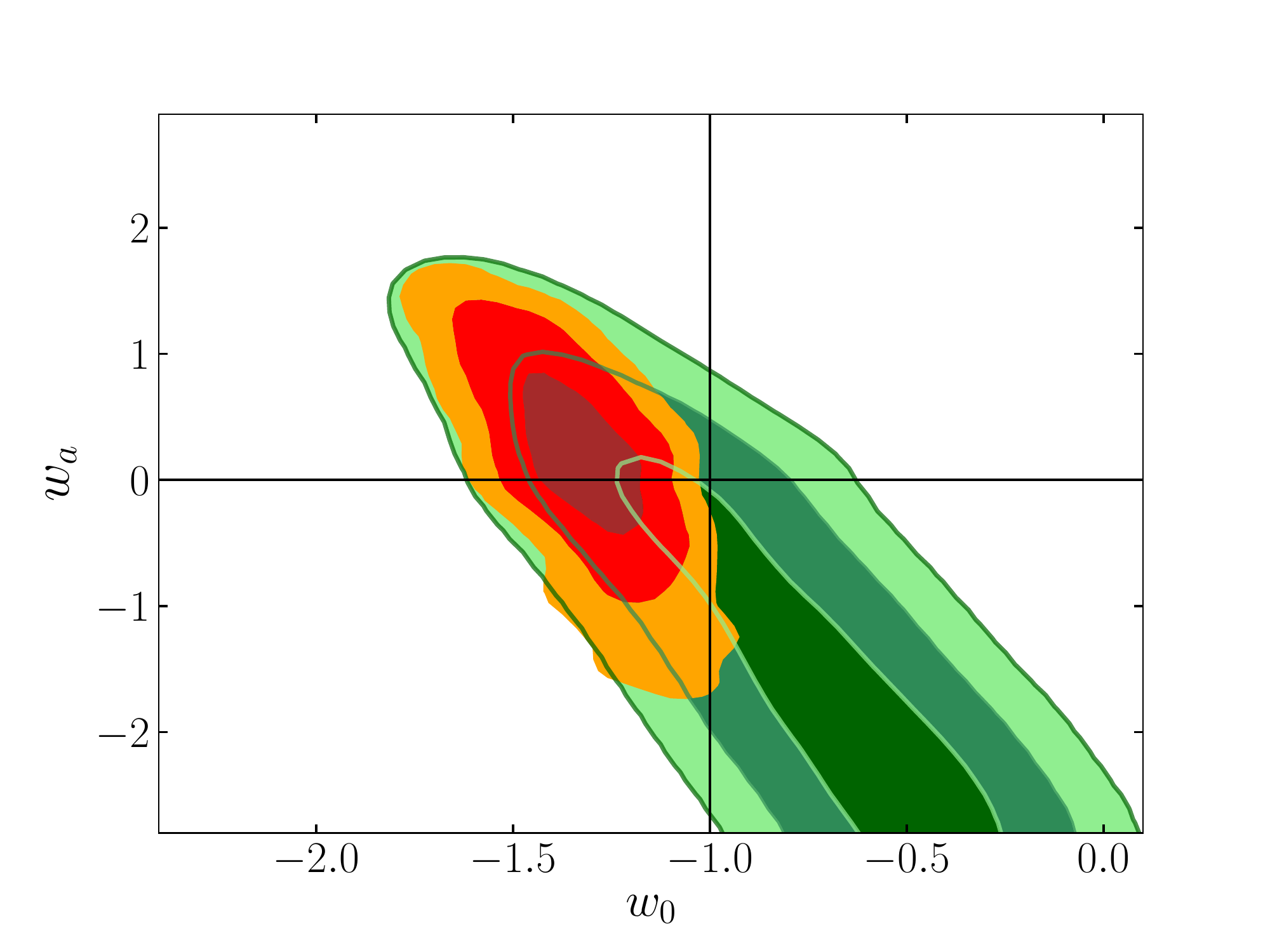}
  \caption{Results from a fit of a $w_0w_a$CDM model to the combined Hubble diagram of supernovae and the ``best'' quasar sample, i.e. removing objects at $z<0.7$ with a photometric determination of the UV flux (see the text for details). The green contours refer to the CMB results from {\it Planck} \citepads{planck2018}. The orange (3$\sigma$), red (2$\sigma$), and brown (1$\sigma$) contours are obtained by adding the constraints from the Hubble diagram of supernovae and quasars.}
  \label{hubblediag}
\end{figure}
Based on the analysis presented in the previous sections, three points deserve further consideration regarding cosmological fits:
\begin{itemize}
	\item We can use the full quasar sample or add a filter of $z>0.7$ for the sources with photometric determination of the UV flux. As discussed in Section~\ref{dusthost}, the possible uncertainties in the extrapolation from the optical of the quasar UV continuum at low redshift, where the host-galaxy contamination can be important, make the 2500~\AA\ monochromatic luminosities less reliable at $z<0.7$. In particular, this effect is likely to be more severe at lower fluxes/luminosities, where the data quality is also lower. If the continuum slope at UV wavelengths becomes steeper than in the optical, the actual 2500~\AA\ flux would be underestimated and this could explain the higher average values of the slope of the $\Lx-\Lo$ relation at $z<0.7$ (see Figure~\ref{fig:gdz}). This point deserves further investigation, which is deferred to a subsequent paper. To define the optical sample for cosmological applications in a conservative way, we thus prefer to cut the quasar sample at $z>0.7$, with the exception of the local sources discussed in Section~\ref{local}, whose 2500~\AA\ flux is determined  from the UV spectra without extrapolations. 
	The results of the fit of this ``best'' quasar sample with the flat $w_0w_a$CDM model are shown in Figure~\ref{hubblediag}. We can see that considering our data significantly reduces the $w_0-w_a$ parametric space with respect to the CMB analysis only  \citepads{planck2018}\footnote{Baseline $\Lambda$CDM chains with baseline likelihoods:\\ https://wiki.cosmos.esa.int/planck-legacy-archive/index.php/Cosmological\_Parameters}, still being compatible with the latter data at $1\sigma$. At the same time, the $\Lambda$CDM model (recovered for values $w_0=-1$ and $w_a=0$) is in tension with our data at more than 3$\sigma$, in agreement with RL19. 

Finally, we note that the role of quasars at $z<1$--1.3 is mainly to set the absolute calibration with supernovae in cosmological fits, with only a small contribution to the determination of the values and uncertainties of the cosmological parameters, given the much higher statistical weight of the supernovae. Removing quasars at $z<0.7$ should not affect the final results significantly, and the number of quasars at redshifts overlapping with supernovae remains high enough for a precise calibration. In order to test these expectations, we repeated the fit with the whole sample, obtaining the results shown in the first panel of Figure~\ref{hubblediag:appendix}, where the contours are nearly indistinguishable from those of Figure~\ref{hubblediag}.

	\item To fit the Hubble diagram, we can either adopt a fixed value of the slope of the relation (with its uncertainty), $\gamma=0.59\pm0.06$, as determined in Section~\ref{Analysis of the relation with redshift}, or marginalize on $\gamma$ as a free parameter, as discussed in Section~\ref{The Hubble diagram}. In general, the latter choice is more conservative and should be preferred. This is what we did for our reference fit, and also for the cosmographic fit used to analyze the residuals. Yet, it is worth noting that, in case of a mismatch in the shape of the Hubble diagrams of quasars and supernovae in the common redshift interval, leaving $\gamma$ as a free parameter allows us to partly alleviate this problem by slightly ``bending'' the $\Lx-\Lo$ relation in order to obtain a better agreement. In our case, the fit of the ``best'' sample gives $\gamma=0.600\pm 0.015$, consistent within 1.5\,$\sigma$ with the value found from the fits of the relation in narrow redshift bins. In order to check the possible effects of this choice, we repeated the fit with the $w_0w_a$CDM model, again obtaining a contour plot totally consistent with the reference case (second panel in Figure~\ref{hubblediag:appendix}). 
	
	\item The residuals in the Hubble diagram show a moderate, but statistically significant, redshift dependence on the X-ray slope, $\gammax$.  It is important to understand whether this introduces a bias in the fits of the Hubble diagram. We checked this possibility by splitting the sample in two parts, with $\gammax<2.2$ and $\gammax>2.2$, respectively, and repeating the fit with the $w_0w_a$CDM model. The results are shown in the last two panels of Figure~\ref{hubblediag:appendix}. These contours are slightly larger than in the previous cases, as expected given the lower statistics, but no systematic trend is observed. We conclude that, while the possible dependence on $\gammax$ deserves further analysis in order to understand its physical and/or observational origin and to reduce the dispersion of the $\Lx-\Lo$ relation (Signorini et al., in preparation), no systematic effect related to $\gammax$ is introduced in the Hubble diagram of quasars.
\end{itemize}

\section{Study of systematics in the Hubble diagram}
\label{Study of systematics in the Hubble diagram}
Since the main aim of our analysis is to check whether any systematic is present in the residuals of the quasar Hubble diagram, at this stage we avoid the inclusion of Type Ia supernovae. As noted above, when only quasars are involved the $\dm$ values should not be considered as proper absolute distances.
In this Section we thus present an in-depth investigation of possible systematics in the residuals of the quasar Hubble diagram, unaccounted for in the selection of the sample. In particular, we explored whether our procedure {\it (1)} to correct for the Eddington bias (Section\,\ref{eddbias}), {\it (2)} to neglect quasars with possible gas absorptions (Section\,\ref{xrayabs}), and {\it (3)} to select blue quasars based on their SED shape, where dust absorption and host-galaxy contamination are minimised (Section\,\ref{dusthost}), introduces spurious trends in the Hubble diagram residuals as a function of redshift and for different intervals of the relevant parameters. 
For each variable, we divided the sample between the sources that fall below and above the average value of the variable itself, and examined each subset separately as any hidden dependence should lead to a systematic difference between the two.

\subsection{Residuals as a function of the Eddington bias}
\label{residuals-eddbias}
To verify whether our adopted technique to correct for the Eddington bias, based on the assumption that the {\it true} slope of the $\Lx-\Lo$ is $\gamma=0.6$, leaves some hidden trends in the residuals of the Hubble diagram as a function of redshift, we defined an Eddington bias parameter, $\Theta$, as the difference between the expected X-ray monochromatic flux at 2 keV and the sum of the flux sensitivity value at 2 keV ($\fmin$) and the product $\kappa \delta$ (see equation~\ref{fthr}):
\begin{equation}
\label{theta}
\Theta = \log F_{2\,\rm keV,\,exp} - (\log \fmin + \kappa \delta).
\end{equation}
Given the fact that the source has survived the Eddington bias filter (see equation~\ref{fthr}, where $\kappa\delta=0.5$ for SDSS--\chandra and 0.9 for XXL and SDSS--4XMM, respectively), $\Theta$ is always positive: the higher its value, the lower the bias due to the flux limit of the specific X-ray observation. 

The other samples benefit from pointed observations, for which the bias due to the X-ray flux sensitivity is fully negligible.

\begin{figure}
   \centering
   \includegraphics[width=\linewidth]{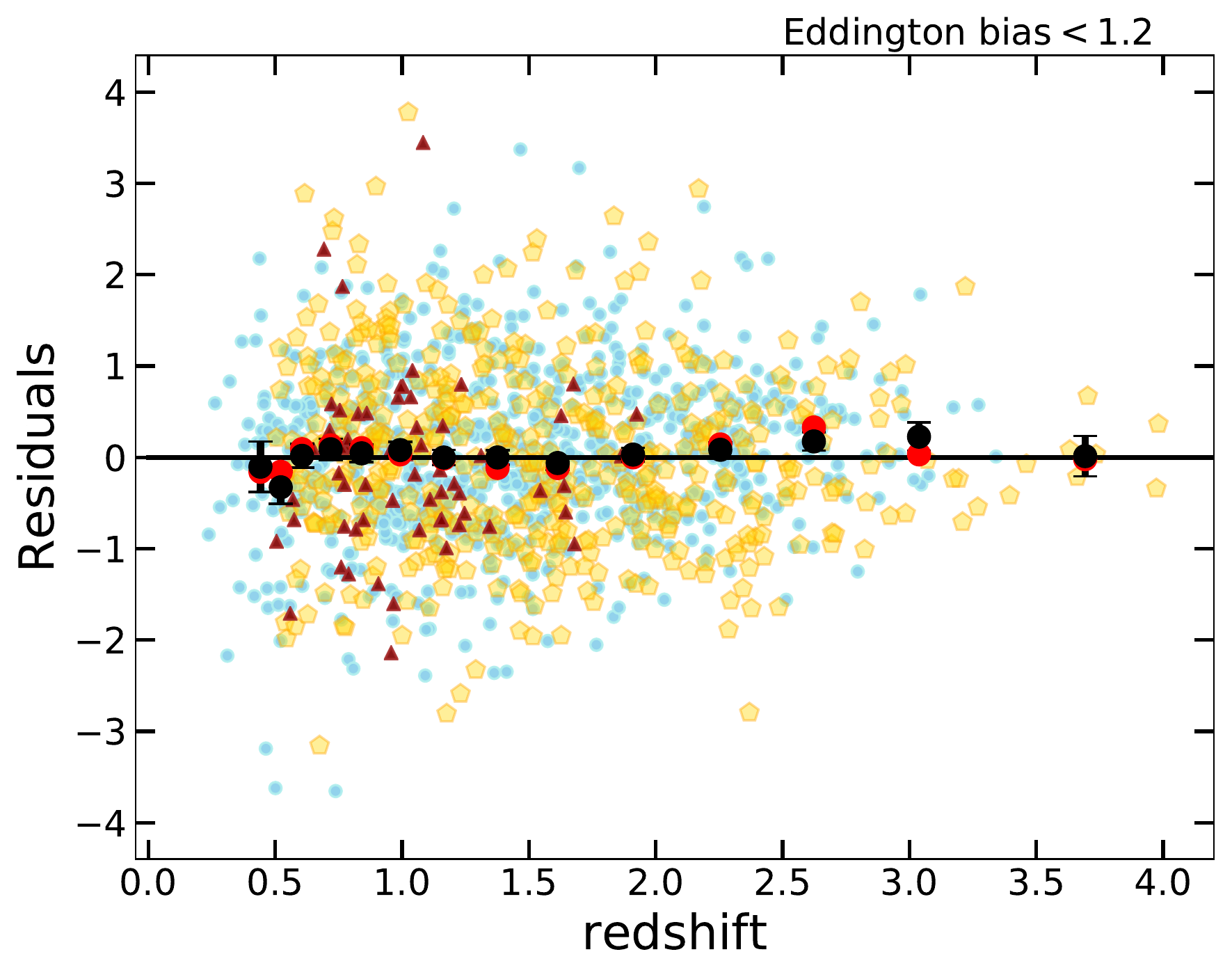}
   \includegraphics[width=\linewidth]{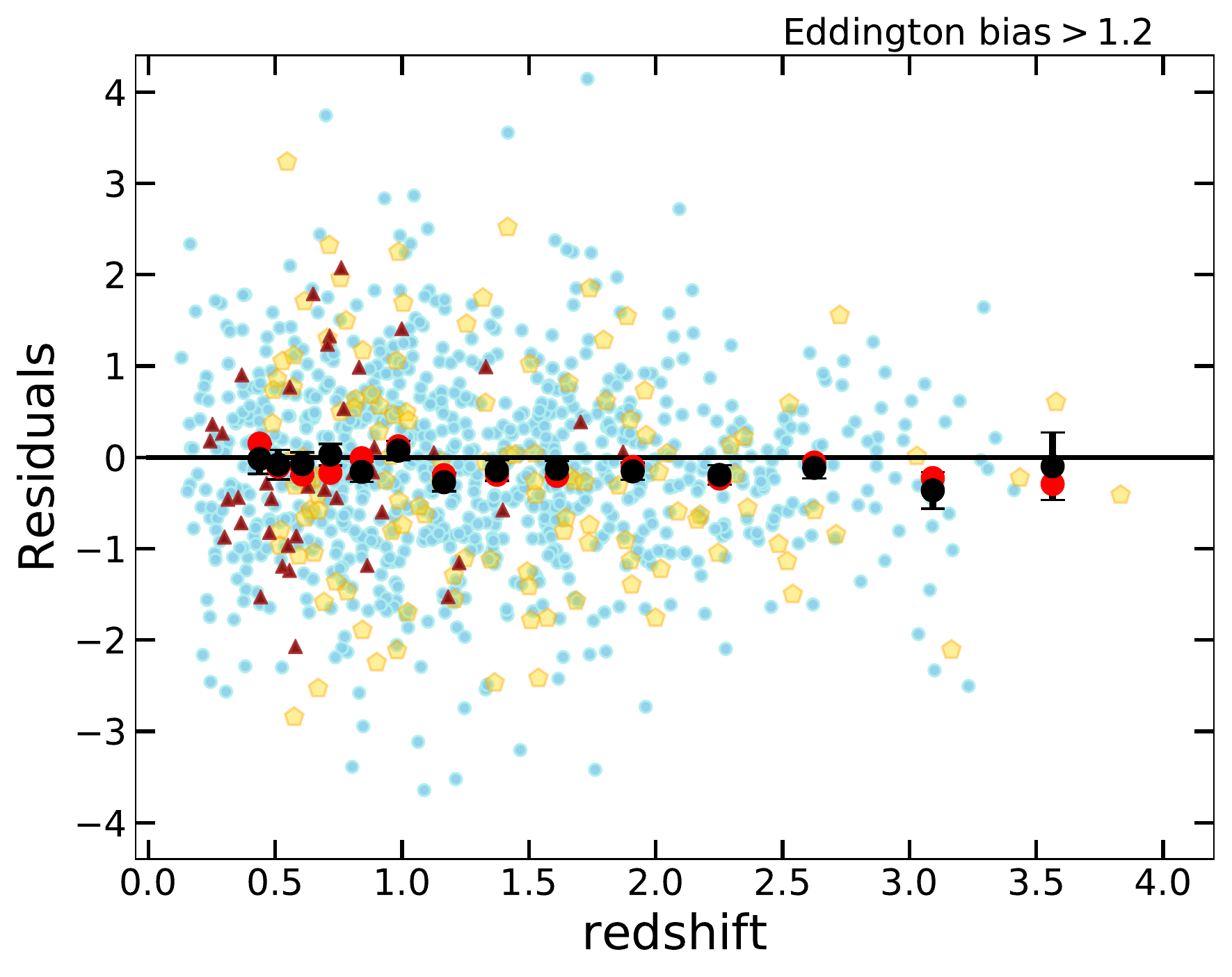}
   \caption{Distribution of the Hubble diagram residuals (\rev{middle} panel of Figure~\ref{hubbleclean}) as a function of redshift for the quasars with $\Theta$ lower (higher) than the average (i.e. $\langle \Theta\rangle=1.2$). The black and red points represent the mean and median of the residuals in narrow redshift intervals, respectively. Symbol keys as in Figure~\ref{loz}.}
              \label{fig:resid_eddbias}
    \end{figure}

We then defined two subsets, above (1069 objects) and below (1289) the mean of this distribution, $\langle \Theta\rangle=1.17\pm0.01$ (with a 1$\sigma$ dispersion of 0.33), and plotted the residuals of the Hubble diagram as a function of redshift in Figure~\ref{fig:resid_eddbias}. The three quasar samples span a rather large redshift interval, showing no clear trend of the residuals with distance and a similar dispersion around zero ($\simeq 1$), with an average residual value of about $-0.07$ ($0.04$) for $\Theta>1.2$ ($\Theta<1.2$). These results imply that our selection of X-ray observations discussed in Section~\ref{eddbias} does not introduce any systematic trends, at least up to $z\simeq4$.

\subsection{Residuals as a function of the photon index}
\label{residuals-gammax}
\begin{figure}
   \centering
   \includegraphics[width=\linewidth]{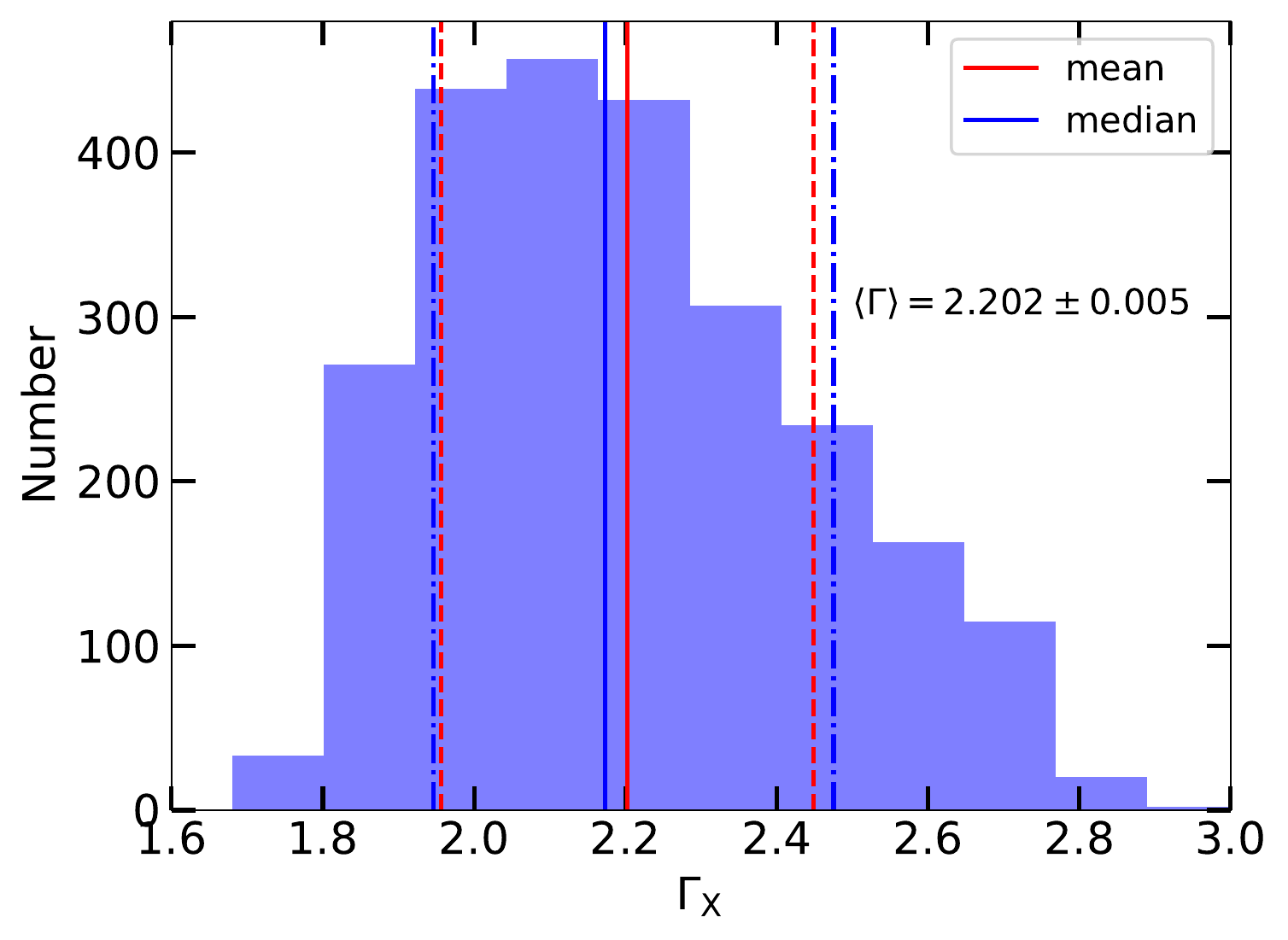}
   \caption{Distribution of the photon index $\gammax$ for the clean final sample. The mean and the statistical error on the mean are also quoted. The red (blue) solid line represents the mean (median) of $\gammax$, with its 1$\sigma$ dispersion around the average marked with the dashed (dot-dashed, 16\% and 84\%) lines.}
              \label{gammax}
    \end{figure}

Gas absorption in the X-ray band and strong outliers with extremely steep $\gammax$ are minimized by neglecting all quasars outside a given interval of photon index values ($1.7\leq\gammax\leq2.8$, see Section~\ref{xrayabs}). In Figure~\ref{gammax} we present the distribution of the $\gammax$ values for the clean quasar sample. The mean (median) value for the sample is $\langle \gammax\rangle \simeq2.202\pm0.005$ ($\langle \gammax\rangle \simeq2.173^{+0.006}_{-0.005}$), with a dispersion of about 0.23 (0.30). Statistical errors are quoted. 
   \begin{figure}
   \centering
   \includegraphics[width=\linewidth]{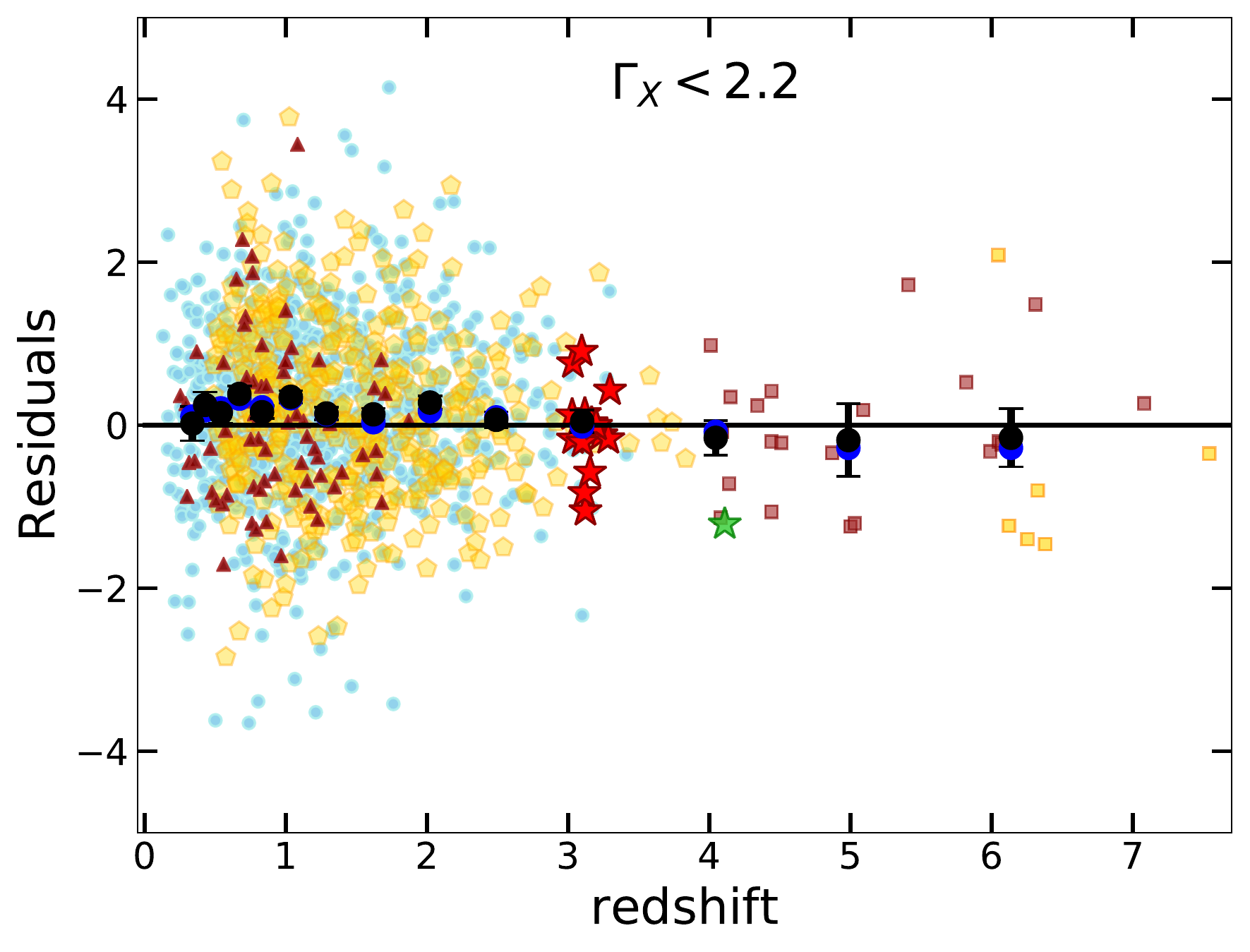}
   \includegraphics[width=\linewidth]{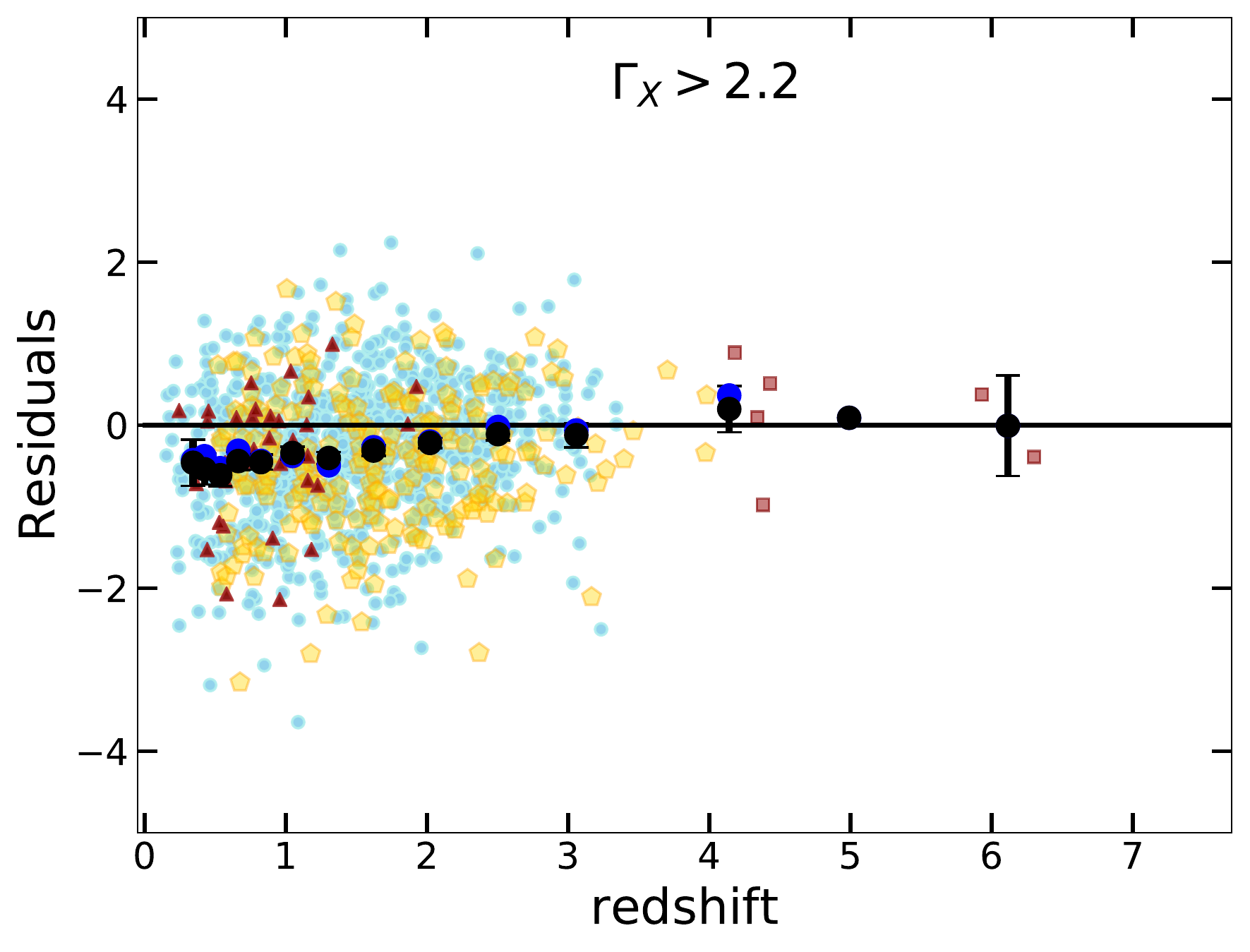}
   \caption{Distribution of the Hubble diagram residuals (\rev{middle} panel of Figure~\ref{hubbleclean}) as a function of redshift for the quasars with $\gammax$ lower (higher) than the average (i.e. $\langle \gammax\rangle=2.2$, see Figure~\ref{gammax}). The black and red points represent the mean and median of the residuals in narrow redshift intervals, respectively. Symbol keys as in Figure~\ref{loz}. 
   }
              \label{resid_gamma}
    \end{figure}

\begin{figure}
   \centering
   \includegraphics[width=\linewidth]{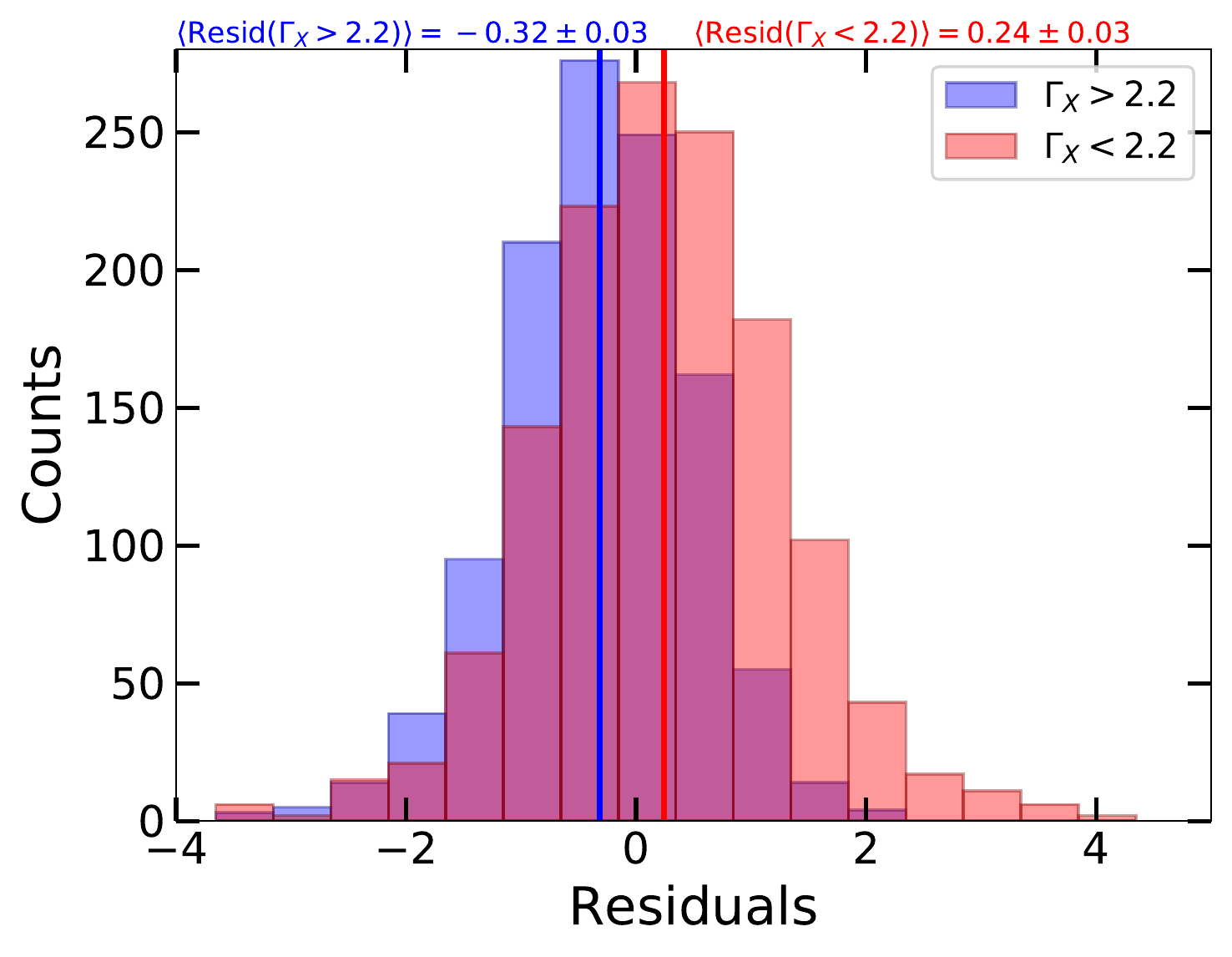}
   \caption{Distribution of the Hubble diagram residuals (\rev{middle} panel of Figure~\ref{hubbleclean}) for the quasars with $\gammax$ higher (blue) and lower (red) than the average (i.e. $\langle \gammax\rangle=2.2$, see Figure~\ref{gammax}). The blue and red solid lines represent the mean of the residuals in each $\gammax$ distribution, respectively (averages are reported on top). 
   }
              \label{resid_gamma_split}
    \end{figure}

The average $\gammax$ is biased towards slightly steeper values with respect to the more typical $\gammax\sim1.9$--2 
(e.g. $\gammax=1.99\pm0.01$ with a dispersion of 0.3, \citealt{scott2011}; see also \citealt{young09,mateos2010}),
which is likely due to our conservative cut at $\gammax=1.7$ and to the presence of a tail of sources with power laws softer than $\gammax=2.6$. The residuals of the Hubble diagram do not show a significant trend as a function of redshift when the sample is split in two subsets with $\gammax$ higher and lower than 2.2 (Figure~\ref{resid_gamma}). None the less, the marginalized distribution of residuals shown in Figure~\ref{resid_gamma_split} presents an offset with respect to zero, with an average value for the two subsamples of $-0.32\pm0.03$ and $0.24\pm0.03$ for $\gammax$ higher and lower than 2.2, respectively. 
The dispersion around the average values is 0.82 (1.05) for $\gammax>2.2$ ($\gammax<2.2$). 
 
 As we pointed out in Section~\ref{xrayabs}, our photometric $\gammax$ values may not be always accurate for individual objects, but they are reliable in a {\it statistical sense} over a large sample of quasars. Consequently, the main drawback of not using the spectroscopic values is likely to increase the dispersion, rather than to introduce a strong systematic with redshift that may affect the cosmological analysis. Moreover, even considering the presence of a small redshift trend in Figure~\ref{resid_gamma} for $z<2$, this is counterbalanced by the use of Type Ia supernove in the same redshift interval. 
 In fact, the statistical significance of the deviation of the quasar Hubble diagram from the standard $\Lambda$CDM reported in our previous works starts to be important for $z>2$, where all the residuals do not present any obvious systematic with redshift. 

\subsection{Residuals as a function of SED colours}
\label{residuals-colours}
As a further check, we have explored whether the SED colours produce some systematic trends with redshift in the residuals of the Hubble diagram. We have selected two subsamples in different regions of the $(\Gamma_1,\Gamma_2)$ plane (see Figure~\ref{redqso}). The first one considers all the quasars within the circle having a radius of 0.45 (i.e. $\sqrt{(\Gamma_1-0.82)^2 + (\Gamma_2-0.40)^2}\leq0.45$; 1271 quasars), which corresponds to a colour excess $\ebv$ of 0.04 (using an SMC-like reddening law). The second is the contiguous annulus with outer radius of 1.1 (1087 quasars) corresponding to an $\ebv\simeq 0.1$.
  As the majority of the quasars in the final sample are drawn from the SDSS--DR14 catalogue, the subsample with a colour excess lower than 0.04 should represents the bluest objects, whilst the sample in the outer annulus with $0.04\leq \ebv\leq0.1$ could show some pattern in the residuals in case of a colour-related systematic with redshift. 

\begin{figure}
   \centering
   \includegraphics[width=\linewidth]{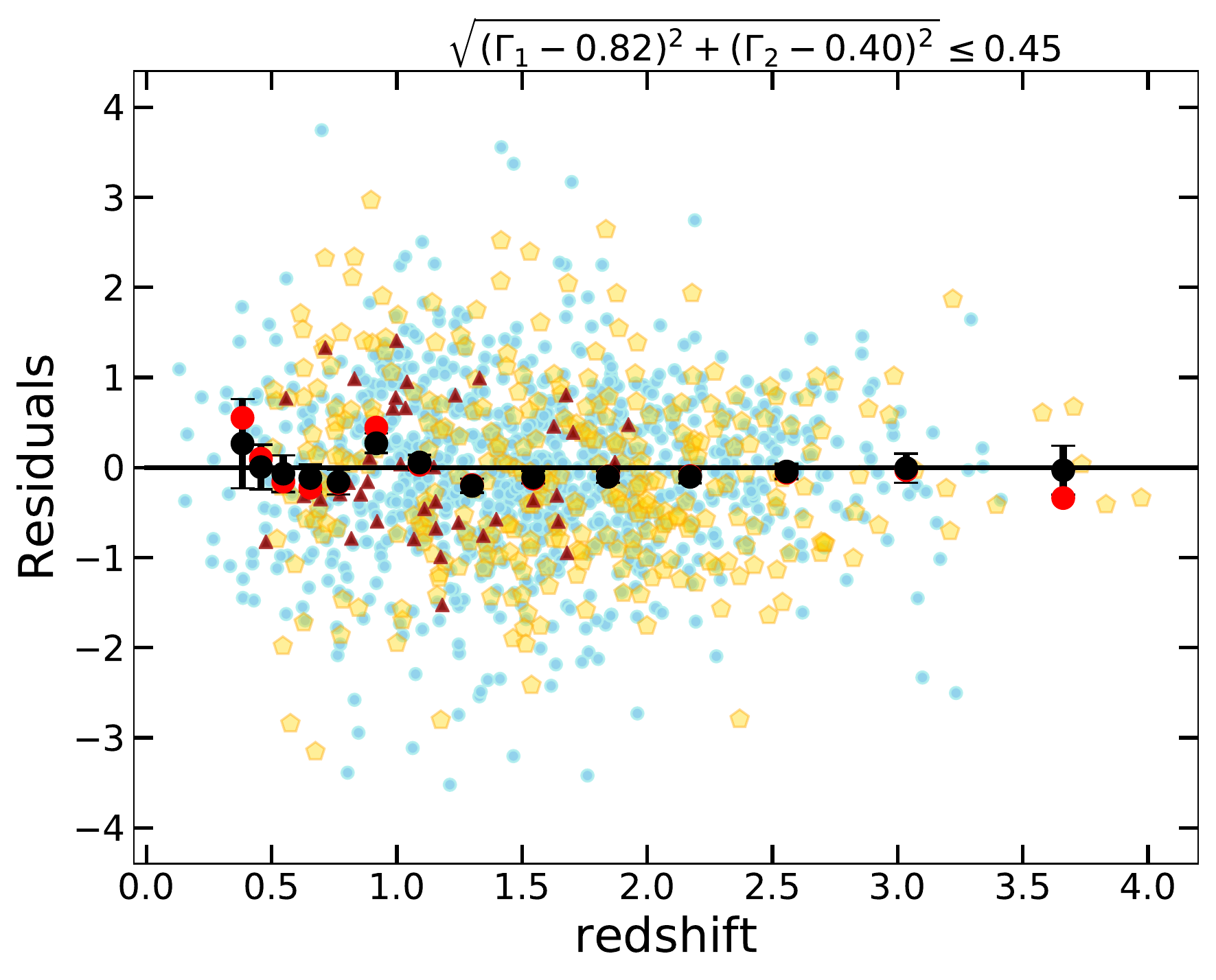}
   \includegraphics[width=\linewidth]{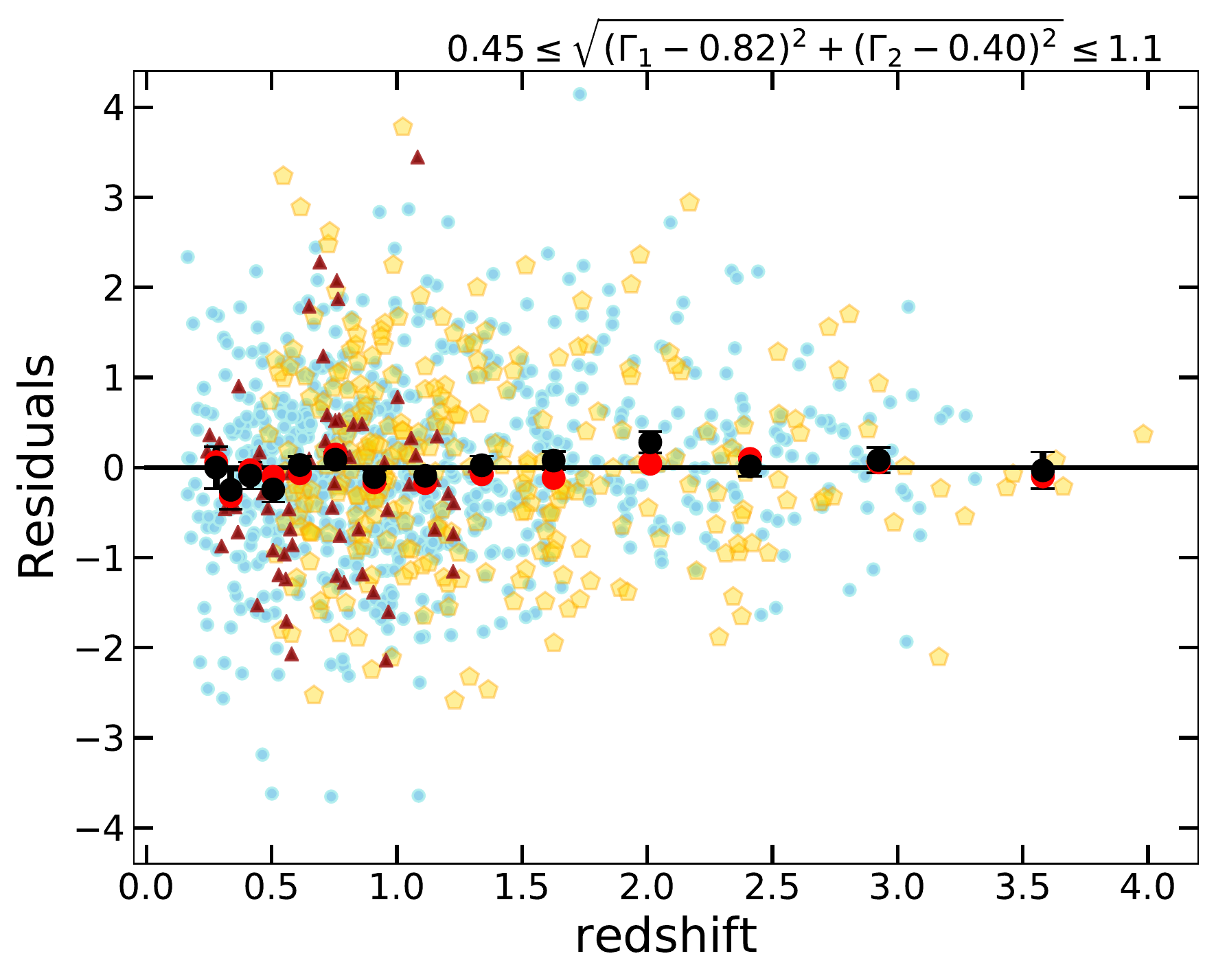}
   \caption{Distribution of the Hubble diagram residuals (\rev{middle} panel of Figure~\ref{hubbleclean}) as a function of redshift for the quasars with $\sqrt{(\Gamma_1-0.82)^2 + (\Gamma_2-0.40)^2}\leq0.45$, i.e. $\ebv\leq0.04$ (top panel, the bluest quasars in the clean sample) and $0.45\leq\sqrt{(\Gamma_1-0.82)^2 + (\Gamma_2-0.40)^2}\leq1.1$, i.e. $0.04\leq\ebv\leq0.1$ (bottom panel) in the clean SDSS--4XMM, XXL and SDSS--\chandra samples. The black and red points represent the mean and median of the residuals in narrow redshift intervals, respectively. Symbol keys as in Figure~\ref{loz}.}
              \label{fig:resid_colours}
    \end{figure}

Figure~\ref{fig:resid_colours} presents the distribution of the residuals as a function of redshift for the two subsamples defined as above for the clean SDSS--4XMM, XXL and SDSS--\chandra samples, since we can construct homogeneous SEDs (and thus retrieve $\Gamma_1$ and $\Gamma_2$ slopes) for them all using the SDSS photometry from the DR14 quasar catalogue. It is clear that there is no obvious trend of the residuals with redshift in either subsample, implying that our colour selection does not introduce any redshift-dependent bias in the Hubble diagram. Moreover, in both cases there is no difference in the dispersion of the residuals around zero ($\simeq 1$), with average values of about $-0.04$ for $\ebv\leq0.04$ and $0.03$ for $0.04\leq\ebv\leq0.1$). This further proves that even the inclusion of sources with possible (yet modest) contamination from dust and/or host galaxy does not lead to any systematics with redshift up to $z\simeq4$.

\subsection{Is dust/gas absorption driving the deviation in the quasar Hubble diagram?}
\label{dust}
   \begin{figure}
   \centering
   \includegraphics[width=\linewidth]{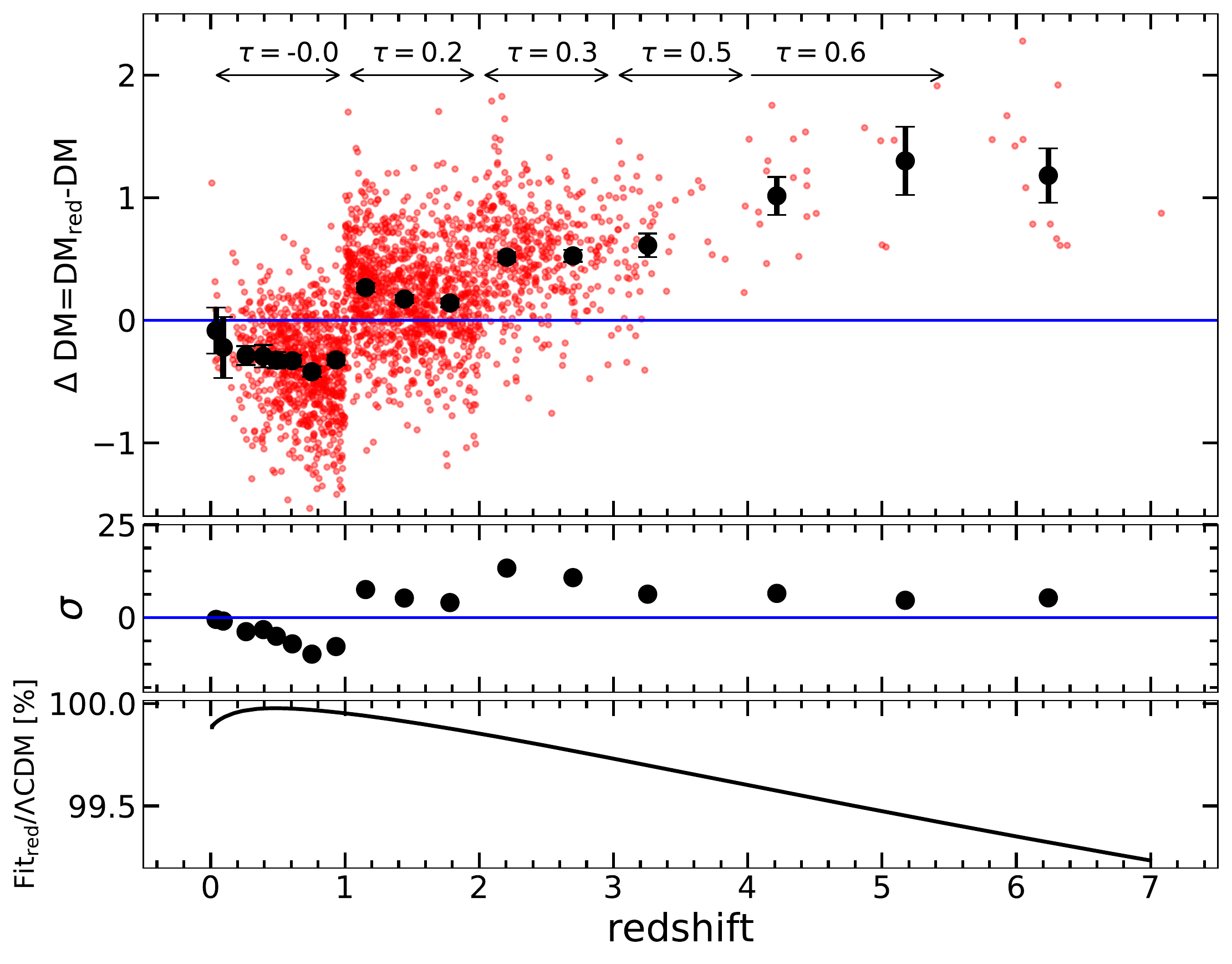}
   \caption{Analysis of the possible contribution of dust at 2500 \AA. Top: distribution of the difference of the distance moduli ($\Delta\dm$, red points; $\Delta\dm=0$ represents the $\Lambda$CDM) as a function of redshift. The parameter $\Delta\dm$ is defined as $\dm_{\rm red}-\dm$, where $\dm$ is computed from the quasar data as described in \S\,\ref{The Hubble diagram}. The $\dm_{\rm red}$ values are estimated by correcting the observed $\Fo$ for an additional intrinsic (and redshift-dependent) extinction. $\tau$ is the optical depth at the rest-frame 2500 \AA\ that would be needed to ascribe the discrepancy to an underestimated dust extinction in the UV flux. The black points represent the average (and 1$\sigma$ uncertainties) of $\Delta\dm$ in the same narrow redshift intervals used in Figures~\ref{fig:gdz} and \ref{hubbleclean}. Middle: deviation in $\sigma$ of the average points with respect to $\Lambda$CDM. Bottom: ratio between the cosmographic fit obtained from the data, where $\Fo$ is corrected for additional extinction, and the $\Lambda$CDM. The cosmographic fit perfectly matches $\Lambda$CDM for the chosen values of dust reddening (see \S\,\ref{dust}).}
              \label{fig:dust}
    \end{figure}
Among the possible residual (and redshift-dependent) observational systematics in the Hubble diagram, we must also consider the presence of an additional contribution of dust reddening in the UV band. As we move to higher redshifts, the rest-frame optical/UV spectra shift to higher (shorter) frequencies (wavelengths), where the dust absorption cross-section is higher. This might lead to an underestimate of $\Fo$, which would imply an intrinsically larger value of the luminosity distance (and thus $\dm$) than the one we measured. 

We thus evaluated the amount of dust extinction required to make the cosmographic fit shown in Figure~\ref{hubbleclean} coincide with a standard flat $\Lambda$CDM model with $\om=0.3$ (i.e. the black dashed line in Figure~\ref{hubbleclean}). 
We defined five redshift intervals with increasing values of $\ebv$, specifically, $\ebv=0$, 0.1, 0.15, 0.2, and 0.25 at $z=0$--1, $z=1$--2, $z=2$--3, $z=3$--4, and $z>4$. We then assumed the standard extinction law by \citetads{prevot84} with $R_V=3.1$ and corrected $\Fo$ by the amount dictated by the chosen $\ebv$ in each redshift interval, and we finally fitted the ``reddening corrected'' distance modulus ($\rm DM_{\rm red}$)--redshift relation.

Figure~\ref{fig:dust} shows the resulting distribution of the difference of the distance moduli, $\Delta\dm$ (red points) between the $\rm DM_{\rm red}$ values computed above and the observed $\rm DM$ in Figure~\ref{hubbleclean} (see \S\,\ref{The Hubble diagram}), as a function of redshift. The cosmographic fit obtained from the DM$_{\rm red}-z$ relation perfectly matches the $\Lambda$CDM curve (bottom panel). 
We also report the values of optical depth, $\tau$, at the rest-frame 2500 \AA\ that would be needed to entirely ascribe the discrepancy to underestimated dust extinction on the 2500 \AA\ quasar fluxes. 

We note that the negative $\Delta\dm$ values at $z<1$ (where no dust extinction correction is applied) are simply caused by the different overall cosmographic fit obtained for the DM$_{\rm red}-z$ relation with respect to the observed one. Therefore, the $\Delta\dm$ values in this redshift range should not be taken at face value.

At 2500 \AA, it is $k(\lambda)\simeq7$, which corresponds to an increase in flux by a factor of $\simeq2.6-5$ for $\ebv=0.15-0.25$ at $z\geq2$. 
The bolometric luminosities of the $z>4$ quasars in the clean sample are in the range $10^{46.5-48.1}$ erg s$^{-1}$, assuming a bolometric correction of 2.75 at 2500 \AA\ \citepads{2013ApJS..206....4K}. If the extra correction were effectively required, this would imply an {\it intrinsic} bolometric luminosity for these sources of the order of $10^{47.3-49}$ erg s$^{-1}$. For example, the highest redshift quasar in the sample, ULAS J134208.10$+$092838.61, would have a bolometric luminosity by at least a factor of 4 higher than reported the literature (i.e. $1.6\times10^{47}$ erg s$^{-1}$, \citeads{banados2018}). 
This would also imply a larger black-hole mass, thus a heavier seed, which needs to be interpreted within current models of black-hole growth \citep[e.g. ][]{latif2013,dijkstra2014,pacucci2015}.

For the spectroscopic \xmm $z\simeq3$ sample, the $\Fo$ values would increase by a factor of at least 3.5, thus shifting $\lbol$ into the range $10^{48-48.8}$ erg s$^{-1}$. We note that the average stack of their SDSS spectra does not suggest any significant levels of dust absorption when compared to other composites obtained from bright, blue quasar spectra (see Figure~2 in \citeads{nardini2019}). As a result, the presence of any additional dust component does not seem to be justified.

For completeness, we have also employed a different reddening law (i.e. \citeads{F99}), and modified both the $\ebv$ values and the redshift intervals (always verifying that the cosmographic fit of the modified Hubble diagram is consistent with the standard $\Lambda$CDM), finding equivalent results. 

In conclusion, the intrinsic UV fluxes required to offset the observed tension between the cosmographic fit of the Hubble diagram and the $\Lambda$CDM through redshift-dependent dust absorption should be, on average, brighter by 0.4--0.8 mag. 
This would imply a steeper extrapolated slope than observed at longer wavelengths, which would be hard to reconcile with a standard $\alpha-$disc model \citep{1976MNRAS.175..613S}. 

We note that concurrent, and redshift-dependent, absorption is found at X-ray energies. The observed 0.5--2 keV energy band moves towards higher rest-frame energies as the redshift increases, where the effect of any gas absorption should become progressively negligible. 
Spectra with flat photon indices ($\gammax<1.7$) are removed from the sample, so we expect a minimal degree of residual gas absorption in the X-rays. In the high-redshift ($z>6$) quasars for which a spectral analysis was carried out (see e.g. \citeads{vito2019}), the level of intrinsic gas absorption is always suggested to be rather low (of the order of $\times10^{23}$ cm$^{-2}$ or less). 
Assuming the presence of a local column density with $N_{\rm H}=10^{23}$ cm$^{-2}$ at $z=6$, the exact correction to the inferred rest-frame 2 keV flux density depends on the details (flux, observed slope, data quality) of the single observation, but it is expected to be within a factor of 2, i.e. comparable to the typical measurement uncertainty. Such high $N_{\rm H}$ values are not required in any of the X-ray spectral fits for the high-redshift quasars that belong to the clean sample. By shifting the data to higher $\dm$ values, this correction would have the effect of increasing the departure from the $\Lambda$CDM (i.e. opposite to the one for the dust absorption in the UV).

\rev{
\subsection{On the possible dependence of the $\Lx-\Lo$ relation on black-hole mass and accretion rate}
\label{residuals-fwhm}
One can argue that the observed $\Lx-\Lo$ relation might be a secondary manifestation of some other, more fundamental relations, involving, for instance, a possible dependence on black-hole mass and accretion rate (usually parametrised by the Eddington ratio, $\ledd$, defined as the ratio of the bolometric and the Eddington luminosity). 
Yet, when estimated from single-epoch spectroscopy, both parameters are derived quantities, i.e. a combination of continuum luminosity and emission-line FWHM. Therefore, the presence or absence of a correlation between the residuals of the Hubble diagram and $\mbh$ and/or the Eddington ratio could be misleading and potentially hide systematics. 

Since the $\Lx-\Lo$ relation already has a dependence on the nuclear luminosity (which we have discussed in detail in Section~\ref{uvfluxcheck}), we can explore possible additional correlations with the FWHM of a given emission line. This issue was already investigated in \citetads{lr17}, who found that not only is the dependence of the $\Lx-\Lo$ relation on the FWHM (of \ion{Mg}{ii}) statistically significant, but also that such a dependence has the effect of further reducing the dispersion of the $\Lx-\Lo$ correlation. 
However, building a statistically significant quasar sample that both covers a wide redshift range (as the one presented here) and relies on a measurement of FWHM from a single line (to ensure homogeneity in the estimate of the line parameters, see \S\,2.1\ in \citeads{lr17}) is impractical, as it would require a series of dedicated spectroscopic near-infrared campaigns to probe the same line (e.g. \ion{Mg}{ii}) at high redshifts ($z>2.5$). 
The statistical relevance of the supplementary dependence of the $\Lx-\Lo$ relation on the FWHM, as well as on other parameters inferred from spectroscopy (e.g. continuum slope, iron content, line strength) will be further investigated in a forthcoming publication, entirely dedicated to the physics of the relation. 

While here we prefer to consider the simple $\Lx-\Lo$ relation for cosmological purposes, as the present analysis aims at verifying that our selection criteria do not introduce systematics in the quasar Hubble diagram, we can still check whether the sources in the clean sample with a measurement of the FWHM from \ion{Mg}{ii} (which provides the widest possible redshift coverage in BOSS) show any systematic trend and/or a reduced scatter in the residuals. We thus cross-matched the clean sample with the catalogue of spectral quasar properties of \citet{rakshit2020}, and selected the quasars with a broad (FWHM\,$>$\,2000 km s$^{-1}$) component of the \ion{Mg}{ii} emission line, finding 1,858 quasars (77\% of the clean sample).

The distribution of the residuals as function redshift for this sample is shown in Figure~\ref{fig:resid_fwhm}.
The amplitude of the scatter of the residuals is similar to the other cases discussed in Sections\,\ref{residuals-eddbias}, \ref{residuals-gammax}, and \ref{residuals-colours}. Moreover, there is no trend with redshift up to $z\simeq2.5$. These results further prove that our selection criteria are already effective in selecting an optimal sample for cosmology, without introducing strong systematics.

   \begin{figure}
   \centering
   \includegraphics[width=\linewidth]{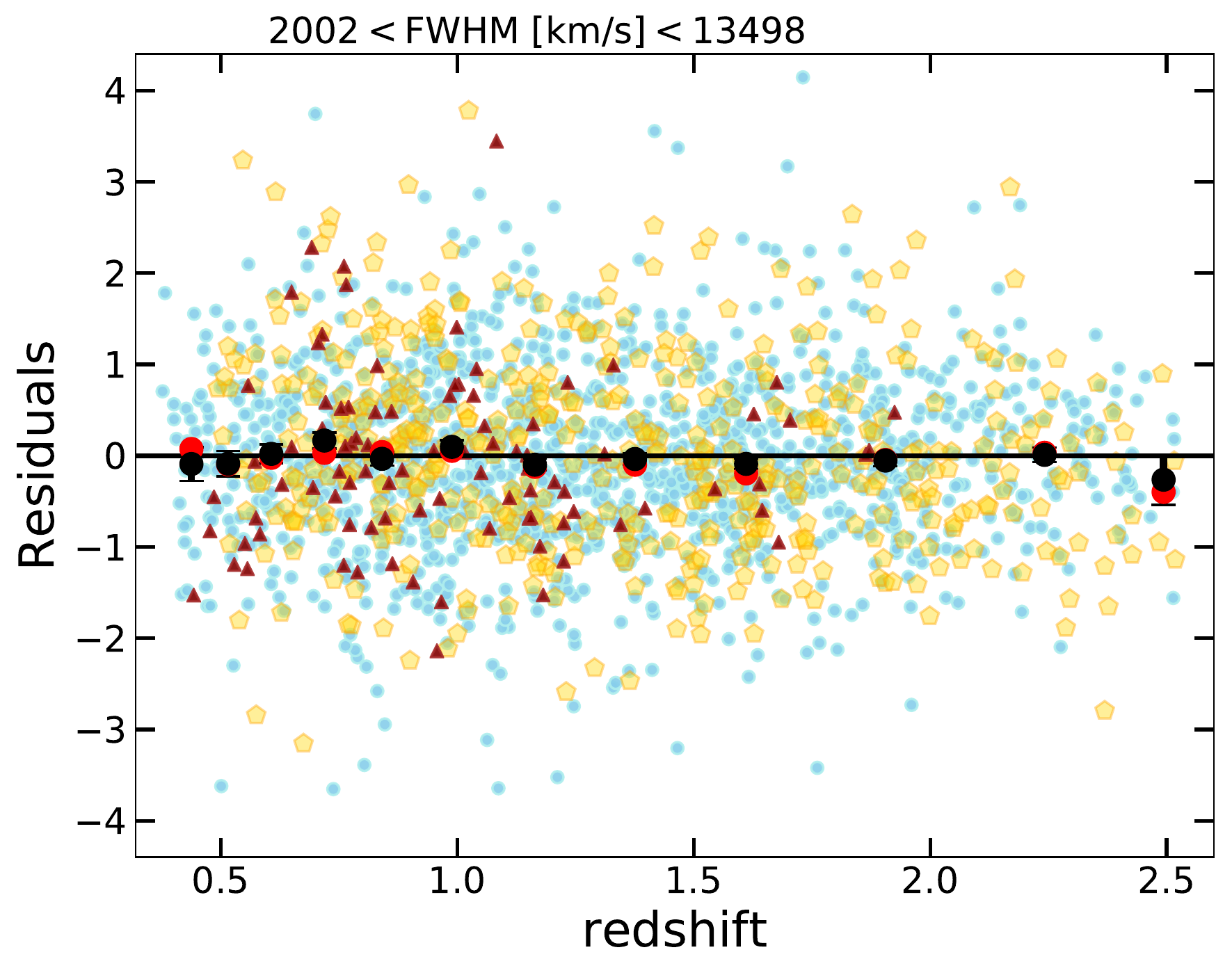}
   \caption{\rev{Distribution of the Hubble diagram residuals (\rev{middle} panel of Figure~\ref{hubbleclean}) as a function of redshift for the quasars with a broad (FWHM\,$>$\,2000 km s$^{-1}$) component of the \ion{Mg}{ii} emission line in the clean sample. The black and red points represent the mean and median of the residuals in narrow redshift intervals, respectively. Symbol keys as in Figure~\ref{loz}.}}
              \label{fig:resid_fwhm}
    \end{figure}
}

\section{Discussion and conclusions}
\label{Conclusions}
Our group has presented a new technique that makes use of the observed non-linear relation between the 2500 \AA\ and the 2 keV emission in quasars to provide an independent measurement of their distances, thus turning quasars into {\it standardizable} candles. Our method extends the distance modulus--redshift relation (or the so-called {\it Hubble-Lema\^itre diagram}) of supernovae Ia to a redshift range still poorly explored ($z>2$; e.g. \citeads{rl15,rl19,lusso19}), and it relies upon the evidence that most of the observed dispersion in the $\Lx-\Lo$ relation is not intrinsic to the relation itself but due to observational issues. When an optimal selection of {\it clean} sources (i.e. for which we can measure the {\it intrinsic} UV and X-ray quasar emission) is possible, the dispersion in the $\Lx-\Lo$ relation drops to $\simeq$0.2 dex \citepads{lr16,lr17}.

We have previously demonstrated that the distance modulus--redshift relation of quasars at $z < 1.4$ is in agreement with that of supernovae Ia and with the concordance $\Lambda$CDM model \citep{rl15,rl19,lusso19}, yet a deviation from the $\Lambda$CDM emerges at higher redshift, with a statistical significance of about 4$\sigma$. If we interpret the latter result by considering an evolution of the dark energy equation of state in the form $w(z)=w_0+w_a\times z/(1+z)$, the data suggest that the dark energy density is increasing with time \citep{rl19,lusso19}. 

However, our technique may still have some limitations, and we need to verify that the observed deviation from the $\Lambda$CDM at redshift $>2$ is not driven by either systematics in the quasar sample selection or the cosmological procedure adopted to fit the distance modulus--redshift relation.

The aim of this manuscript is thus to discuss, on the one hand, all the criteria required to select a homogeneous sample of quasars for cosmological purposes and, on the other hand, the specific procedures adopted to compute the UV and X-ray fluxes and spectral slopes from the available photometry. 
We identified the quasars that can be used for a cosmological analysis, examined the key steps in fitting the distance modulus--redshift relation, and considered the possible systematics in the quasar Hubble-Lema\^itre diagram. In particular, we investigated in depth the residuals of the quasar Hubble diagram, in order to unveil any systematics unaccounted for in the selection of the sample. We explored whether our procedure {\it (1)} to correct for the Eddington bias, {\it (2)} to neglect quasars with possible gas absorption, and {\it (3)} to select blue quasars based on their SED shape, where dust absorption and host-galaxy contamination are minimised, introduces spurious trends in the Hubble diagram residuals as a function of redshift and for different intervals of the relevant parameters.

Our main results are the following:
   \begin{itemize}
      \item We verified that the $\Lx-\Lo$ relation (i.e. slope and dispersion) for the final ``best'' quasar sample does not evolve with redshift.      
      \item We confirmed that, while the quasar Hubble diagram is well reproduced by a standard flat $\Lambda$CDM model (with $\om=0.3$) up to $z\sim1.5$, a statistically significant deviation emerges at higher redshifts, in agreement with our previous works (e.g. \citeads{rl15,rl19,lusso19}). 
      \item  We found that none of the adopted filters introduce strong systematics in the Hubble diagram residuals, and specifically where the quasars become the only contributors and the deviation from the standard $\Lambda$CDM is more significant, i.e. at $z>1.5$. 
   \end{itemize}
Even if our analysis shows that both the quasar selection criteria and the cosmological fitting technique are robust, we can already envisage several further improvements, especially in the quasar sample selection (e.g. by using spectra instead of photometry in both X-rays and UV to better measure the X-ray and UV fluxes and slopes). None the less, given the remarkable absence of systematics in the residuals, the main effect of these refinements will be on the reduction of the dispersion, thus allowing a better estimate of the cosmological parameters.

With currently operating facilities, dedicated observations of well-selected high-$z$ quasars (similarly to what our group has done at $z\simeq3$) will greatly improve the test of the cosmological model and the study of the dispersion of the $\Lx-\Lo$ relation, especially at $z\simeq4$. 
The {\it extended Roentgen Survey with an Imaging Telescope Array} (eROSITA, \citeads{2012SPIE.8443E..1RP, 2012arXiv1209.3114M}), flagship instrument of the ongoing Russian {\it Spektrum-Roentgen-Gamma} (SRG) mission, will represent an extremely powerful and versatile X-ray observatory in the next decade. The sky of eROSITA will be dominated by the AGN population, with $\sim$3 million AGN with a median redshift of $z\sim1$ expected by the end of the nominal 4-year all-sky survey at the sensitivity of $F_{0.5-2\,\rm keV} \simeq 10^{-14}$ erg s$^{-1}$ cm$^{-2}$, for which extensive multiwavelength follow-up is already planned. 
\rev{Concerning the constraints on the cosmological parameters (such as $\om$, $\ol$, and $w$) through the Hubble diagram of quasars, we predict that the 4-year eROSITA all-sky survey alone, complemented by redshift and broadband photometric information, will supply the largest quasar sample at $z<2$ (average redshift $z\simeq1$), but a relatively small population should survive the Eddington bias cut at higher redshifts (see e.g. \citeads{medvedev2020}), thus being available for cosmology. Indeed, eROSITA will sample the brighter end of the X-ray luminosity function \citep[][but see also section 6.2 in \citeads{comparat2020}]{lusso2020}.
None the less, the large number of eROSITA quasars at $z\simeq1$ will be pivotal for both a better cross-calibration of the quasar Hubble diagram with supernovae and a more robust determination of $\ol$, which is  sensitive to the shape of the low redshift part of the distance modulus--redshift relation.}

In the mid/long term, surveys from {\it Euclid} and LSST in the optical/UV, and {\it Athena} in the X-rays, will also provide samples of millions of quasars. 
With these samples it will be possible to obtain constraints on the observed deviations from the standard cosmological model, which will rival and complement those available from the other cosmological probes.

\begin{acknowledgements}
We thank the referee for their constructive comments and suggestions which have significantly improved the clarity of the paper. We also thank M. Millon and F. Courbin for reading the manuscript and providing useful comments.
We acknowledge financial contribution from the agreement ASI-INAF n.2017-14-H.O. 
\end{acknowledgements}

%
%



\bibliographystyle{aa}
\bibliography{bibl} 


%

\begin{appendix} 

\section{Detailed sample selection summary}
\label{Detailed sample selection summary}
Table~\ref{tbl:sel} provides a detailed summary of the statistics of the various subsamples for a given selection. 
\longtab[1]{
\begin{landscape}
\begin{longtable}{l | cccccccc}
\caption{Sample statistics of the various subsamples in each give selection.}\\
\label{tbl:sel}\\
\hline
Selection & \multicolumn{8}{c}{Samples}\\ 
                & \xmm  & \xmm  & High-$z$ & XXL & SDSS--4XMM & SDSS--\chandra & Local AGN & Total \\
                & $z\simeq3$ & $z\simeq4$ & &  &  & &  &  \\
\endhead
\hline
\endfoot
\endlastfoot
Initial sample (1)                                          & 29 & 1 & 64$^a$ & 840 & 13,800 & 7,036 & 15 & 21,785\\                    
Main sample (2)                                          & 29 & 1 & 64 & 840 & 9,252  & 2,392 & 15 & 12,593 \\                    
Dust \& Host correction (\S\,\ref{dusthost}) & 29 & 1 & 64 & 535 & 7,319  & 1,136 & 15 & 9,099\\                    
X-ray absorption (\S\,\ref{xrayabs})            & 14$^\dagger$ & 1 & 35$^b$ & 259 & 2,576  & 994 & 13 & 3,921\\                    
Eddington bias (\S\,\ref{eddbias})              & 14 & 1 & 35 & 106 & 1,644  & 608 & 13 & 2,421\\                                       
\end{longtable}
\tablefoot{
(1)~These number counts refer to the sample statistics before correcting for overlaps amongst the subsamples.
(2)~Sample statistics after accounting for overlaps and some preliminary selection (e.g RL, BAL, off-axis angle in the X-rays). The order of priority decreases from the leftmost to the rightmost column. 
$^a$~See \S\,\ref{high redshift sample}. $^b$~35 quasars = 29 sources from \citetads{salvestrini2019} and 6 from \citetads{vito2019}. \rev{$^\dagger$ Seven X-ray weak quasars are amongst the 15 objects excluded (see \citeads{nardini2019} for details).}
}
\end{landscape}
}

\section{Fit of the relation $\Fx-\Fo$ in redshift intervals}
\label{Fit of the relation in redshift intervals}
Figure~\ref{fig:fxfuv} presents the fits of the $\Fx-\Fo$ relation in narrow redshift bins as discussed in Section~\ref{Analysis of the relation with redshift}.
The best-fit parameters (slope and dispersion) of the $\Fx-\Fo$ relation and their uncertainties, the number of objects in each bin and the average redshift are also shown.

   \begin{figure}
   \centering
   \includegraphics[width=\linewidth]{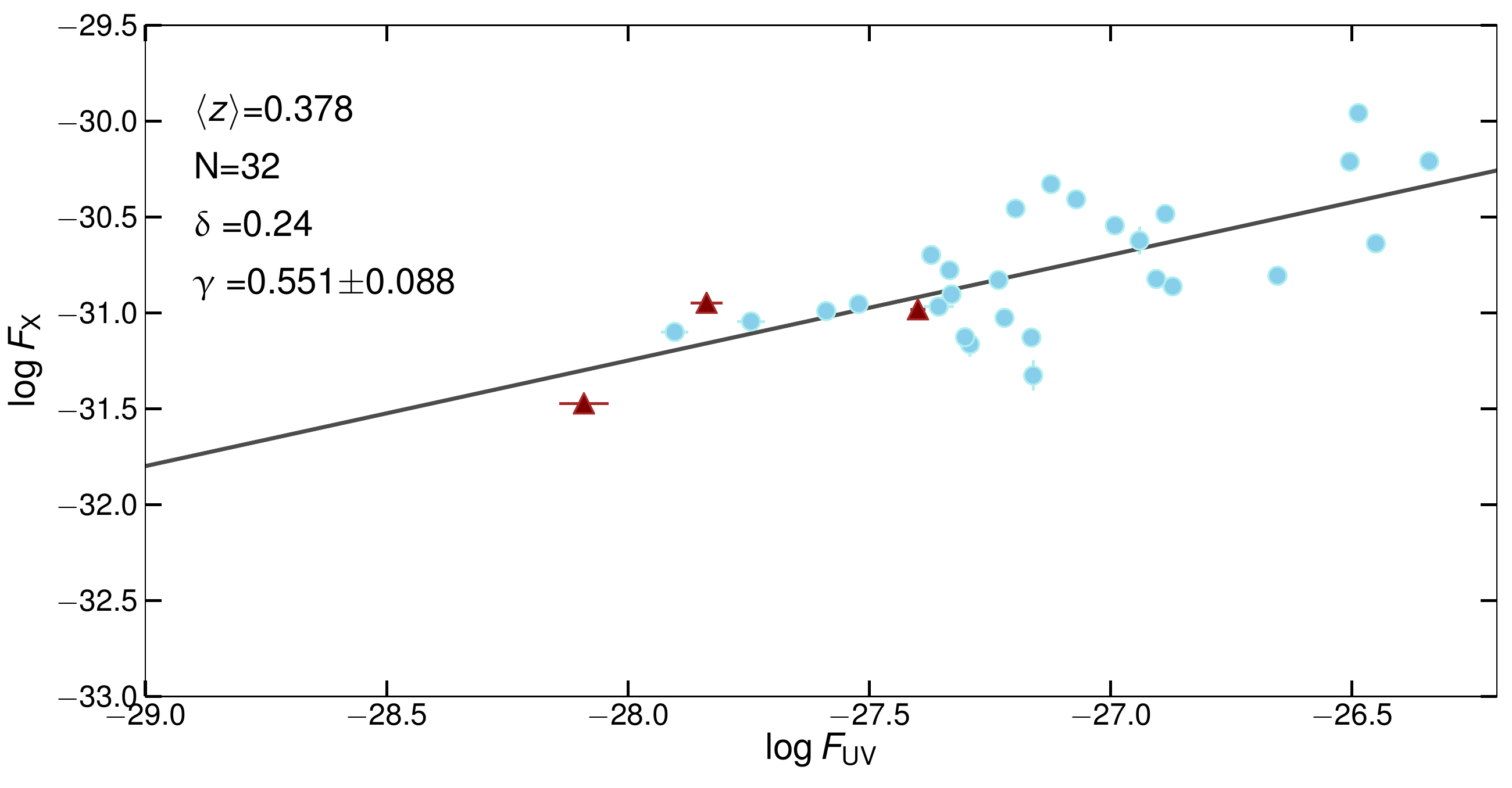}
   \includegraphics[width=\linewidth]{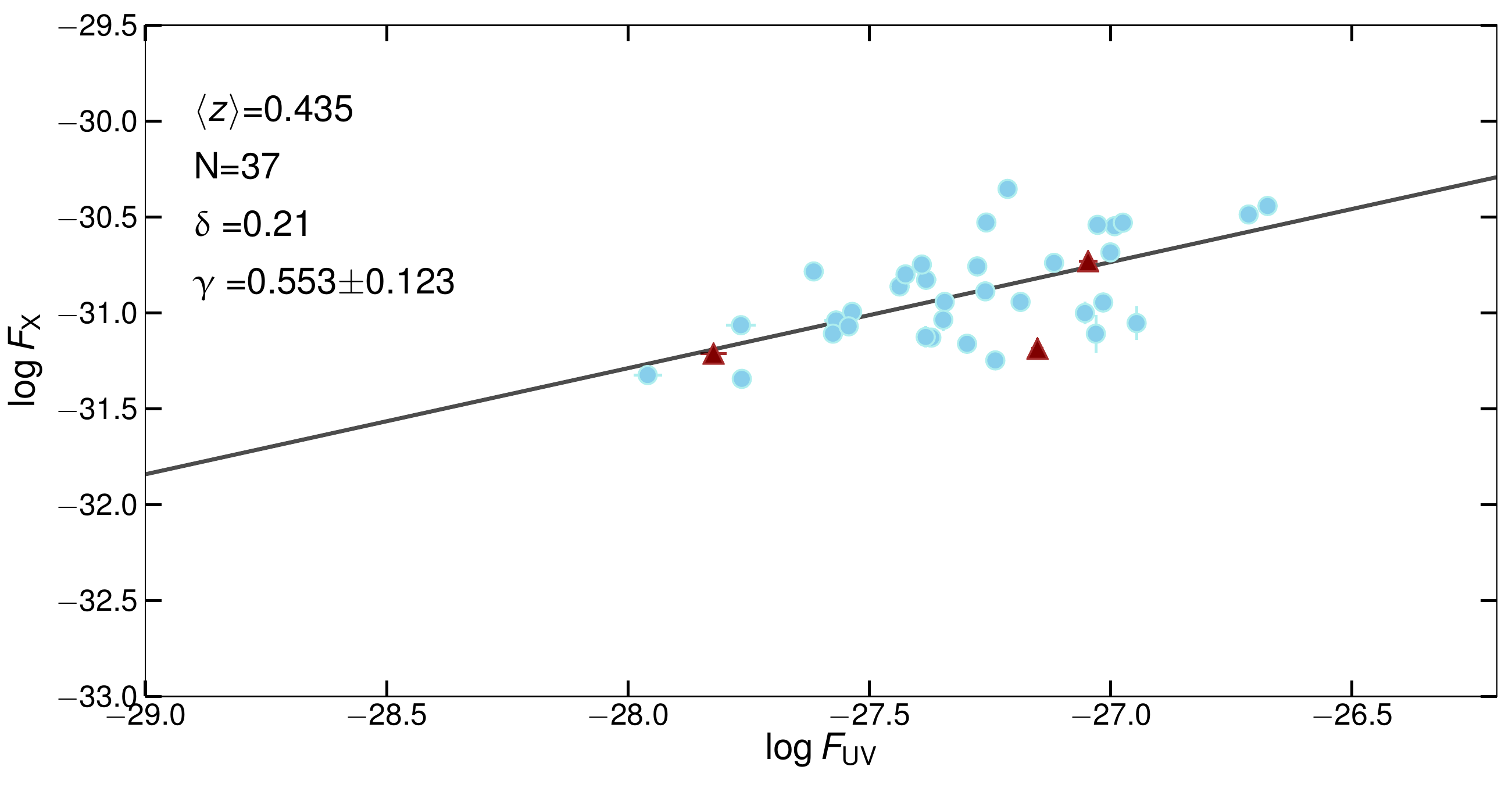}
   \includegraphics[width=\linewidth]{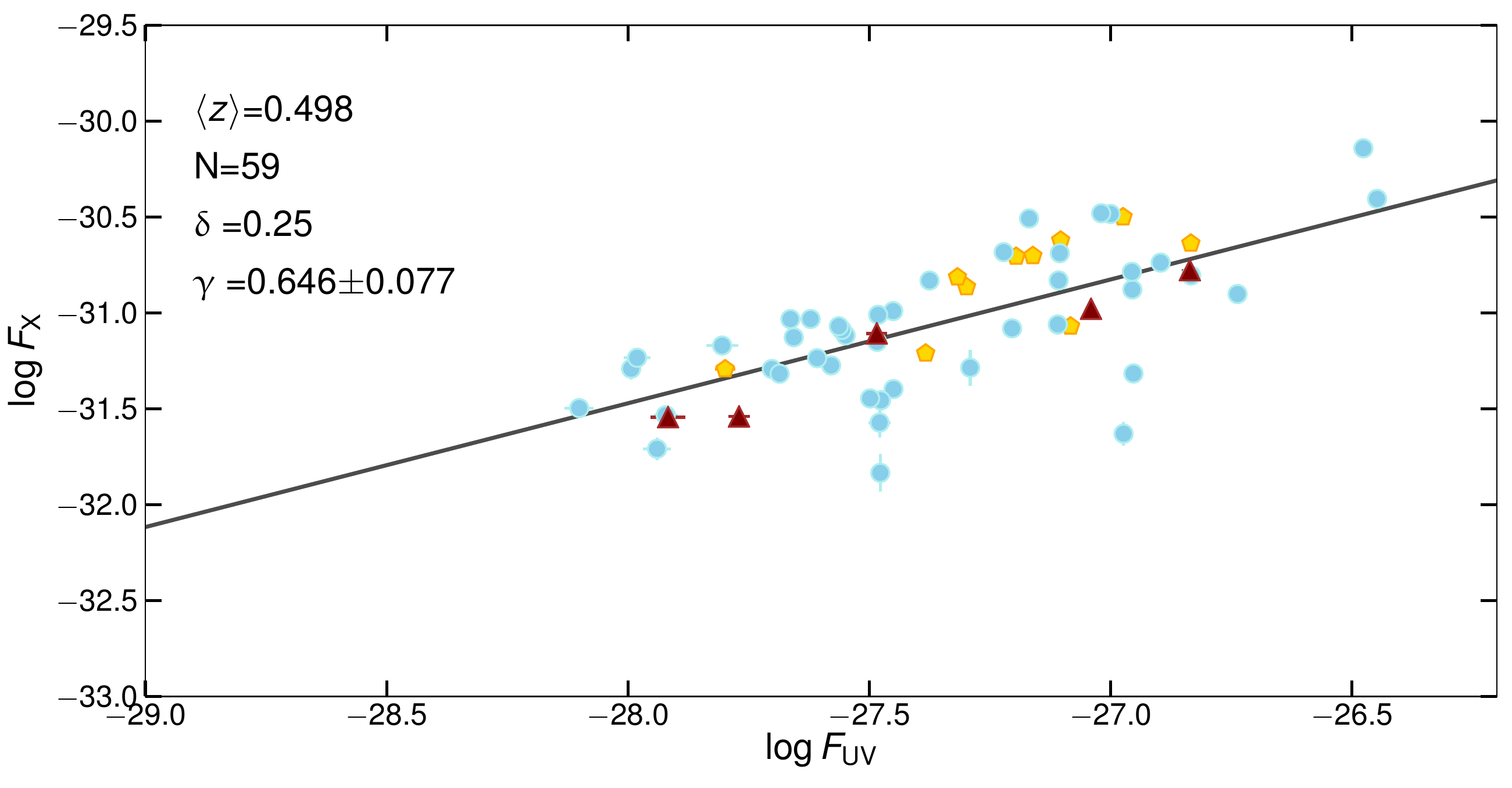}
    \caption{$\Fx-\Fo$ relation in narrow redshift bins as discussed in Section~\ref{Analysis of the relation with redshift}. The best-fit parameters (slope and dispersion) of the $\Fx-\Fo$ relation and their uncertainties, the number of objects in each bin, and the average redshift are also reported. Symbol keys as in Figure~\ref{loz}.}
              \label{fig:fxfuv}
    \end{figure}
   \begin{figure}
   \centering
   \ContinuedFloat
   \captionsetup{list=off,format=cont}
   \includegraphics[width=\linewidth]{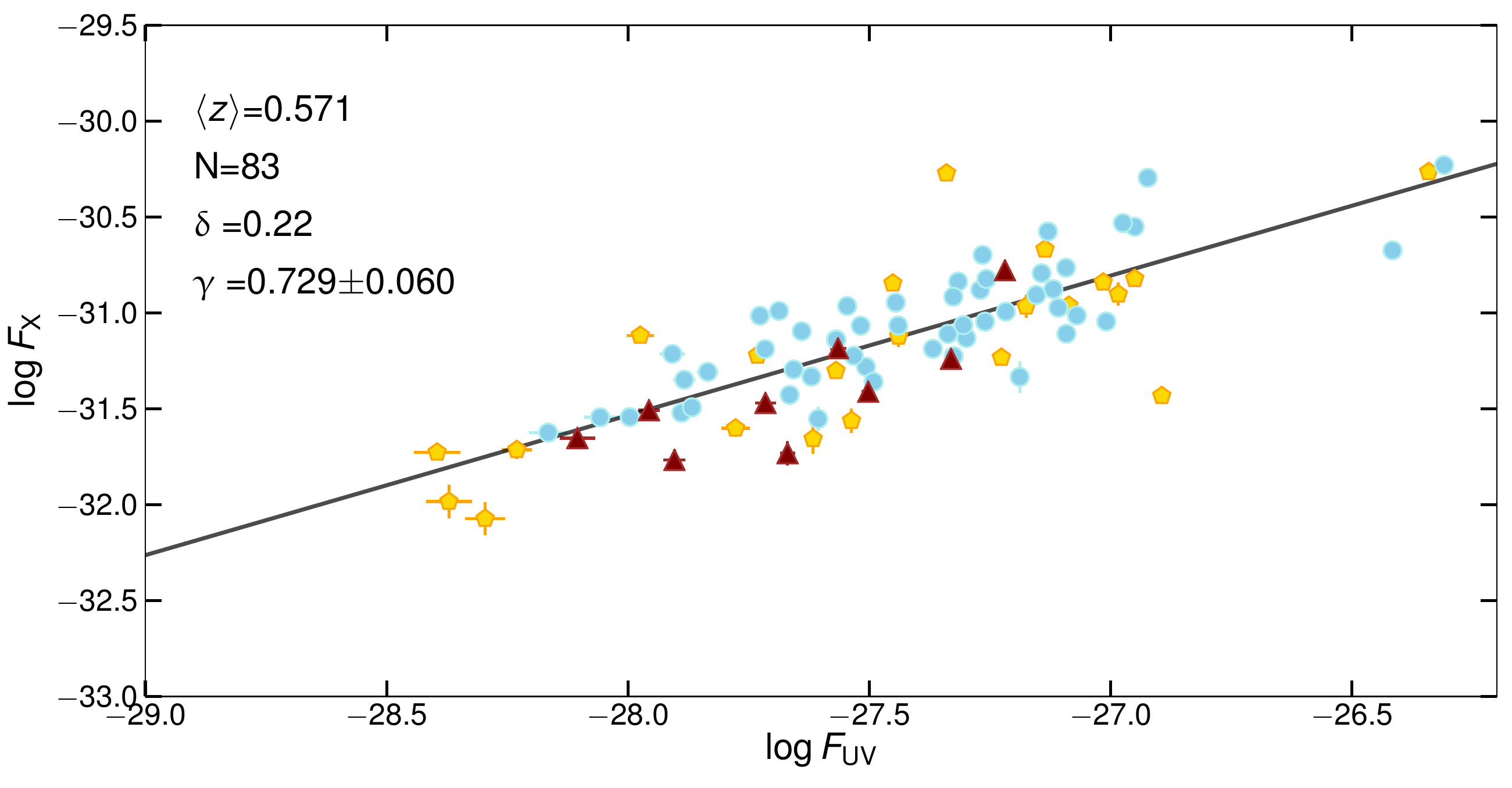}
   \includegraphics[width=\linewidth]{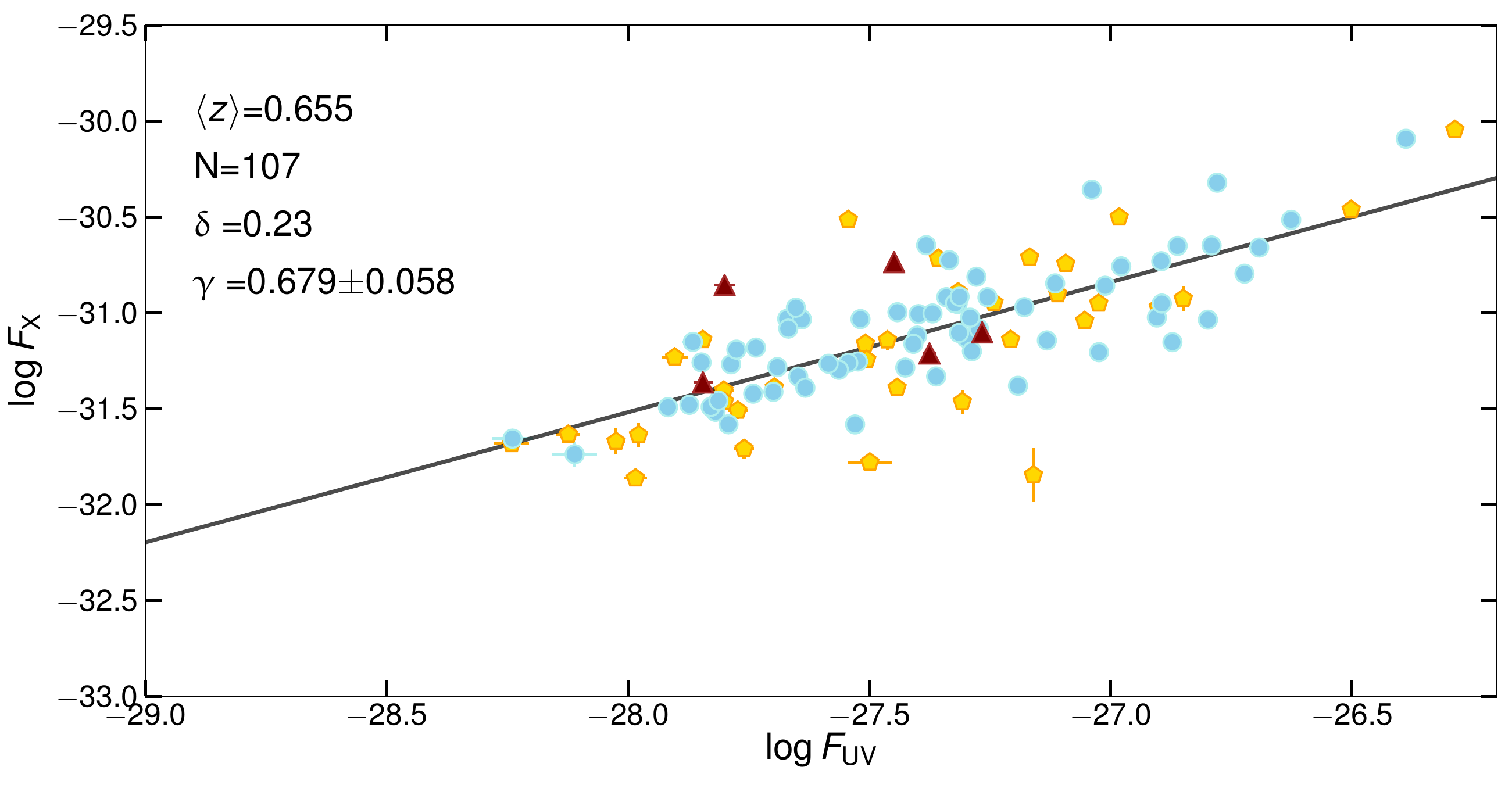}
   \includegraphics[width=\linewidth]{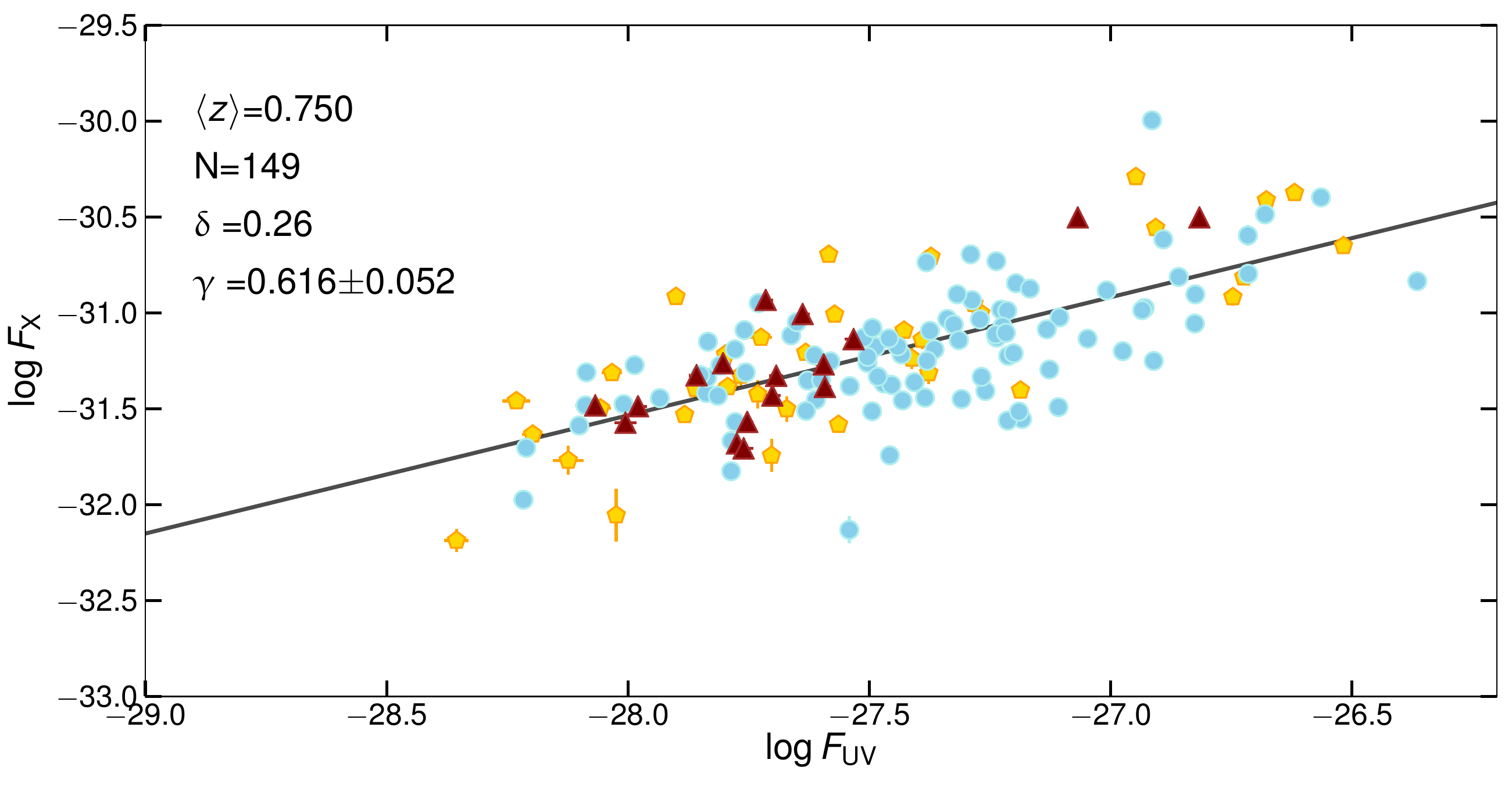}
   \includegraphics[width=\linewidth]{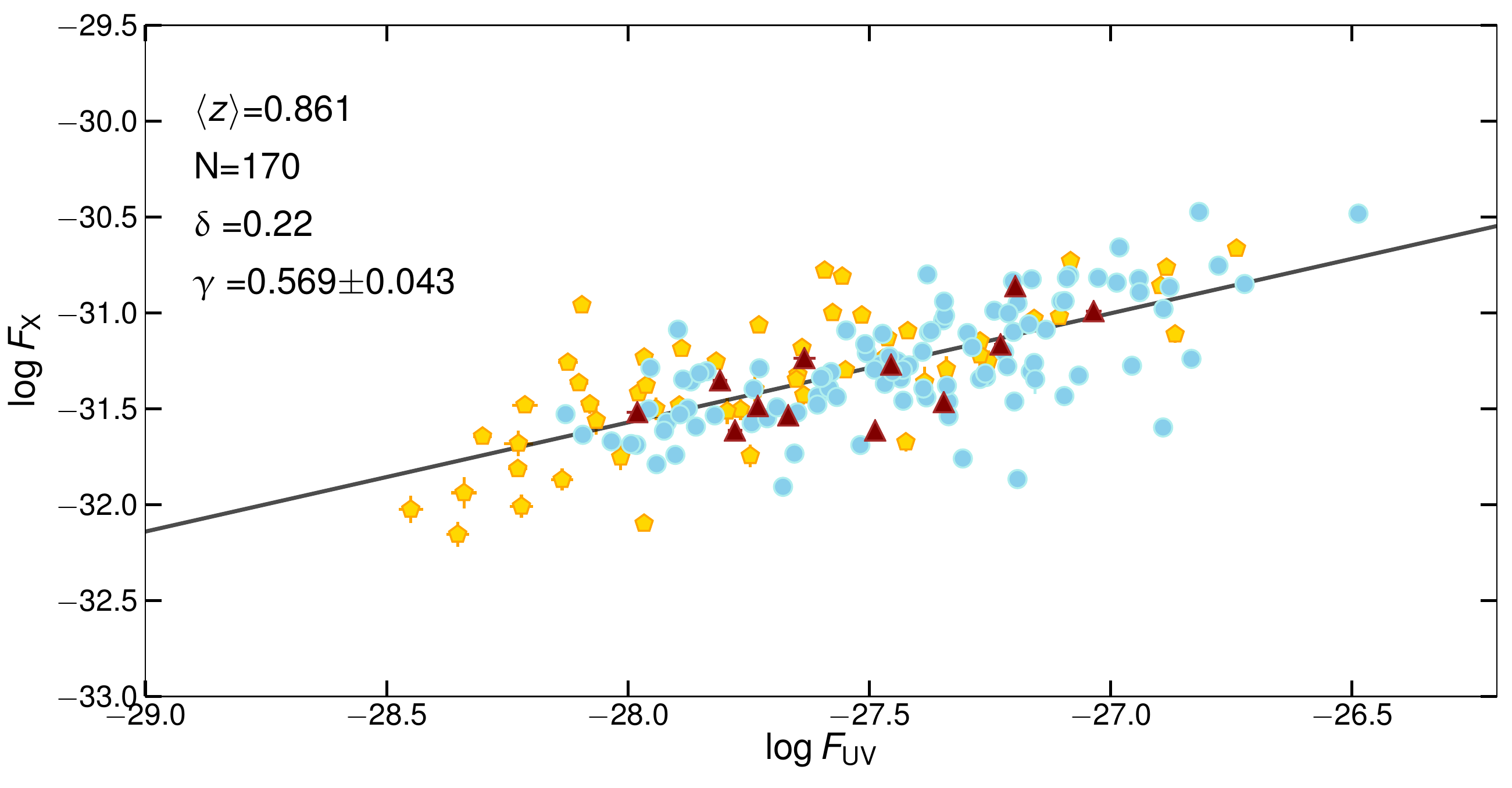}
   \includegraphics[width=\linewidth]{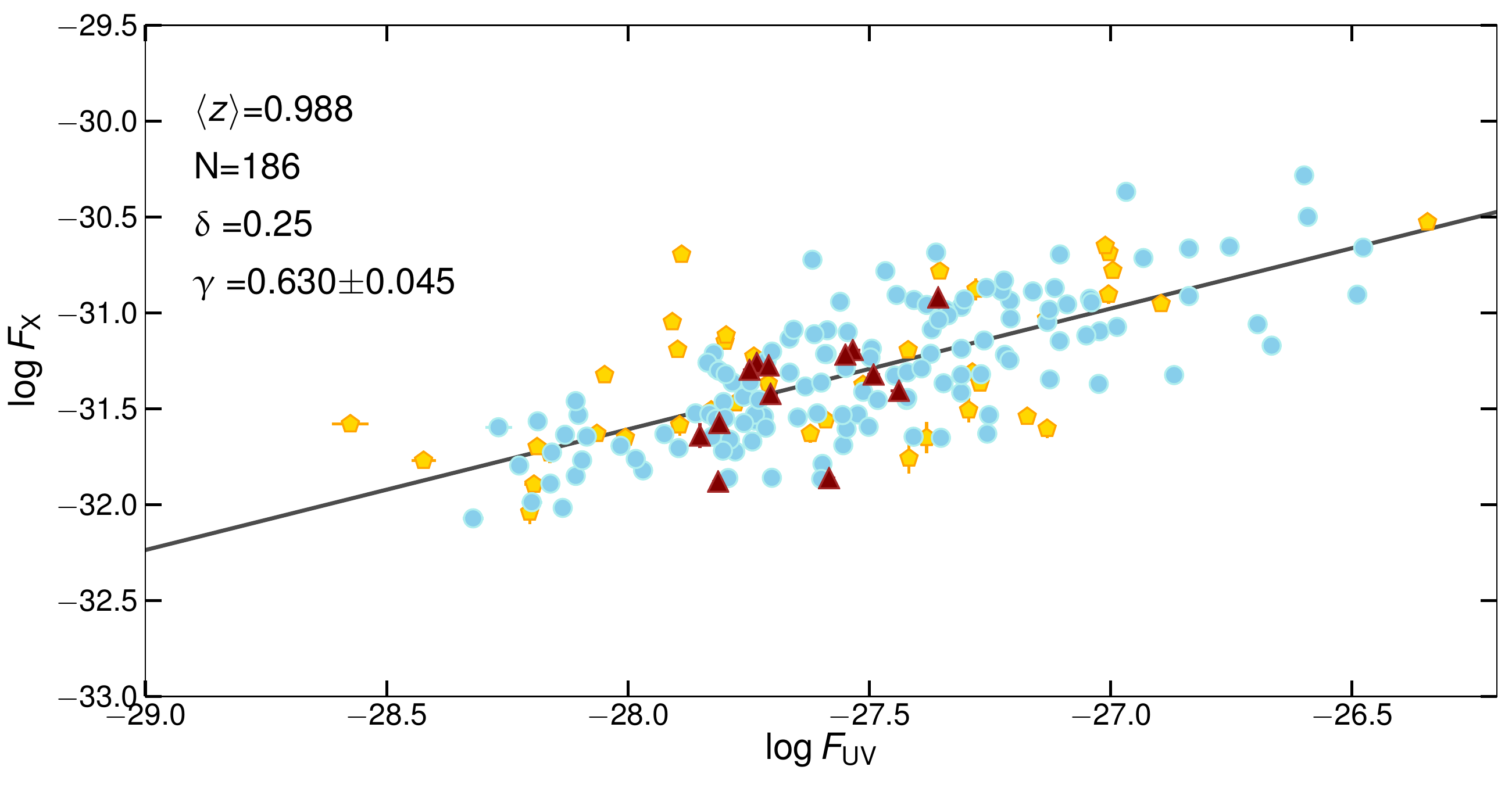}
   \caption{Continued.}
              \label{fig:fxfuv}
    \end{figure}

   \begin{figure*}
   \ContinuedFloat
   \captionsetup{list=off,format=cont}
   \centering
   \includegraphics[width=0.49\linewidth]{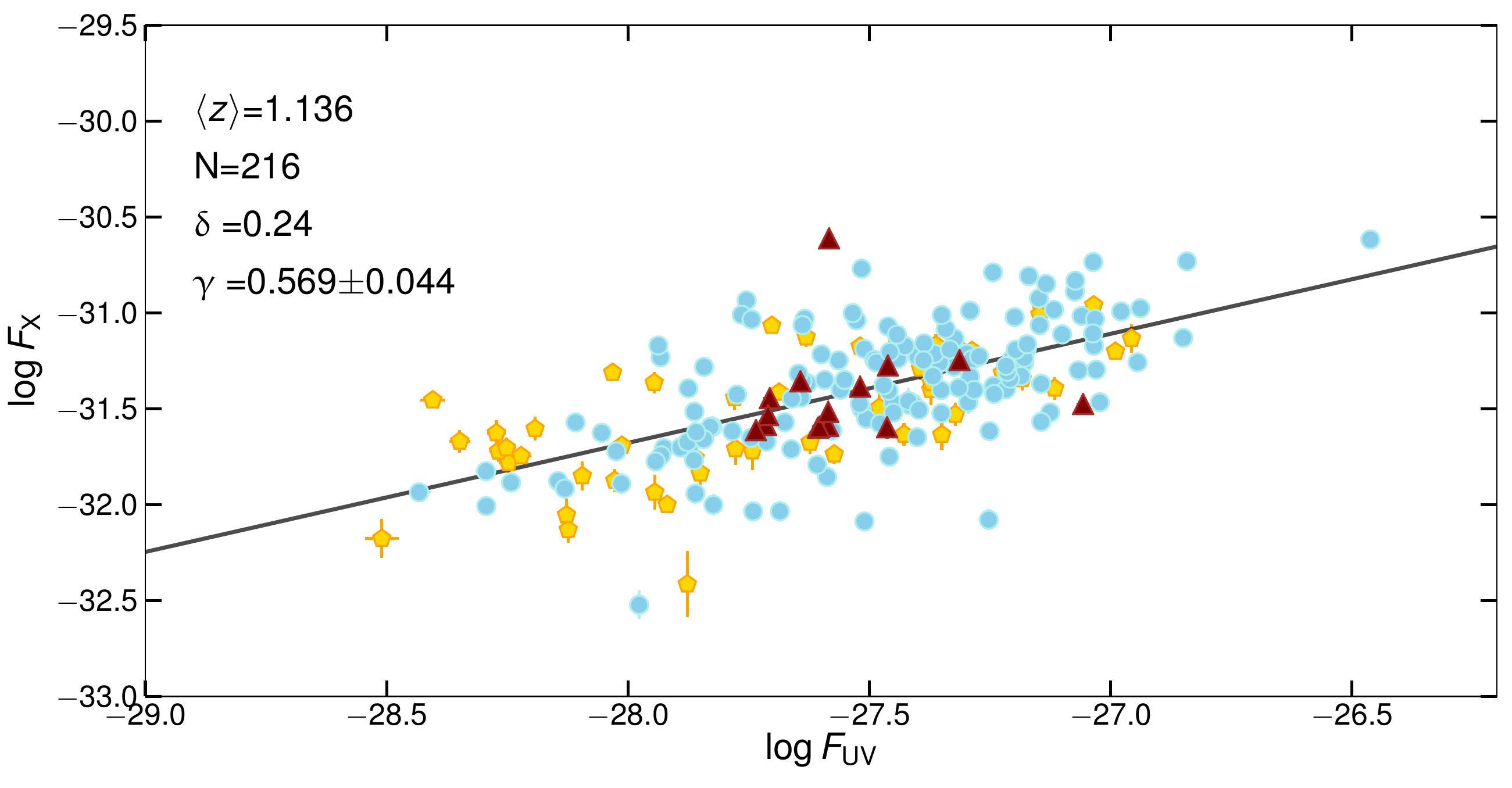}
   \includegraphics[width=0.49\linewidth]{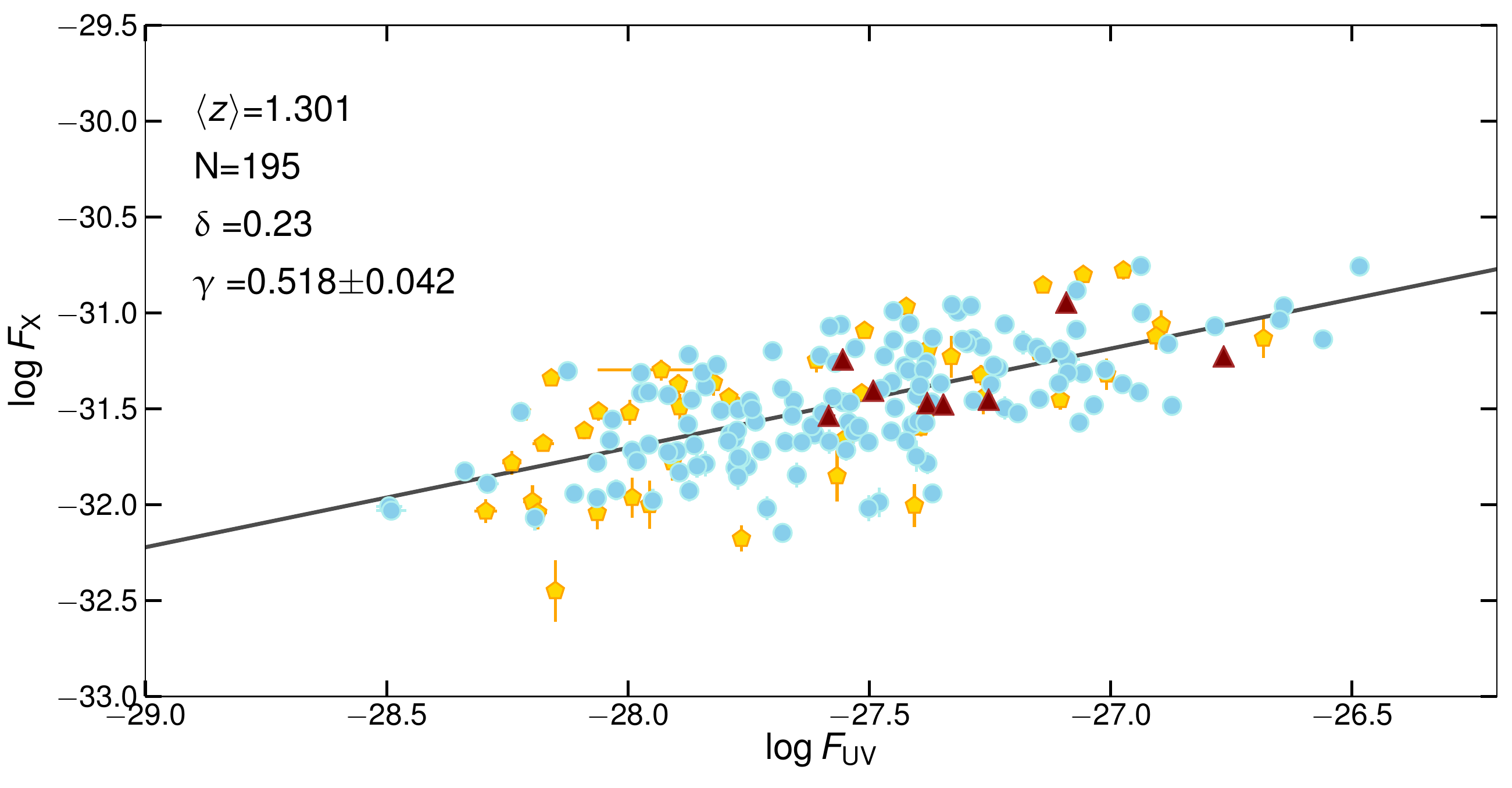}
   \includegraphics[width=0.49\linewidth]{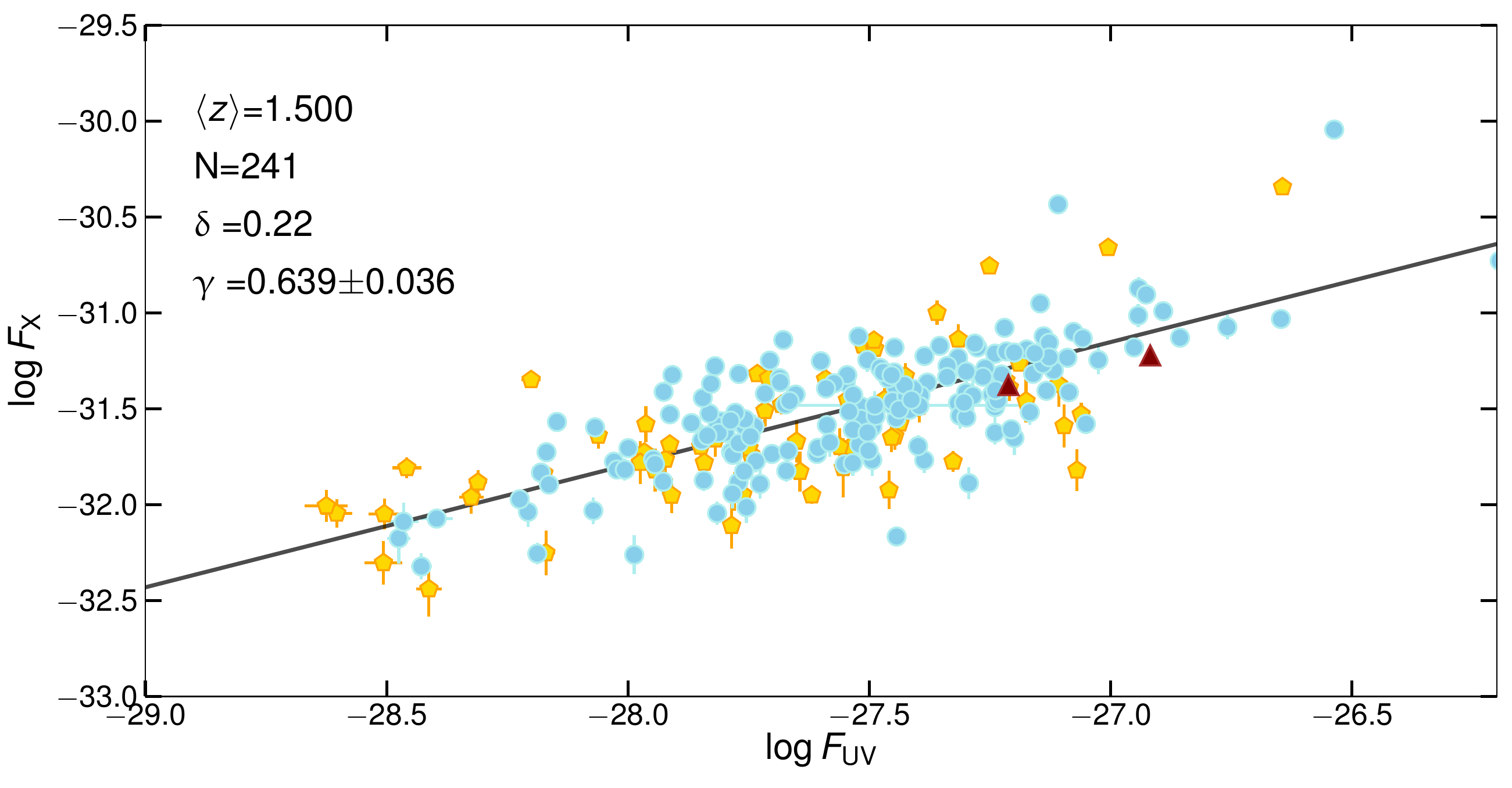}
   \includegraphics[width=0.49\linewidth]{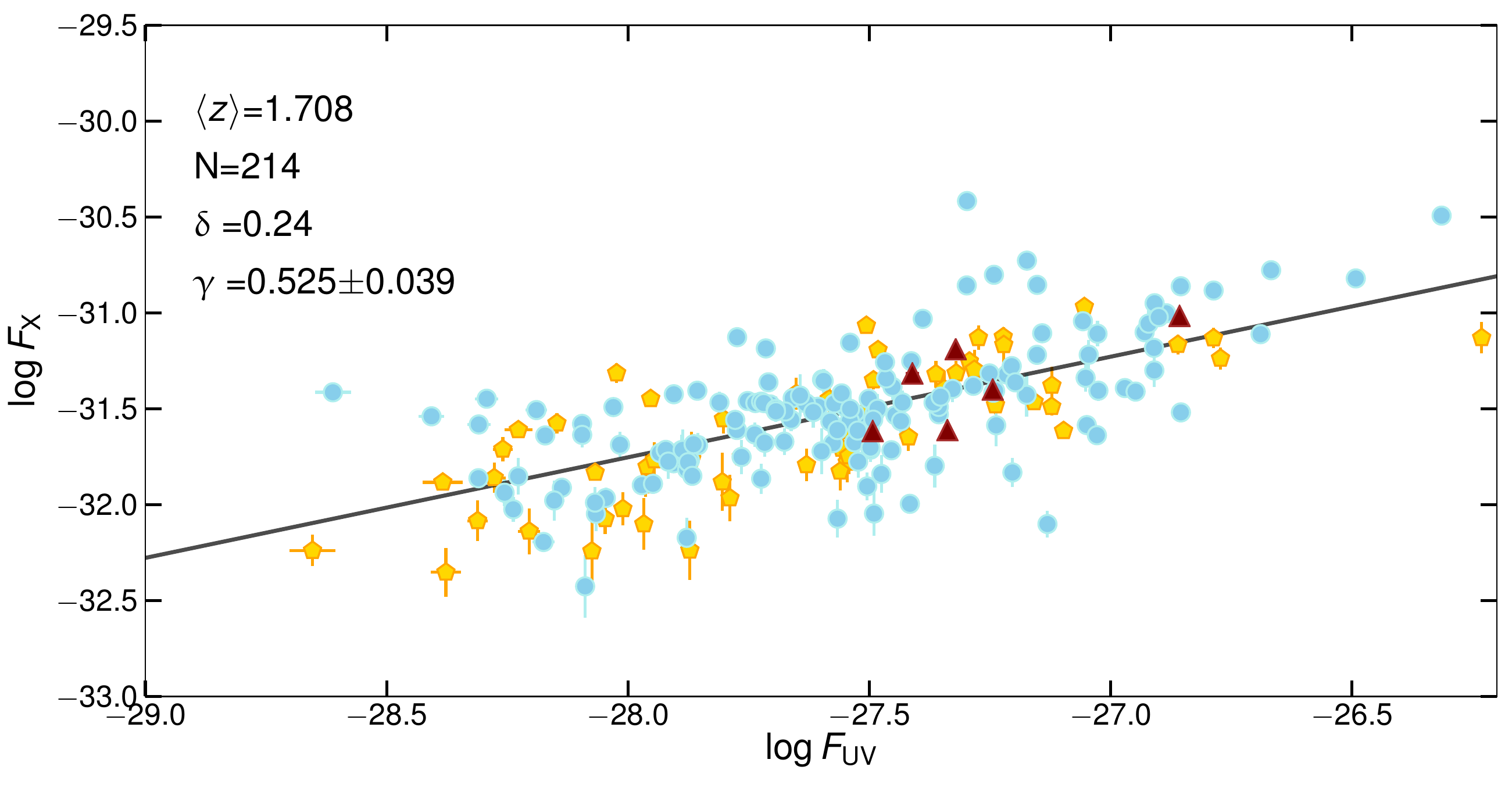}
   \includegraphics[width=0.49\linewidth]{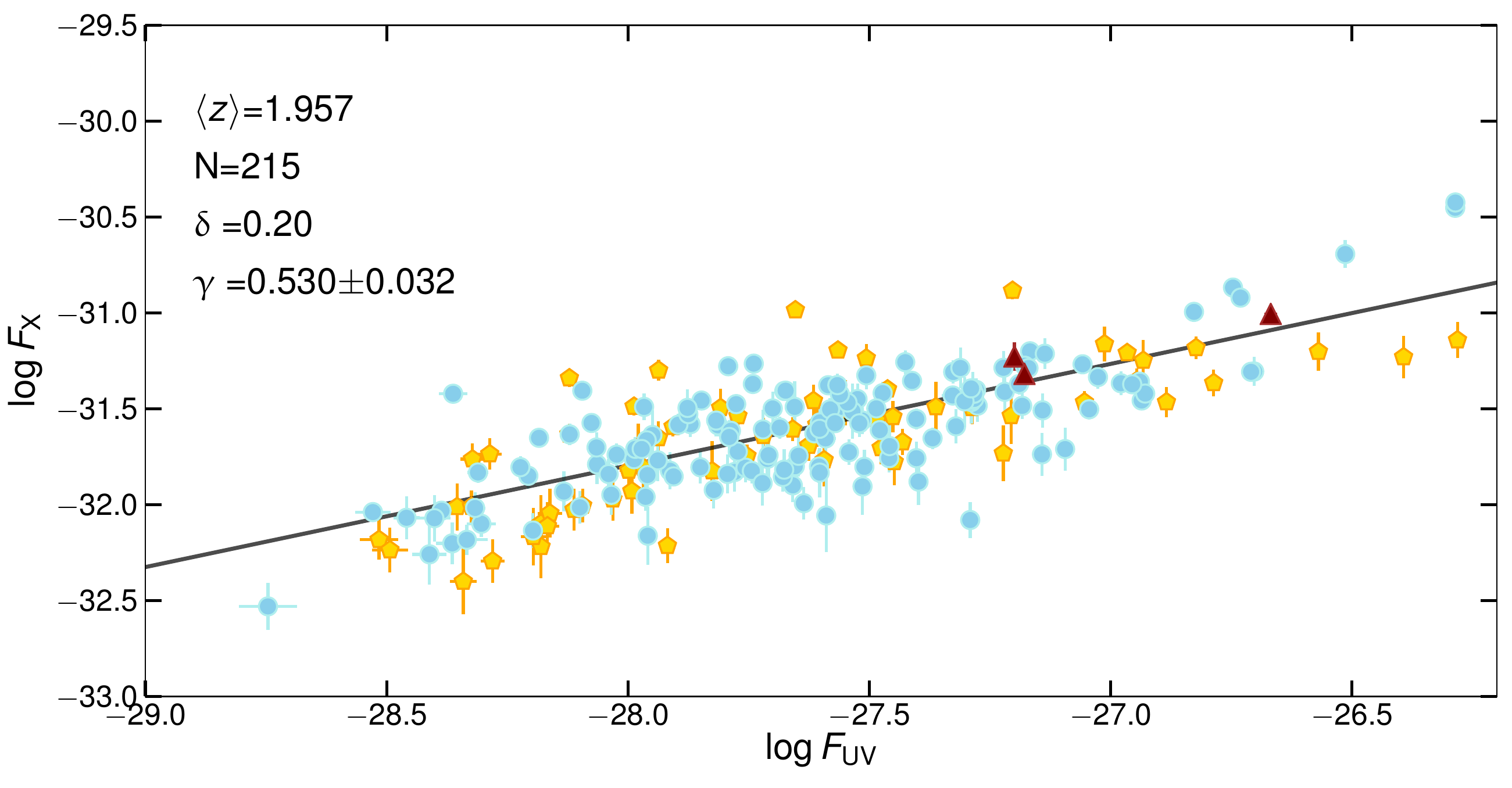}
   \includegraphics[width=0.49\linewidth]{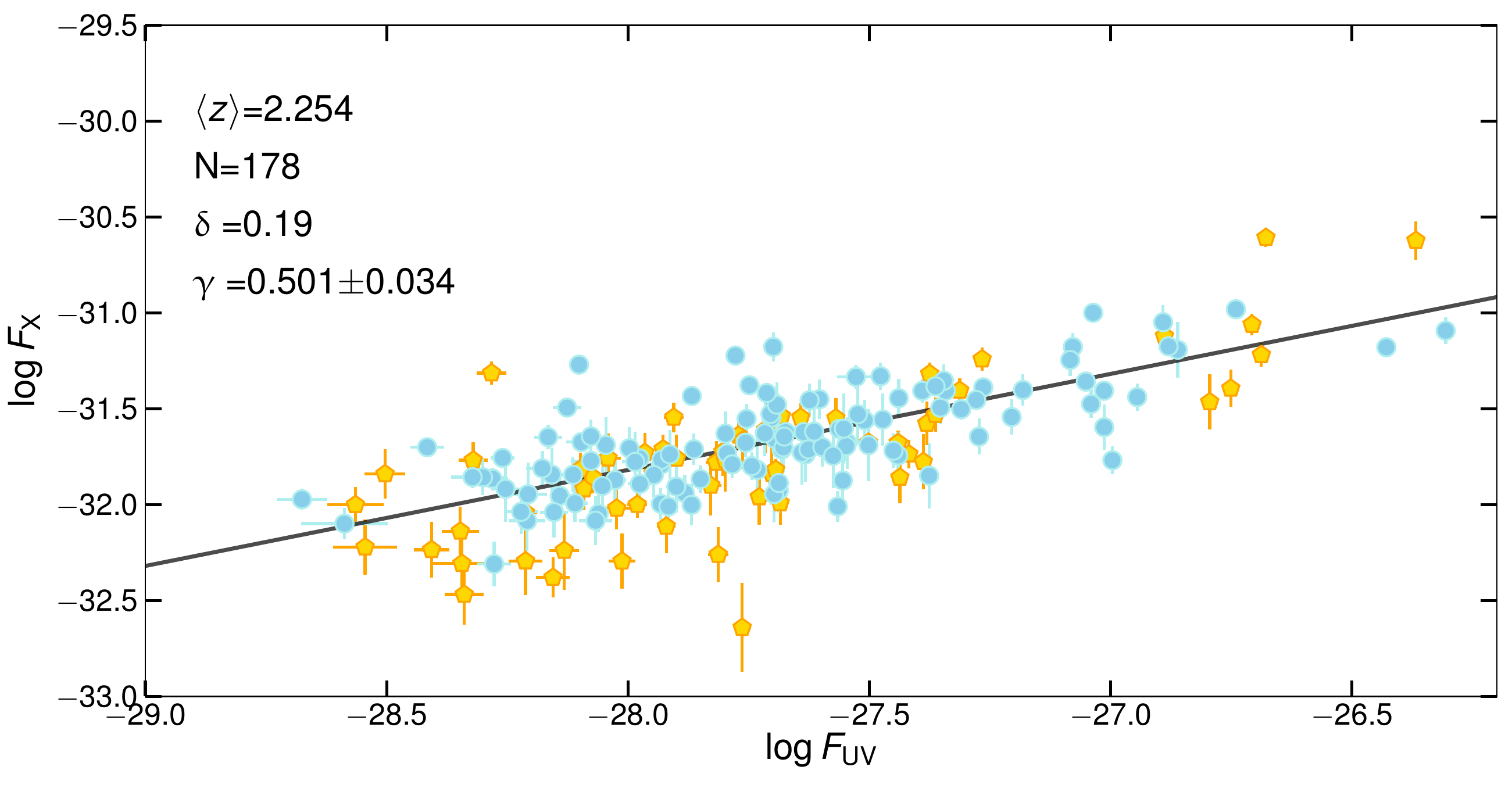}
   \includegraphics[width=0.49\linewidth]{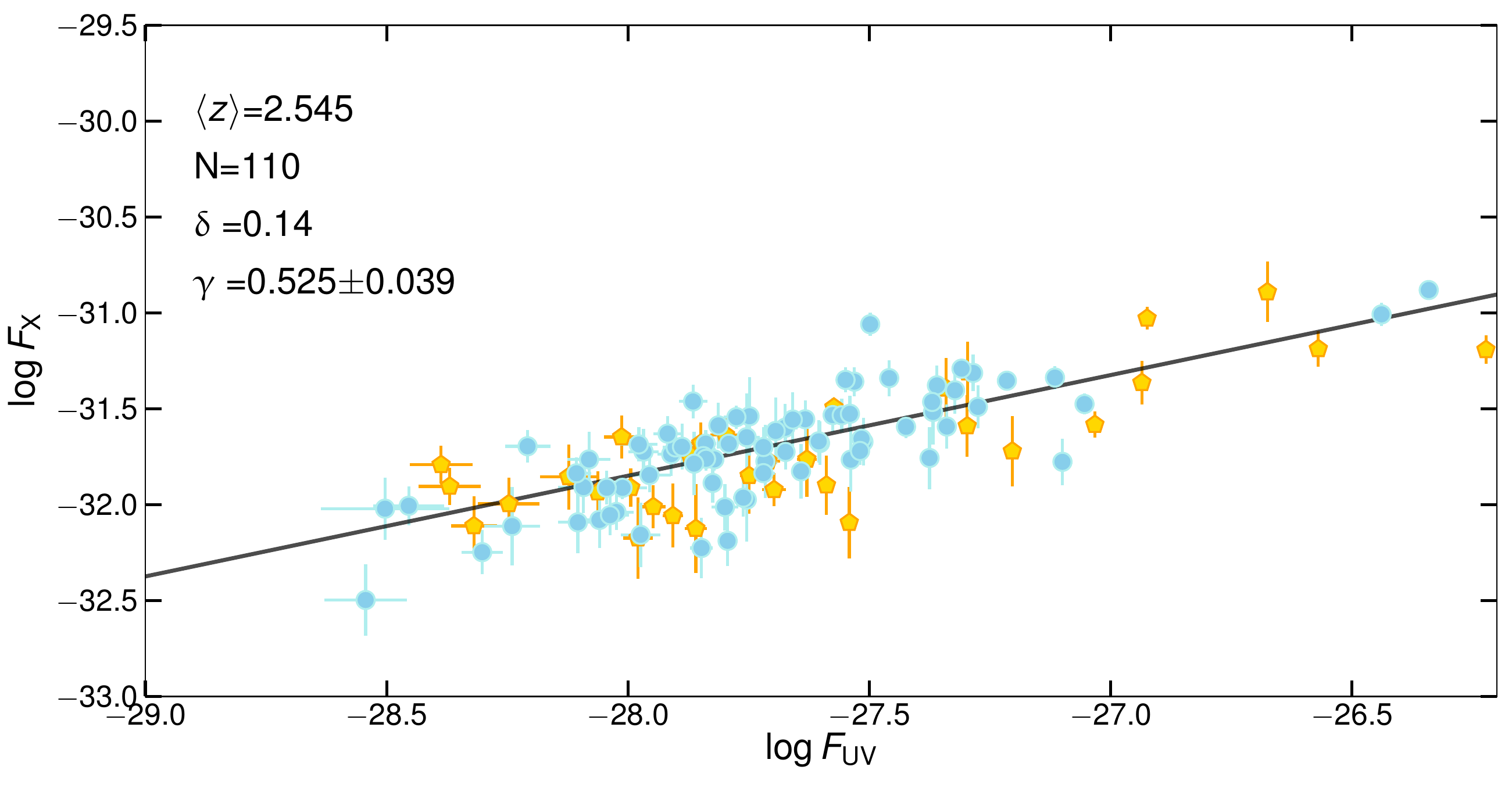}
   \includegraphics[width=0.49\linewidth]{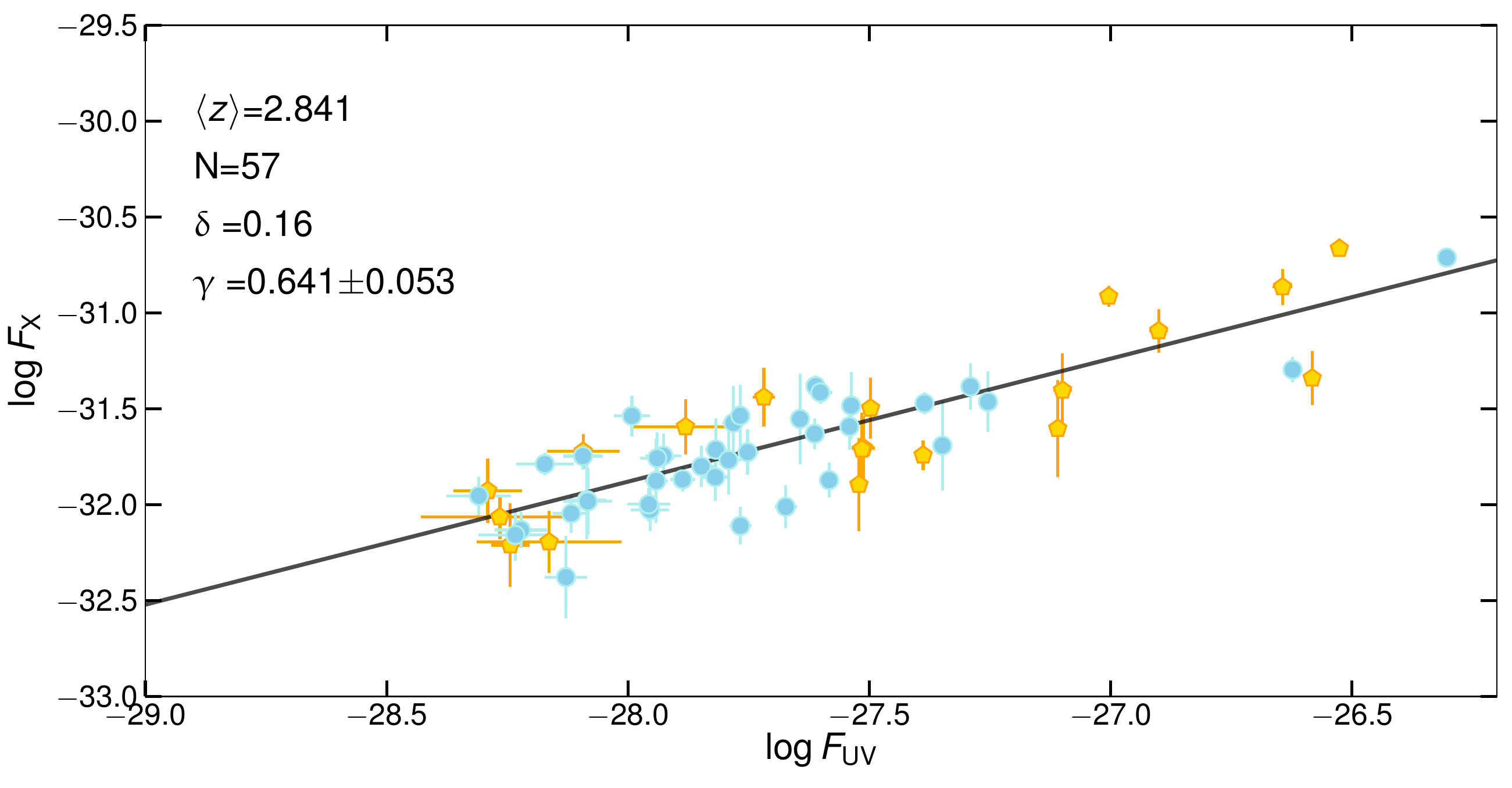}
   \includegraphics[width=0.49\linewidth]{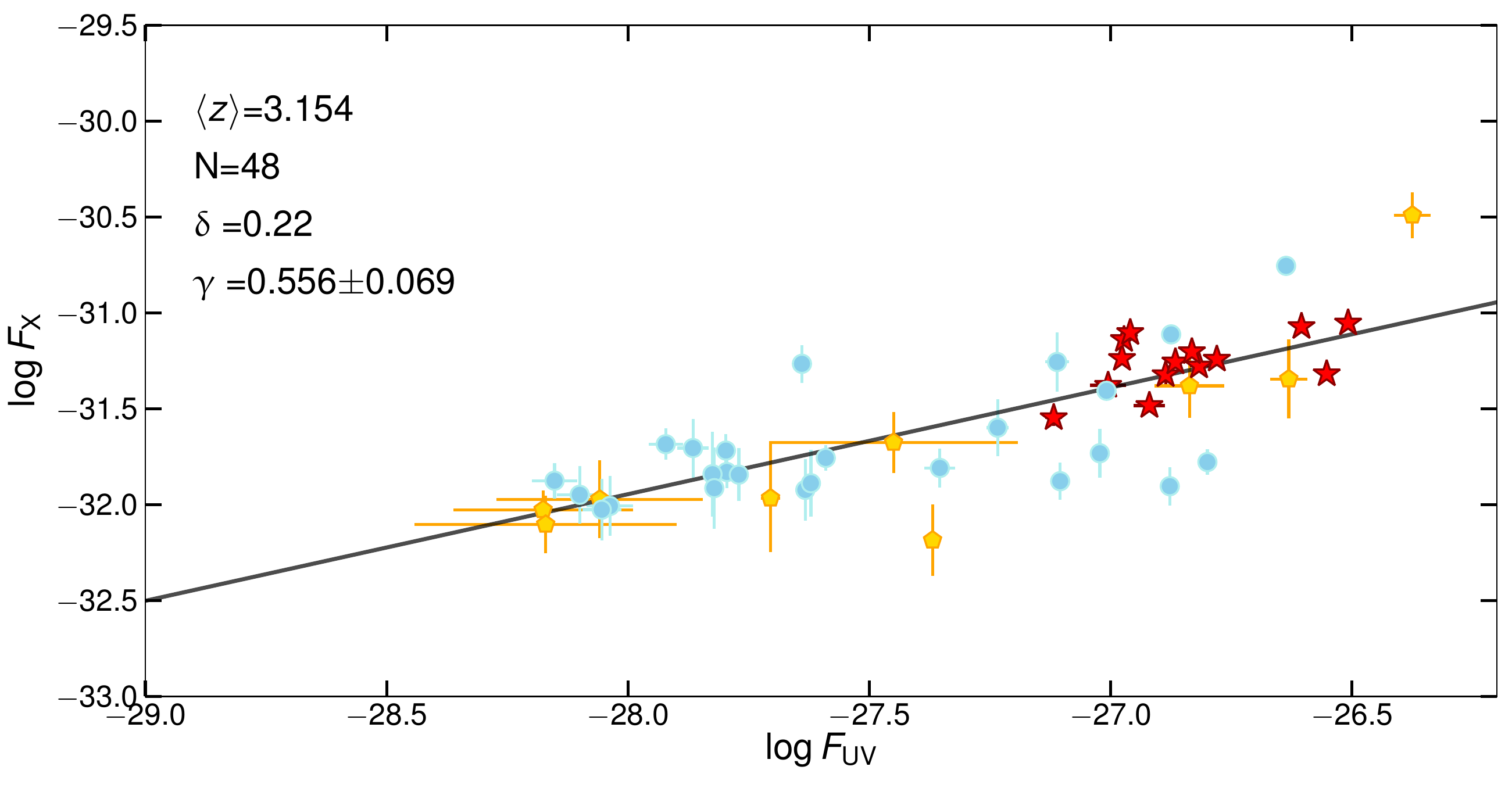}
   \includegraphics[width=0.49\linewidth]{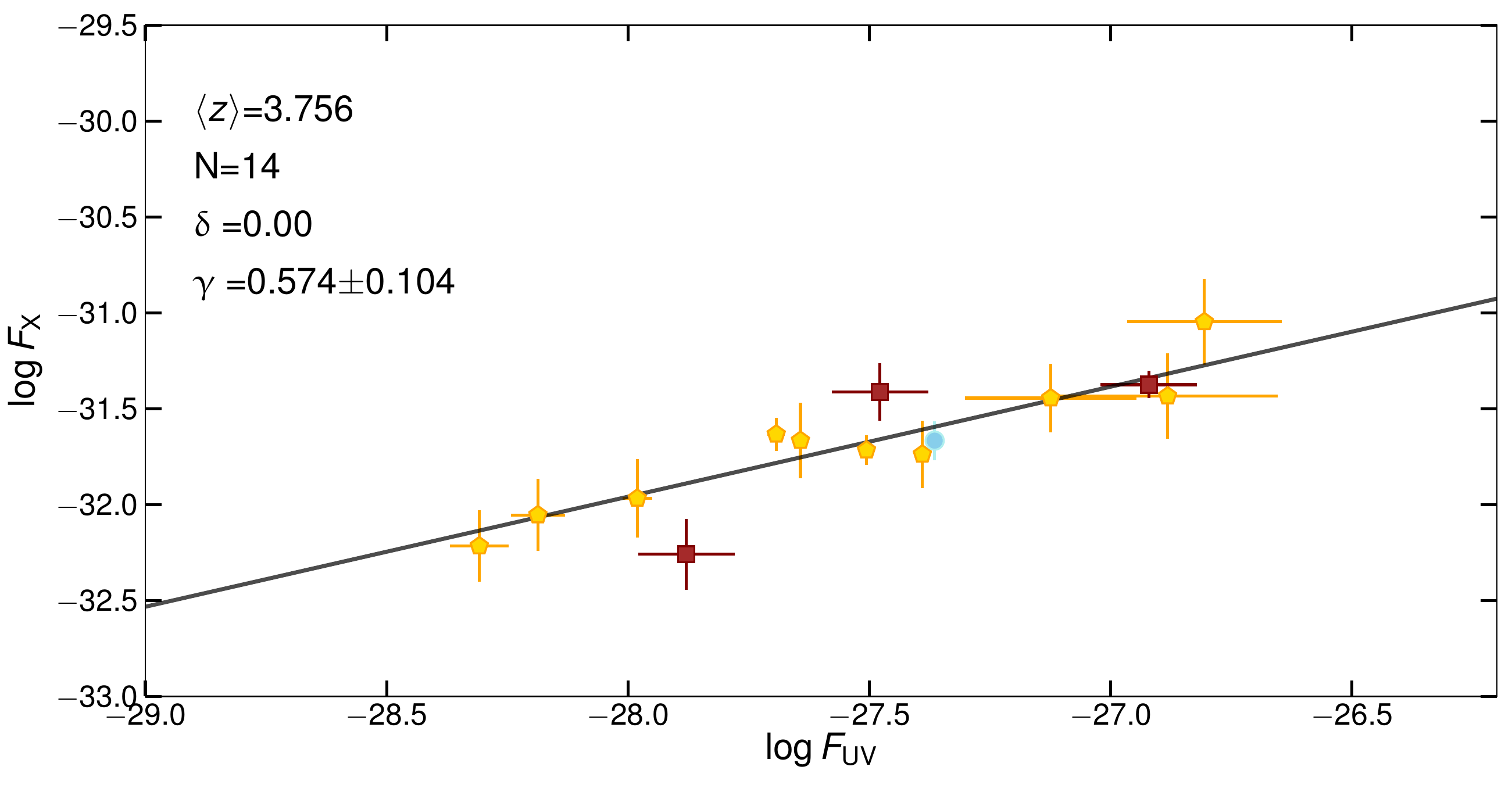}
   \caption{Continued.}
              \label{fig:fxfuv}
    \end{figure*}
\section{Additional cosmological fits of the Hubble diagram}
\label{Additional cosmological fits of the Hubble diagram}
\begin{figure*}
 \centering
  \includegraphics[width=\linewidth,clip]{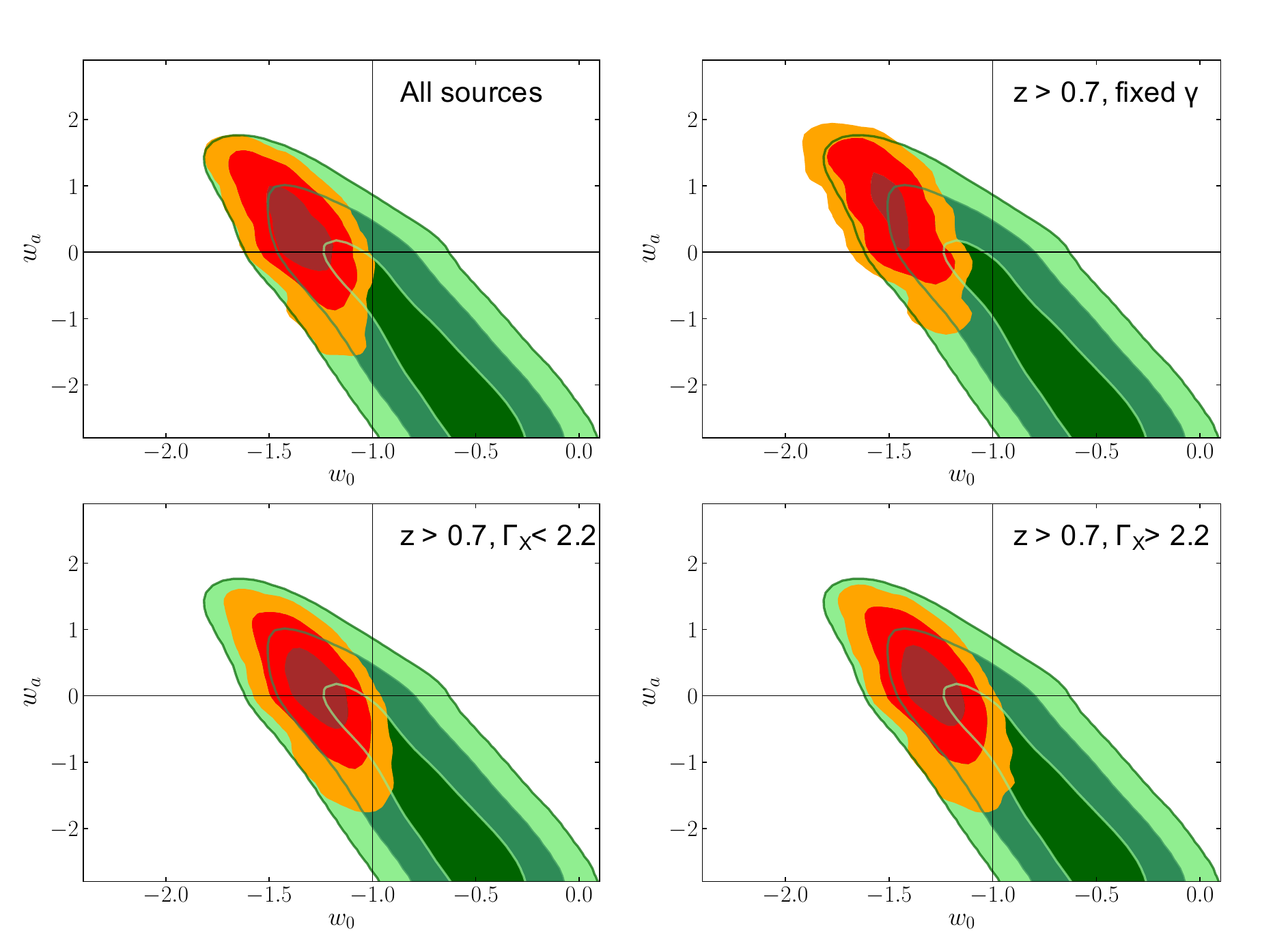}
	\caption{Results from a fit of a $w_0w_a$CDM model to our Hubble diagram of supernovae and quasars in four different cases. Top left: whole quasar sample, with the slope of the $\Lx-\Lo$ relation as a free parameter; top right: quasars at $z>0.7$ with a fixed $\Lx-\Lo$ slope, $\gamma=0.59$, i.e. the average value from the fit of the relation in narrow redshift intervals, as shown in Section~\ref{Analysis of the relation with redshift}; bottom panels: results for two subsamples with $\gammax<2.2$ and $\gammax>2.2$. Contours are at the 1$\sigma$, 2$\sigma$, and 3$\sigma$ confidence levels. The green contours refer to the CMB results from {\it Planck} \citepads{planck2018}. The orange, red, and brown contours are obtained by adding the constraints from the Hubble diagram of supernovae and quasars.}
  \label{hubblediag:appendix}
\end{figure*}
Figure~\ref{hubblediag:appendix} presents the results of the fit with the flat $w_0w_a$CDM model of the quasar sample in four different cases: 
{\it (1)} the full sample of $\sim$2,400 quasars where the slope, $\gamma$, of the $\Fx-\Fo$ relation is a variable parameter, {\it (2)} the ``best'' $z>0.7$ quasar sample with $\gamma$ fixed to the average value of the entire sample estimated in narrow redshift bins (i.e. $\gamma=0.59$, see \S~\ref{Analysis of the relation with redshift}), {\it (3)} the ``best'' $z>0.7$ quasar sample with $\gammax$ lower than the average photon index as discussed in \S~\ref{residuals-gammax} (see Figure~\ref{gammax}) and finally {\it (4)} the ``best'' $z>0.7$ quasar sample with $\gammax$ higher than the average photon index. In both fits with the {\it (3)} and {\it (4)} samples $\gamma$ is left free. This analysis confirms that the $\Lambda$CDM model is in tension with our data at more than 3$\sigma$, in agreement with the results in RL19, irrespectively of the sample selection. 

\end{appendix}
\end{document}